\documentclass[12pt]{article} 
\usepackage{amssymb}
\usepackage{amsmath} 
\usepackage{cite}
\usepackage{axodraw}
\usepackage{hyperref}
\input xy
\xyoption{all}
\usepackage{rotfloat}   

\def\sqr#1#2{{\vcenter{\vbox{\hrule height.#2pt
            \hbox{\vrule width.#2pt height#1pt \kern#1pt
                  \vrule width.#2pt}\hrule height.#2pt}}}}

\def\sqra#1#2#3{{\vcenter{\vbox{\hrule height.#2pt
            \hbox{\vrule width.#2pt height#1pt \kern5pt 
#3
                  \vrule width.#2pt}\hrule height.#2pt}}}}

\numberwithin{equation}{section}

\numberwithin{table}{section}

\begin{document} 

\begin{center}

{\large\bf Topological operators, noninvertible symmetries and decomposition}

\vspace*{0.2in}

Eric Sharpe

Department of Physics MC 0435\\
850 West Campus Drive\\
Virginia Tech\\
Blacksburg, VA  24061

{\tt ersharpe@vt.edu}

\end{center}

In this paper we discuss noninvertible 
topological operators in the context of one-form symmetries 
and decomposition of two-dimensional quantum field
theories, focusing on two-dimensional orbifolds with and without
discrete torsion.
As one component of our analysis, 
we study
the ring of dimension-zero operators 
in two-dimensional
theories exhibiting decomposition.
From a commutative algebra perspective, the rings are naturally associated
to a finite number of points, one point for each universe in the
decomposition.  Each universe is canonically associated to
a representation, which defines a projector,
an idempotent in the ring of dimension-zero operators.
We discuss how bulk Wilson lines act as defects bridging universes,
and how Wilson lines on boundaries of two-dimensional theories
decompose, and compute actions of projectors.
We discuss one-form symmetries of the rings, and related properties.
We also give general formulas for projection operators, 
which
previously were computed on a case-by-case basis.
Finally, we propose a characterization of noninvertible higher-form
symmetries in this context in terms of representations.  In that
characterization, non-isomorphic universes appearing in decomposition
are associated with noninvertible one-form symmetries.

\begin{flushleft}
August 2021
\end{flushleft}

\newpage

\tableofcontents

\newpage

\section{Introduction}

In recent years, there has been a great deal of interest in
more general notions of symmetries,
such as ``one-form symmetries'' and other higher symmetries,
and more recently, ``non-invertible''
symmetries, see 
e.g.~\cite{Komargodski:2020ved,Rudelius:2020orz,Heidenreich:2021tna,Komargodski:2020mxz,Thorngren:2019iar,Ji:2019jhk,Nguyen:2021yld,Nguyen:2021naa,Bhardwaj:2017xup,Brunner:2013ota,Brunner:2013xna,Brunner:2014lua}. 
Briefly, these symmetries are often described by defects such as
topological Gukov-Witten
operators \cite{Gukov:2006jk,Gukov:2008sn}.
It has been noted that all these notions extend ordinary notions of
symmetries, and they have been applied to 
e.g.~the completeness hypothesis \cite{Polchinski:2003bq}.  
In particular, these symmetries arise in discrete gauge theories with
trivially-acting subgroups, and existence of such subgroups is
on its face at odds with the statement that all matter representations
should appear in the spectrum, see 
e.g.~\cite{Craig:2018yvw,Rudelius:2020orz,Heidenreich:2021tna}.

Now,
two-dimensional gauge theories in which a subgroup of the gauge group 
acts trivially
(discussed in \cite{Pantev:2005rh,Pantev:2005wj,Pantev:2005zs})
are examples of theories with (possibly noninvertible)
higher-form symmetries.
One of the most important properties of 
such theories
is decomposition, the statement that these theories 
are equivalent to (``decompose into'')
disjoint unions of other
quantum field theories, known in this context as ``universes.''
This was first described in
\cite{Hellerman:2006zs}, and has since been applied in many different
areas, including 
Gromov-Witten theory 
(see e.g.~\cite{ajt1,ajt2,ajt3,tseng1,gt1,xt1}),
phases of gauged linear sigma models 
(see 
e.g.~\cite{Caldararu:2007tc,Hori:2011pd,Halverson:2013eua,Hori:2013gga,Wong:2017cqs}),
elliptic genera (see e.g.~\cite{Eager:2020rra}),
and 
anomaly resolution in orbifolds (see e.g.~\cite{Robbins:2021lry,Robbins:2021ibx,Robbins:2021xce}).
(See also  
e.g.~\cite{Sharpe:2019ddn,Tanizaki:2019rbk,Cherman:2020cvw,Komargodski:2020mxz,Robbins:2021ylj,Delmastro:2021otj} 
for other recent work.)  

The structure of decomposition is tied to existence of
one-form symmetries.
Briefly, if the universes are copies of one another, one would
naturally label the one-form symmetry invertible,
and if some universes are distinct, then for reasons we shall
discuss, the one-form symmetry could be naturally called
noninvertible.
Studying this solely from the topological Gukov-Witten
operators (dimension-zero twist fields)
can be confusing:  in the
ring of dimension-zero twist
fields, there exist many noninvertible elements, such as  
e.g.~projectors\footnote{
Idempotents built from dimension-zero twist fields that project states
and operators onto corresponding universes.
},
regardless of whether
there is a noninvertible one-form symmetry.
The universes of decomposition are
associated with representations of the trivially-acting subgroup,
and we will propose that those representations can be used to
distinguish invertible from noninvertible one-form symmetries.
If the representation is
nontrivial and one-dimensional, then it signals existence of an 
(invertible) one-form
symmetry and, hand-in-hand, multiple identical universes.
If the representation associated with a universe is of higher dimension,
then it indicates a noninvertible symmetry, and,
except in special cases, implies that the corresponding universe is
distinct.

We give a couple of examples in orbifolds (without
discrete torsion) below:
\begin{enumerate}
\item 
If the trivially-acting normal subgroup lies within
the center (so that one has an ordinary one-form symmetry), 
the theory decomposes
into copies of one other theory (with possible theta angle or
$B$ field shifts).
For example, consider an orbifold $[X/{\mathbb Z}_n]$ where
the ${\mathbb Z}_n$ acts trivially on $X$.  In this case, decomposition says
\begin{equation}
{\rm QFT}\left( [X/{\mathbb Z}_n] \right) \: = \:
{\rm QFT}\left( \coprod_n X \right).
\end{equation}
Both sides have a $B {\mathbb Z}_n$ (one-form) symmetry, the
dimension-zero symmetry generators (twist fields) associated to elements of
the trivially-acting
${\mathbb Z}_n$ are invertible, and the representations of
${\mathbb Z}_n$ associated to each copy of $X$ are one-dimensional.
\item If the trivially-acting normal subgroup does not
lie within the center,  
then in general
the theory decomposes into multiple different theories,
and there are often noninvertible 
topological Gukov-Witten operators (dimension-zero twist fields)
associated
to conjugacy classes.
For example, consider an orbifold $[X/{\mathbb H}]$ by the
eight-element group of quaternions ${\mathbb H}$,
in which the subgroup $\langle i \rangle \cong {\mathbb Z}_4$ acts
trivially.  In this case, decomposition says
\cite[section 5.4]{Hellerman:2006zs}
\begin{equation}
{\rm QFT}\left( [X/{\mathbb H}] \right) \: = \:
{\rm QFT}\left( X \, \coprod \, [X/{\mathbb Z}_2] \, \coprod \,
[X/{\mathbb Z}_2] \right).
\end{equation}
Both sides have a $B {\mathbb Z}_2$ symmetry, but not a 
$B {\mathbb Z}_4$, and instead of four copies of a theory, one gets
two copies of one theory plus one more distinct theory.
Furthermore, one of the dimension-zero twist fields associated to
a conjugacy class is noninvertible, and the representation of
${\mathbb Z}_4$ associated to $X$ is two-dimensional.
\end{enumerate}

We will also see decomposition reflected explicitly in the
rings of dimension-zero operators 
(linear combinations of the twist fields associated to conjugacy
classes).
These rings possess projection operators, projecting 
onto one-dimensional spaces associated with the distinct universes.
Furthermore, these (commutative) rings also have a geometric interpretation
via commutative algebra.
They describe complete intersections of hypersurfaces
in vector spaces, which are a set of
points -- one point for each universe -- located on a space of order
parameters\footnote{
To be clear, decomposition is not spontaneous symmetry breaking,
and universes are not superselection sectors,
although we borrow the language of `order parameters' here.
See e.g. \cite{Tanizaki:2019rbk} for a recent detailed discussion of this point.
}, corresponding to vevs of dimension-zero twist fields.  
We will explore formal properties
of those rings, and also describe residual one-form symmetry actions on
these rings.

We begin in section~\ref{sect:genl} by outlining general aspects of
rings of dimension-zero operators in orbifolds with
trivially-acting normal subgroups of the orbifold group.
After reviewing decomposition and its relation to properties of the
trivially-acting subgroup, we compute the fusion algebra of 
dimension-zero  
twist fields associated to conjugacy classes, give a systematic construction
of projectors onto universes of decomposition, and discuss the geometry
implicit in the commutative rings of dimension-zero operators,
remarking on the semilocality and semisimpleness of those rings.  
We also discuss
both bulk and boundary Wilson lines (Chan-Paton factors)
and how projectors act on them.  In particular, mathematically
bundles and sheaves on gerbes decompose into bundles and sheaves on
universes of decomposition, and by explicitly computing the action of
projectors on boundary Wilson lines, we can recover that mathematical
decomposition of bundles and
sheaves explicitly.  We also discuss one-form symmetries
(invertible and noninvertible)
in this context, and subtleties in characterizing the presence of 
noninvertible one-form symmetries from the ring of dimension-zero
twist fields  We propose a characterization in terms of the representations
associated to the universes.

In section~\ref{sect:ex:disjoint} we study the example of a sigma model
on a disjoint union.  This is not (presented as) a gauge theory,
but it is a prototype for the results of decomposition, and in particular
has an invertible one-form symmetry.  As a result, it has a ring of
dimension-zero operators, which include projection operators as part
of a more general subspace of noninvertible dimension-zero operators.

In section~\ref{sect:ex:orbs} we discuss several examples in orbifolds.
In subsection~\ref{sect:bandedex}, we discuss orbifolds with a 
trivially-acting central
${\mathbb Z}_n$  -- banded abelian gerbes, with a $B {\mathbb Z}_n$
symmetry.  These are physically
equivalent to sigma models on disjont unions as above, and we discuss
the rings of dimension-zero operators in greater detail.
In subsection~\ref{sect:ex:nonband:z4}, we discuss an orbifold with
a trivially-acting non-central ${\mathbb Z}_4$.  Here the (invertible)
one-form
symmetry is only $B {\mathbb Z}_2$, not $B {\mathbb Z}_4$, and the
decomposition is more complicated than just copies of one space,
which is reflected in the existence of a noninvertible symmetry.
We discuss the ring of dimension-zero operators in detail, explicitly
construct projectors, explicitly compute the action of projectors on
Wilson lines, and so forth.
In section~\ref{sect:ex:nonbanded:z2z2}, we discuss an orbifold with
a trivially-acting non-central ${\mathbb Z}_2 \times {\mathbb Z}_2$,
which we analyze in the same fashion.
In this example there is no invertible one-form symmetry, only
noninvertible symmetries.
In section~\ref{sect:ex:nonabelian:d4}, we discuss an orbifold with
a trivially-acting nonabelian group, analyzing its ring of dimension-zero
operators.  In section~\ref{sect:ex:nonabelian:d4dt} we turn on
discrete torsion in the same example, and study its effects.  
Briefly, turning on discrete torsion has the effect of complicating the
dictionary between one-form symmetries and invertible operators.

In section~\ref{sect:ex:gauge} we study two-dimensional supersymmetric
gauge theories with trivially-acting subgroups, specifically,
a family of analogues of the supersymmetric ${\mathbb P}^n$ model
and their mirrors.  Since the trivially-acting subgroup is in the center,
these theories have full one-form symmetries.  As in other cases, we discuss
rings of dimension-zero operators.

In section~\ref{sect:4d}, we discuss analogous phenomena in four-dimensional
theories exhibiting decomposition.

In appendix~\ref{app:casimirs} we work out technical results concerning
the action of twist fields on Wilson lines.
In appendix~\ref{app:char} we collect several character identifies
regarding orthogonality and normalization of characters, which are used
elsewhere in the text.
In appendix~\ref{app:induced} we use induced representations to give some
technical results on restrictions and extensions of representations of
finite groups to subgroups.
Finally, in appendix~\ref{app:miscgpcoh} we collect some miscellaneous
results on group cohomology, to make this paper's discussion of discrete
torsion self-contained.

In this paper we focus on orbifolds with and without
discrete torsion.  Now, it was recently argued in
\cite{Robbins:2021lry,Robbins:2021ibx,Robbins:2021xce} that
two-dimensional orbifold in which a subgroup of the orbifold group
acts trivially admit additional modular-invariant degrees of freedom,
beyond discrete torsion, which were labelled ``quantum symmetries.''
A version of decomposition for orbifolds with quantum symmetries
was discussed in \cite{Robbins:2021lry,Robbins:2021ibx,Robbins:2021xce},
and we leave further details of quantum symmetries to future work.

\section{General aspects}
\label{sect:genl}

In this section we discuss several general features of rings of
dimension-zero operators appearing in two-dimensional quantum field theories
(typically with one-form symmetries), Wilson lines, symmetries,
and the relation to decomposition.  In subsection~\ref{sect:dictionary}
we review decomposition; in subsection~\ref{sect:rings} we discuss
the rings of dimension-zero operators:  computation of the fusion algebra,
the basis of projectors, and the geometry.  In subsection~\ref{sect:Wlines}
we discuss the action of those dimension-zero operators on Wilson lines
at the boundary.  Finally, in section~\ref{sect:symm} we discuss
one-form and noninvertible symmetries of decomposition
and their realization in terms of
the dimension-zero operators and Wilson lines.
In later sections we discuss specific examples.

\subsection{Dictionary and decomposition}
\label{sect:dictionary}

It will be helpful to begin by correlating math and physics nomenclature.
To that end, let us quickly review.
Orbifolds and gauge theories with trivially-acting normal subgroups
(and hence one-form symmetries)
were first discussed in
\cite{Pantev:2005rh,Pantev:2005wj,Pantev:2005zs}, as part of a program
of making sense of string compactifications on certain generalizations of
manifolds known as ``stacks.''  Briefly, gauge theories correspond to
stacks, and gauge theories with trivially-acting
subgroups correspond to special stacks known as ``gerbes,''
essentially, fiber bundles whose fibers are `groups' of one-form
symmetries.  As a result of that geometry, a sigma model on a gerbe
admits a one-form symmetry (or a noninvertible analogue), corresponding
to translations along the fibers.

Such gauge theories have a natural classification:
\begin{enumerate}
\item Cases in which the trivially-acting normal subgroup lies within the
center of the gauge group.  In these cases, at least in the absence
of twisting, the trivially-acting subgroup defines a(n invertible)
one-form symmetry.
\item Cases in which the trivially-acting normal subgroup is abelian,
but does not lie within the center of the gauge group.
Here, the one-form symmetry is obstructed, and as we shall see in
examples, this
leads to what we will propose to label noninvertible one-form symmetries.
\item Cases in which the trivially-acting normal subgroup is nonabelian.
Here, one does not expect a one-form symmetry at all, in the absence
of twisting, unless of course that nonabelian group has a center.
Here, again, one gets invertible symmetries.
\end{enumerate}
These cases have a mathematical understanding, in terms of a classification
of gerbes.  As discussed in 
e.g.~\cite{Pantev:2005rh,Pantev:2005wj,Pantev:2005zs},
the first case corresponds to
`banded abelian' gerbes; the second, to `nonbanded abelian' gerbes;
and the third, to nonabelian gerbes.
We summarize this dictionary in table~\ref{table:dict}.

\begin{table} 
\begin{center}
\begin{tabular}{c|c}
Math & Physics \\ \hline
Stack & Gauge theory \\
Gerbe & Gauge theory with trivially-acting normal subgroup \\
Banded abelian gerbe & Trivially-acting subgroup is in center \\
Nonbanded abelian gerbe & Trivially-acting subgroup abelian but not in center \\
Nonabelian gerbe & Trivially-acting subgroup not abelian
\end{tabular}
\caption{A dictionary \cite{Pantev:2005wj,Hellerman:2006zs}
between math and physics language for descriptions
of two-dimensional gauge theories (including orbifolds).
\label{table:dict}
}
\end{center}
\end{table}

In most of this paper, we will focus on orbifolds, so let us elaborate
for such theories.  Given an orbifold by a finite group $\Gamma$, say,
with trivially-acting subgroup $K \subset \Gamma$, the orbifold is
physically equivalent to a disjoint union, with components 
of the disjoint union
determined as follows.

Let $\hat{K}_{\iota^* \omega}$ denote the set of isomorphism classes of
irreducible projective representations of $K$, twisted by the restriction
of discrete torsion $\omega \in H^2(\Gamma,U(1))$ to $K$ (along the inclusion
map $\iota: K \hookrightarrow \Gamma$).  The number of components
of 
the orbifold $[X/\Gamma]_{\omega}$ is given by the number of orbits of
the action of $G = \Gamma/K$ on $\hat{K}_{\iota^* \omega}$.
We define the action as follows \cite{Hellerman:2006zs,Robbins:2021ylj}.
Given an isomorphism class $[ \rho ] \in \hat{K}_{\iota^* \omega}$
and $q \in G$, we take
\begin{equation}
q \cdot [ \rho ] \: = \: [ L_q \rho ]
\end{equation}
where $L_q \rho$ is another irreducible projective representation of
$K$, also twisted by $\iota^* \omega$, defined by
\begin{equation}
(L_q \rho)(k) \: = \: \frac{ \omega( s(q)^{-1} k, s(q) ) }{
\omega(s(q), s(q)^{-1} k) } \rho( s(q)^{-1} k s(q) )
\end{equation}
for $k \in K$, $s: G \rightarrow \Gamma$ a section, and products
are taken in $\Gamma$.  The various components (corresponding to orbits) 
may also
have discrete torsion, determined as described in
\cite{Hellerman:2006zs,Robbins:2021ylj}.

Briefly, the orbifold
$[X/\Gamma]_{\omega}$ is equivalent to a disjoint union of theories,
one for each
orbit of $G$ on $\hat{K}_{\iota^* \omega}$,
with the theory corresponding to a given orbit being an orbifold of 
$X$ by the stabilizer of the orbit, with some discrete torsion
given by an analysis described elsewhere.
This is known as ``decomposition,'' and the various components of the
disjoint union, corresponding to orbits of the $G$
action on $\hat{K}_{\iota^* \omega}$, are known
as ``universes.''  (See \cite{Hellerman:2006zs,Robbins:2021ylj} for
numerous additional details.)

There also exist more modular-invariant phases than just discrete torsion
in the case a nontrivial subgroup of $\Gamma$ acts trivially;
see \cite{Robbins:2021lry,Robbins:2021ibx,Robbins:2021xce} 
for more information on such orbifolds and their corresponding
decomposition.  We leave
such more general theories for future work.

In the special case that
the $\Gamma$ orbifold does not contain any discrete torsion ($\omega=0$),
then decomposition reduces to \cite{Hellerman:2006zs}
\begin{equation}
{\rm QFT}\left( [X/\Gamma] \right) \: = \:
{\rm QFT}\left( \left[ \frac{X \times \hat{K} }{G} \right]_{\hat{\omega}}
 \right),
\end{equation}
where $\hat{K}$ is the set of
ordinary irreducible representations of $K$, and $\hat{\omega}$ is some possible
discrete torsion, which is described in detail in \cite{Hellerman:2006zs}.

If in addition 
$K$ is in the center of $\Gamma$, so that the $\Gamma$ orbifold has
a $BK$ (one-form) symmetry, then the action of $G$ on $\hat{K}$ is
trivial, and the theory decomposes into a disjoint union of identical
copies (modulo choices of discrete torsion $\omega$):
\begin{equation}
{\rm QFT}\left( [X/\Gamma] \right) \: = \: 
{\rm QFT}\left( \coprod_{ k \in \hat{K} } [X/G]_{\hat{\omega}(k)} \right).
\end{equation}
Here, each $[X/G]_{\hat{\omega}(k)}$ is a universe, so-named because
in a string compactification on such an orbifold, each
summand $[X/G]_{\hat{\omega}(k)}$ would give rise to its own separate
low-energy theory.

If $K$ is not central (but $\omega$ still vanishes), 
then instead of just copies of $[X/G]$, one will
get various covers of $[X/G]$.  (The sum of the number of copies and the
orders of the covers will always equal the number of elements in
$\hat{K}$.)

For example, as we shall see in section~\ref{sect:ex:nonband:z4},
if $\Gamma = {\mathbb H}$, the eight-element group of unit quaternions,
with trivially-acting subgroup $K = \langle i \rangle \cong {\mathbb Z}_4$,
then
\begin{equation}
{\rm QFT}\left( [X/{\mathbb H}] \right) \: = \:
{\rm QFT}\left( X \, \coprod \, [X/{\mathbb Z}_2] \, \coprod \,
[X/{\mathbb Z}_2] \right).
\end{equation}

If in the decomposition, there are $n$ universes which are identical
(up to choices of discrete torsion / $B$ fields / theta angles),
then the theory has a $B {\mathbb Z}_n$ symmetry.  If some of the universes
are different, then the theory has a noninvertible symmetry.
We can make this more precise as follows.  Each universe is associated
to an orbit of the $G$ action on $\hat{K}_{\iota^* \omega}$, the set of
isomorphism classes of $\iota^* \omega$-projective irreducible representations
of $K$.  Given an orbit, if we let $R_1, \cdots, R_{\ell}$ denote
representations of each isomorphism class appearing in the orbit, then
we can associate the universe to the representation
\begin{equation}
R \: = \: R_1 \oplus \cdots \oplus R_{\ell},
\end{equation}
and so we see that we can also associate universes to (isomorphism classes of)
(possibly projective) representations
of the trivially-acting part of the
gauge group.
As we will see in examples, universes related by a
$B {\mathbb Z}_n$ symmetry are associated to one-dimensional irreducible
representations $R$.  Universes associated to a higher-rank representation
$R$ are related by a noninvertible symmetry.

In the rest of this paper we will elaborate on that dictionary.
We will look at the structure of both the ring of dimension-zero operators
as well as its action on Wilson lines and the corresponding symmetries.

\subsection{Ring of dimension-zero operators}
\label{sect:rings}

Consider an orbifold $[X/\Gamma]$, perhaps with discrete torsion,
in which a normal subgroup $K \subset \Gamma$ acts trivially.  
In this section we will study the ring of dimension-zero operators.
We will
describe two
bases for the dimension-zero operators:
\begin{itemize}
\item Twist fields, constructed from ($\omega$-compatible) conjugacy
classes of trivially-acting group elements, and their products (forming a 
fusion algebra)
in subsection~\ref{app:computation},
\item Projectors, corresponding to (projective) representations
of the trivially-acting subgroup $K$ of the gauge group, in
subsection~\ref{sect:projectors}. These project onto states and operators
in the universes of decomposition, and we will give
general expressions.
\end{itemize}
Mathematically, the vector space of dimension-zero operators is the
center of the (twisted) group algebra, and it is a standard result that the
twist field and projector constructions we will review here both form
bases for that space.  Essentially as a result of the existence of the
(complete orthogonal) set of projectors, the vector space of dimension-zero
operators is equivalent to ${\mathbb C}^n$, where $n$ is the number of
universes in the decomposition. 

In subsection~\ref{sect:geom},
we will also describe the corresponding geometry:
each ring $R$ of dimension-zero operators will have the form
\begin{equation}
{\mathbb C}[x_0, \cdots, x_m] / I,
\end{equation}
which describes a complete intersection in ${\mathbb C}^{m+1}$,
a space of order parameters,
and that complete intersection will always be a set of points,
essentially because of the existence of the projectors.

\subsubsection{Twist fields and their products}
\label{app:computation}

In this section, we describe the topological (dimension zero) twist fields
and their products\footnote{
In principle, a mathematically rigorous description of such products
in considerably greater generality can be found in
\cite{Chen:2000cy}.  In this section we describe only the special case
of
trivially-acting group elements.
}.

Consider an orbifold $[X/\Gamma]$ in which a normal subgroup
$K \subset \Gamma$ acts trivially.
Suppose that there is discrete torsion in the $\Gamma$ orbifold, represented
by a cocycle $\omega$.  Formally, let $\tau_g$ denote the operator
associated to $g \in K$.  (Formally, $\tau_g$ 
generates a branch cut or topological defect line in the sense of
e.g. \cite{Frohlich:2006ch,Chang:2018iay,yujitasi2019}.  It need not be a twist field
in the usual sense, as it is not gauge invariant and
does not commute in general.)
These operators obey a multiplication given
by
\begin{equation} \label{eq:tau-mult}
\tau_g \tau_h \: = \: \omega(g,h) \tau_{gh}.
\end{equation}
Associativity is guaranteed by the cocycle condition on $\omega$.
Rescaling the $\tau$'s alters $\omega$ by a coboundary,
so this product structure is naturally associated to the cohomology
class of $\omega$, rather than $\omega$ itself.
(Technically, for any group $G$,
the linear combinations of $\tau_g$ for $g \in G$ with this
multiplication define a twisted group algebra, denoted
${\mathbb C}[G]_{\omega}$, see e.g. \cite{conlon}.)

Now, not all of the $\tau_g$ are
relevant:
only conjugation-invariant combinations contribute to the
pertinent dimension-zero operators, which we label $\sigma_{[g]}$,
where $[g]$ denotes a conjugacy class of $g$ (in $\Gamma$).  
The $\sigma_{[g]}$ are what one would ordinarily call twist fields.
The algebra of
$\sigma$'s is naturally commutative, which at root follows from
the observation:
\begin{equation}
g h \: = \: (g h g^{-1} ) g.
\end{equation}
The left-hand-side is a representative of the product $[g][h]$,
and the right-hand-side is a representative of the product $[h][g]$,
hence one expects
\begin{equation} 
\sigma_{[g]} \sigma_{[h]} \: = \: \sigma_{[h]} \sigma_{[g]},
\end{equation}
so that the $\sigma$'s are commutative, whereas the $\tau$'s are not.
(Technically, just as the $\tau_g$ generate the twisted group
algebra, the $\sigma_{[g]}$ generate the center of the
twisted group algebra, see e.g. \cite[section 6.3]{serrerep}.)

In detail, under conjugation by an element $h \in \Gamma$,
\begin{eqnarray}
\tau_g & \mapsto & \tau_h \tau_g \tau_{h^{-1}}
\: = \: \tau_h \left( \omega(g,h^{-1}) \tau_{g h^{-1}} \right),
\\
& & \: = \: \omega(g, h^{-1}) \omega(h, g h^{-1}) \tau_{h g h^{-1}},
\\
& &  \: = \: \omega(h,g) \omega(h g, h^{-1}) \tau_{h g h^{-1}},
\end{eqnarray}
using the cocycle condition.
(Note that since $K$ is normal in $\Gamma$,
$h g h^{-1} \in K$ for all $g \in K$ and $h \in \Gamma$.)
We assume the cocycle is normalized\footnote{
See appendix~\ref{app:miscgpcoh}.
}
so that
\begin{equation}
\omega(1,g) \: = \: 1 \: = \: \omega(g,1),
\end{equation}
\begin{equation}
\omega(g,g^{-1}) \: = \: 1 \: = \: \omega(g^{-1},g).
\end{equation}
The $\sigma_{[g]}$ are then conjugation-invariant combinations of the
$\tau_{g'}$ for $g'$ conjugate to $h$.

Furthermore, the reader should note that although the dimension-zero
fields arise from elements of $K$, we need to consider conjugation by
elements of $\Gamma \supset K$.  Therefore, we will discuss conjugacy classes
in $\Gamma$, such that the conjugacy classes contain elements in $K$ only.

In addition, we will see that only certain conjugacy classes in $\Gamma$
contribute.  Beyond restricting to conjugacy classes whose
elements are in $K$, we will also restrict to conjugacy
classes (in $\Gamma$) of ``$\omega$-regular'' elements (of $K$).  An $\omega$-regular
element of $\Gamma$ is defined to be an element such that, for all
$h \in \Gamma$ commuting with $g$,
\begin{equation}
\omega(g,h) \: = \: \omega(h,g).
\end{equation}

For any $\omega$-regular conjugacy class $[g]$ whose elements all lie in
$K \subset \Gamma$, 
one defines \cite[section 3]{cheng}
\begin{equation}  \label{eq:sigma-defn}
\sigma_{[g]} \: = \: \frac{1}{|\Gamma|} \sum_{h \in \Gamma}
\frac{\omega(h,g) \omega(hg,h^{-1}) }{ \omega(h,h^{-1}) }
\tau_{h g h^{-1}}.
\end{equation}
This is a dimension-zero twist field.
(In fact, in mathematics, this is a standard construction of one of two
bases for the center of the twisted group algebra, as discussed in
e.g. \cite{cheng}.)
In the case that discrete torsion is trivial, this reduces to
\begin{equation} \label{eq:sigma-defn-nodt}
\sigma_{[g]} \: = \: \frac{1}{|[g]|} 
\sum_{h \in [g]} \tau_h.
\end{equation}
(The reader should note that additional twist fields of nonzero dimension
may exist in a theory, but for our purposes we are only interested in
topological operators.)

The factors of $\omega$ make the fact that $\sigma$'s commute with
$\tau$'s more obscure, so let us check that more carefully.
\begin{eqnarray}
\sigma_{[g]} \tau_y & = & \frac{1}{|\Gamma|}
\sum_{h \in \Gamma} \frac{ \omega(h,g) \omega(hg,h^{-1}) }{ \omega(h,h^{-1}) }
\tau_{hgh^{-1}} \tau_y,
\\ 
& = & \frac{1}{|\Gamma|}
\sum_{h \in \Gamma}
\frac{ \omega(h,g) \omega(hg,h^{-1}) \omega(hgh^{-1},y)
}{ \omega(h,h^{-1}) } \tau_{hgh^{-1}y},
\\
& = & \frac{1}{|\Gamma|}
\sum_{h \in \Gamma}
\frac{
\omega(h,g) \omega(hg,h^{-1}) \omega(hgh^{-1},y)
}{
\omega(y,y^{-1}hgh^{-1}y) 
\omega(h,h^{-1})
}
\tau_y \tau_{y^{-1}hgh^{-1}y}.
\end{eqnarray}
It can be shown that
\begin{equation}
\frac{
\omega(h,g) \omega(hg,h^{-1}) \omega(hgh^{-1},y)
}{
\omega(y,y^{-1}hgh^{-1}y) \omega(h^{-1},h)
}
\: = \: 
\frac{
\omega(y^{-1} h, g) \omega(y^{-1} h g, h^{-1} y)
}{
\omega(h^{-1}y, y^{-1} h)
},
\end{equation}
by multiplying in
\begin{equation} 
\frac{
(d\omega)(y,y^{-1},hgh^{-1}y) (d\omega)(hg, h^{-1}, y)
(d\omega)(h^{-1},y,y^{-1}h)
}{
(d\omega)(y^{-1},hg,h^{-1}y) (d\omega)(y^{-1},h,g)
(d\omega)(y,y^{-1},h)
},
\end{equation}
hence
\begin{equation}
\sigma_{[g]} \tau_y \: = \: \tau_y \sigma_{[g]},
\end{equation}
as claimed.

Thus, we see that $\sigma_{[g]}$ is central in the twisted group algebra,
and in fact, it can be shown \cite[section 3]{cheng} that the collection of
$\sigma_{[g]}$ for $\omega$-regular conjugacy classes (in $\Gamma$, for
classes whose elements lie in $K$)
form a basis for the center of the twisted group algebra.
The reader should note that the number of irreducible projective
representations matches the number of $\omega$-regular conjugacy
classes \cite{cheng,costache,karpilovsky,schur1,schur2,schur3}.  
There is not a canonical isomorphism between
the two, but the number of elements of each set is the same.

The definition of $\sigma_{[g]}$ above is slightly sensitive to the
choice of representative $g$ of the conjugacy class:  
as described in \cite[remark 3.2]{cheng}, changing it will multiply
$\sigma_{[g]}$ by a constant factor.  Specifically, if $g' = h g h^{-1}$
is another representative of the same conjugacy class, then
as shown in \cite[section 3]{cheng},
\begin{equation}  \label{eq:sigma-var}
\sigma_{[hgh^{-1}]} \: = \: 
\frac{ \omega( gh^{-1}, h) }{ \omega(h, gh^{-1} ) }
\sigma_{[g]}.
\end{equation}

Now, let us make some general observations about the conjugacy classes.
First, we will see in examples below that if one drops the
$\omega$-regularity condition, the conjugacy class will not contribute:
the various terms in the sum~(\ref{eq:sigma-defn}) 
will appear with different signs and
cancel out.

Note that since $\tau_1 \mapsto \omega(g,g^{-1}) \tau_1$ under
conjugation by $g$, and we have normalized so that $\omega(g,g^{-1}) = 1$,
the identity should always contribute, and indeed, thanks to the
normalization condition, the identity is always an
$\omega$-regular element.

We have restricted to $\omega$-regular conjugacy classes, and mentioned
that for other conjugacy classes, the terms in~(\ref{eq:sigma-defn}) may
cancel out.  As a consistency check, let us verify that the same does not
happen in $\omega$-regular conjugacy classes.
To see this, note that if $h$ commutes with $g$, then
from our expression above, under conjugation by $h$,
\begin{equation}
\tau_g \: \mapsto \:  \omega(h,g) \omega(h g, h^{-1}) \tau_g.
\end{equation}
However, from the cocycle condition for $(\delta \omega)(g, h, h^{-1})$
and the normalization condition, we have
\begin{equation}
\omega(gh, h^{-1}) \omega(g,h) \: = \: 1,
\end{equation}
hence
\begin{eqnarray}
\omega(h,g) \omega(h g, h^{-1})
& = &
\omega(h,g) \omega(g h, h^{-1}), 
\\
& = &
\frac{ \omega(h,g) }{ \omega(g,h) },
\end{eqnarray}
which equals one by assumption.  Therefore, if for all $h$ that commute
with $g$, $\omega(g,h) = \omega(h,g)$, then under conjugation by $h$,
$\tau_g$ is invariant, and so should contribute to the spectrum of
dimension-zero ground states.  

The reader should note that the fact that the twist fields
$\sigma_{[g]}$ are counted by $\omega$-regular conjugacy classes
is in accord
\cite{Robbins:2021ylj} with the fact that irreducible projective representations
of a finite group $G$ are also counted by $\omega$-regular
conjugacy classes,
as discussed in \cite{cheng,costache,karpilovsky,schur1,schur2,schur3}.
Thus, the number of twist fields $\sigma_{[g]}$ matches the number of
irreducible projective representations, as expected.

So far we have discussed the construction of the twist fields $\sigma_{[g]}$
themselves.  Now, let us turn to their products.
The $\tau$'s obey
a group-like multiplication~(\ref{eq:tau-mult}). 
Results for products of $\sigma$'s can be derived from the product structure
on the $\tau$'s, as the $\sigma$'s are linear combinations of the $\tau$'s.
However, the resulting products of $\sigma$'s are no longer group-like,
but rather are more general ring elements, and in fact the products of
$\sigma$'s define a fusion algebra. 

To clarify these remarks, next we compute the $\sigma$'s and their
products in examples.
For our first example, suppose that $K$ is in the center of $G$,
and there is no discrete torsion.  In this case, every element of
$K$ is its own conjugacy class, and so the set of pertinent dimension-zero
operators is simply a copy of $K$, specifically
$\sigma_g = \tau_g$ for all $g \in K$.

For another example, consider the orbifold $[X/{\mathbb H}]$,
where ${\mathbb H}$ is the eight-element group of unit quaternions,
${\mathbb Z}_4 \cong \langle i \rangle$
acts trivially, and again without discrete torsion.
(This orbifold will be described in greater detail
in section~\ref{sect:ex:nonband:z4}.)
Here, $\pm 1$ are in the center of ${\mathbb H}$,
and so they are their own conjugacy classes.  The elements $\pm i$,
on the other hand, are mapped into one another under conjugation:
there exists $g \in {\mathbb H}$ such that $g (i) g^{-1} = -i$.
Therefore, we have the dimension-zero operators
\begin{equation}
\sigma_{[+1]} \: = \: 1, \: \: \: 
\sigma_{[-1]} \: = \: \tau_{-1}, \: \: \:
\sigma_{[i]} \: = \: (1/2) \left(\tau_i + \tau_{-i} \right).
\end{equation}
For example, since
\begin{equation}
\tau_{-1} \tau_{\pm i} \: = \: \tau_{\mp i},
\end{equation}
we have
\begin{equation}
\sigma_{[-1]} \sigma_{[i]} \: = \: \sigma_{[i]}.
\end{equation}
Similarly, since $\tau_{\pm i}^2 = \tau_{-1}$ and $\tau_{\pm i} \tau_{\mp i}
= 1$, we have
\begin{equation}
\sigma_{[i]}^2 \: = \: \frac{1}{4} \left( \tau_i^2 + \tau_{-i}^2 +
\tau_i \tau_{-i} + \tau_{-i} \tau_i \right)
\: = \: \frac{1}{2} \left( 1 + \sigma_{[-1]} \right).
\end{equation}
In particular, $\sigma_{[i]}$ does not have a group-like multiplication.
In this example, the $\sigma_{[g]}$ form a fusion algebra.

For another example, consider the orbifold $[{\rm point}/ {\mathbb Z}_2
\times {\mathbb Z}_2]$, this time with discrete torsion.
(As $H^2({\mathbb Z}_2 \times {\mathbb Z}_2,U(1)) = {\mathbb Z}_2$,
there is only one nontrivial choice of discrete torsion.)
Without discrete torsion, every element of ${\mathbb Z}_2 \times
{\mathbb Z}_2$ would be its own conjugacy class, since the group is abelian.
With discrete torsion, the story is somewhat more complicated.
Write ${\mathbb Z}_2 \times {\mathbb Z}_2 =
\langle a, b \rangle$, then under conjugation by $b$,
\begin{equation}
\tau_a \: \mapsto \: - \tau_a
\end{equation}
and under conjugation by $a$,
\begin{equation}
\tau_b \: \mapsto \: - \tau_b.
\end{equation}
In this fashion, using equation~(\ref{eq:sigma-defn}), one can show
\begin{equation}
\sigma_{[a]} \: = \: 0 \: = \: \sigma_{[b]} \: = \: \sigma_{[ab]}.
\end{equation}
The only nonzero dimension-zero twist field
is the identity itself -- the sums defining $\sigma$'s associated
to other conjugacy classes cancel out.
Indeed, the other conjugacy classes are not $\omega$-regular, so this example
confirms our earlier statement that only $\omega$-regular conjugacy classes
contribute to twist fields.
This is also 
consistent with the statement that ${\mathbb Z}_2 \times {\mathbb Z}_2$
has only one irreducible projective representation.

Finally, we should comment on the relation to computations of
massless spectra with discrete torsion in 
\cite{Vafa:1994rv}.  There, it was stated that the action of an
element $h \in G$ on a $g$-twisted sector is given by multiplying by a phase
\begin{equation}
\epsilon(g,h) \: = \: \frac{
\omega(g,h) }{ \omega(h,g) }.
\end{equation}
This defines an action because, as a consequence of the cocycle condition,
\begin{equation}
\epsilon(g, h_1 h_2) \: = \: \epsilon(g,h_1) \epsilon(g,h_2).
\end{equation}
We can derive this phase factor from the algebra above by interpreting the
algebra above in terms of OPEs of topological defect lines, as in
e.g. \cite{Chang:2018iay,yujitasi2019}.  
In this context, a trivalent vertex of topological defect
lines is defined by group elements $g$, $h$, and their product $gh$,
which in our case is weighted by the cocycle $\omega$.  The $g$ action
on an $h$-twisted sector is described formally by the intersection of two
orthogonal lines, one for $g$ and one for $h$.  That intersection can be
resolved into a pair of trivalent vertices, each of which contributes a 
factor of $\omega$, which together give $\epsilon$.

Fusion algebras have also been recently studied from other perspectives
in e.g. \cite{Heidenreich:2021tna,Thorngren:2019iar,Thorngren:2021yso}.

\subsubsection{Projectors}
\label{sect:projectors}

So far, we have discussed the dimension-zero twist fields and their
(fusion) products.  Mathematically, those
twist fields form a basis for the center of a twisted group algebra.
Now, such group algebras have another basis consisting
of idempotents (projectors).  In physics, this is a reflection
of decomposition.
In mathematics, it is ultimately
because of Wedderburn's theorem,
see e.g.  \cite[section XVII.3]{lang}.  In any event, this means the
(twisted) group algebra can be written in the form
\begin{equation}
{\mathbb C}[G] \: = \: \oplus_i {\rm End}( V_i),
\end{equation}
where the $\{ V_i \}$ are irreducible (projective) representations
of $G$.  Technically because this is a decomposition as an algebra,
the center is the subset of elements of the form
\begin{equation}
\sum_i c_i I_{V_i}.
\end{equation}
In particular, the identity matrices $I_{V_i}$ are a basis for the ring of
dimension-zero operators as
a vector space.  Since these are idempotents in the center, we can write
the ring of dimension-zero operators in the form
\begin{equation}
\oplus_i {\mathbb C} I_{V_i}
\end{equation}
as an algebra.  In particular, 
this ring is isomorphic
to a sum of copies of ${\mathbb C}$, with a complete set of orthogonal
projectors.  (Geometrically, in subsection~\ref{sect:geom} we will see that
this means the ring describes a finite set of points.)

In this section we will describe those projectors more explicitly.
In fact, expressions for projectors are known in the relevant mathematics
literature, as we will review.  Projectors are in one-to-one correspondence
with (projective) representations, and also coincide with
the projectors onto local operators associated with each universe in
the decomposition of the quantum field theory.

A general expression for projectors is given abstractly in
\cite[section 2.6, exercise 6.4]{serrerep}, 
\cite[prop. 9.21]{cr81} (for the case of vanishing discrete torsion),
and more generally in \cite[chapter 7, theorem 3.1]{karpilovsky}
which is easily adapted to the
present situation.  

Let $R$ be a representation of $K$ associated to a universe -- meaning,
\begin{equation}
R \: = \: \oplus R_i,
\end{equation}
for $R_i$ a set of irreducible (projective) representations of $K$
spanning the isomorphism classes of $\hat{K}_{\iota^* \omega}$
in a single fixed orbit of the action of $G$.
Then, for each irreducible $R_i$,
define\footnote{
Equivalently, instead of summing over all $g$, one might only
sum over $\omega$-regular $g$, meaning those elements $g$ such that
$\omega(g,h) = \omega(h,g)$ for all $h$ commuting with $g$.
The resulting sum is equivalent because
characters of projective representations vanish for 
non-$\omega$-regular elements. 
}
\begin{equation}
\Pi_{R_i} \: = \: \frac{\dim R_i}{|K|} \sum_{k \in K} 
\frac{ \chi_{R_i}\left( k^{-1} \right) }{ \omega(k,k^{-1}) } \tau_k,
\end{equation}
and then finally define
\begin{equation}  \label{eq:genl-def-proj}
\Pi_R \: = \: \sum_i \Pi_{R_i}.
\end{equation}
(In the case that a representation $R_i$ is projective,
characters $\chi_R$ exist, but their properties are slightly obscure.
We will describe them in more detail later in section~\ref{sect:Wlines}.)

Just as the $\sigma_{[g]}$ form a basis for the center of the
twisted group algebra, so too do the projectors above, see e.g.
\cite[prop. 9.14]{cr81}, \cite[section 3]{cheng}.  In particular,
although it may not be obvious, each projector $\Pi_R$ is a linear
combination of $\sigma$'s, and so can be written in terms of the
closed string twist fields.  

We will see later in examples that
the $\Pi_R$ also coincide with the projectors onto universes described
elsewhere in the literature on decomposition.
Specifically, the universe projectors are 
the $\Pi_R$ for distinct $R$
obtained as restrictions to $K$ of
irreducible representations of $G$.
Note that in general,
the restriction of an irreducible representation of $G$ may be a reducible
representation of $K$, as we shall see in examples.

It is straightforward to check that the $\Pi_R$ are projectors.
For example,
for any two irreducible representations $R$, $S$,
\begin{eqnarray}
\Pi_R \Pi_S & = & \frac{(\dim R)(\dim S)}{|K|^2}
\sum_{g,h \in K} \frac{\chi_R(g^{-1}) \chi_S(h^{-1})}{
\omega(g,g^{-1}) \omega(h,h^{-1}) } \omega(g,h) \tau_{gh},
\\
& = & \frac{(\dim R)(\dim S)}{|K|^2}
\sum_{g,k \in K} \chi_S(k^{-1} g) \chi_R(g^{-1}) 
\frac{ \omega(g,g^{-1} k) }{ \omega(g,g^{-1}) \omega(g^{-1} k, k^{-1} g) }
\, \tau_k,
\end{eqnarray}
Using the identity $\omega(h,h^{-1}) = \omega(h^{-1},h)$ (see
appendix~\ref{app:miscgpcoh}), the condition
$(d \omega)(k^{-1},g,g^{-1}k) = 1$ implies that
\begin{equation}
\frac{ \omega(g,g^{-1}k) }{ \omega(g^{-1} k, k^{-1} g) }
\: = \: \frac{ \omega(k^{-1},g) }{ \omega(k,k^{-1}) },
\end{equation}
so we can apply identity~(\ref{eq:char-om1}) to find that
\begin{eqnarray}
\Pi_R \Pi_S 
& = & \frac{(\dim R)(\dim S)}{|K|} \sum_{k \in K} \delta_{R,S}
\frac{1}{\omega(k,k^{-1})}
\frac{\chi_R(k^{-1})}{\dim R} \tau_k,
\\
& = & \delta_{R,S} \Pi_R.
\end{eqnarray}

Next, we check that these projectors are complete.
\begin{eqnarray}
\sum_R \Pi_R & = & 
\sum_R \frac{\dim R}{|K|} \sum_{k \in K} \frac{ \chi_R(k^{-1}) }{
\omega(k,k^{-1}) } \tau_g,
\\
& = & \frac{1}{|K|} \sum_{k \in K} \left(
\sum_R \frac{ (\dim R) \chi_R(k^{-1})}{ \omega(k,k^{-1}) } \right) \tau_g.
\end{eqnarray}
Now, in a sum over all irreducible (projective)
representations of a group $K$,
from~(\ref{eq:orthog2}), we have
\begin{eqnarray}
\sum_R \frac{ (\dim R) \chi_R(k^{-1}) }{ \omega(k,k^{-1}) } & = &
\sum_R \frac{ \chi_R(1) \chi_R(k^{-1}) }{ \omega(k,k^{-1}) }, \\
& = & \left\{ \begin{array}{cl}
0 & k \neq 1, \\
|K| & k = 1.
\end{array} \right.
\end{eqnarray}
The sum above is a 
sum over all irreducible representations of $K$,
hence
\begin{equation}
\sum_R \Pi_R \: = \: 1.
\end{equation}

\subsubsection{Geometry}
\label{sect:geom}

So far we have discussed the twist fields associated to conjugacy classes
of trivially-acting group elements in orbifolds, as well as
projectors.  As discussed, 
these each form a basis for the space
of all dimension-zero operators.
Next, we will study the geometry and commutative algebra of
the ring of such linear combinations,
the ring of dimension-zero operators in a given orbifold with a
trivially-acting subgroup.

In all cases, the rings of dimension-zero operators geometrically describe
a set of points, one for each universe in the decomposition.
Schematically, each such ring ${\cal R}$ is of the form
\begin{equation}
{\mathbb C}[x_0, \cdots, x_n] / I,
\end{equation}
where $I$ is some ideal, and the locus on which $I$ vanishes will
be a set of isolated points.
In the language\footnote{
Such language is not entirely appropriate, as it incorrectly suggests
that the different universes are mere superselection sectors.
In particular, as described in detail in e.g. \cite{Tanizaki:2019rbk},
universes are not just superselection sectors, as for example
decomposition exists
for finite volumes, not just in an infinite volume limit.
} of spontaneous symmetry breaking, we can think of 
$x_0, \cdots, x_n$ above as being order parameters, and the points
at which the ideal $I$ vanishes correspond to vevs of order
parameters corresponding to each universe.

If the theory has a one-form symmetry, it is reflected in a symmetry
in the ring of zero-dimensional operators as phases of the form
\begin{equation}
x_i \: \mapsto \: \xi x_i,
\end{equation}
which leave the ideal $I$ invariant.  In orbifolds, these phases look
identical to quantum symmetries.  We shall discuss this in detail in
examples.

In any theory exhibiting a nontrivial decomposition,
the ring of dimension-zero operators will always
have noninvertible operators -- for example, the projectors
which project onto subspaces associated with the
universes, as we described in subsection~\ref{sect:projectors}. 
In general, there will be additional noninvertible elements.
The locus of noninvertible operators
is always codimension-one in the space of order parameters.
This is because we are working in a quantum-mechanical system, and these
are operators acting on a finite-dimensional Hilbert space.  Such operators
can be represented by finite-dimensional matrices, so the noninvertibility
criterion\footnote{
Strictly speaking, we should show that the inverse is in the ring.
Let $T$ be the matrix corresponding to an element of the ring,
and $f$ its characteristic polynomial.  By the Cayley-Hamilton theorem,
$f(T) = 0$.  Since we are assuming the determinant is nonzero,
the constant term of $f(T)$ is nonzero, so the inverse of $T$ is a 
polynomial in $T$.  Taking the same polynomial in the original ring element
gives the inverse of that element.  We would like to thank T.~Pantev
for making this observation.
} is for the determinant to vanish, which always gives a 
codimension-one condition.
This fact will complicate our efforts to characterize noninvertible
symmetries in theories admitting a decomposition, as we shall
discuss in subsection~\ref{sect:symm}.

These rings of dimension-zero operators can be
understood conveniently in the mathematical language of commutative
algebra (see e.g. \cite{atmac} for a very readable introduction),
which is pertinent to the geometry.
Let ${\cal R}$ denote the ring of dimension-zero operators appearing in a 
two-dimensional conformal field theory, then ${\cal R}$ is a commutative
ring with identity.  Saying that ${\cal R}$ ``is associated to a set of points''
can be understood more formally as saying that Spec $R$ is a disjoint
union of reduced\footnote{
`Reduced' is a technical term in commutative algebra, which distinguishes
ordinary points from `fat' points which are collisions of multiple points.
} points, or equivalently that ${\cal R}$ has finitely many maximal
ideals and no nilpotent elements.  
(A nilpotent element would correspond 
to an operator whose eigenvalues all vanish, which
seems unlikely, and we assume that each individual
universe has a unique dimension-zero operator.)

Furthermore, each universe is supported at a maximal ideal ${\mathfrak m}$ 
such that
all the projectors but one are in ${\mathfrak m}$.
Equivalently,
\begin{equation}  \label{eq:piincap}
\Pi_i \: \in \: \cap_{j \neq i} {\mathfrak m}_j.
\end{equation}

In some sense, the ideals $\Pi_i$ and the maximal ideals
${\mathfrak m}_j$ (corresponding to points on spaces of order parameters
at which universes are supported) are `dual' to one another.
We have already seen that projectors are in the intersections of 
maximal ideals corresponding to other points.  Conversely,
given the projectors, the point at which any universe is supported
is given by the vanishing locus of all other projectors:
\begin{equation}
\cap_{j \neq i} \{ \Pi_j = 0 \} \: = \: \{ \mbox{ $i$th point } \}.
\end{equation}

Given any ring ${\cal R}$ and maximal ideal ${\mathfrak m} \subset {\cal R}$,
the quotient ${\cal R}/{\mathfrak m}$ is a field
\cite[chapter 1]{atmac}.  In the present circumstances,
given the decomposition 
\begin{equation}  \label{eq:ringdecomp}
{\cal R} \: = \: \sum_k (\Pi_k),
\end{equation}
of the ring ${\cal R}$,
for a maximal ideal ${\mathfrak m}$ corresponding to a projector
$\Pi$, we have that
\begin{equation}
{\cal R}/{\mathfrak m} \: = \: \sum_k (\Pi_k)/{\mathfrak m} \: = \:
( \Pi )/{\mathfrak m} \: \cong \: {\mathbb C},
\end{equation}
where we have used~(\ref{eq:piincap}).

This also has a simple understanding in terms of localization
(in the sense of commutative algebra).
Briefly, ${\cal R}_{\mathfrak m}$ for any prime (here, maximal) ideal 
${\mathfrak m}$ denotes a ring of fractions
$S^{-1} {\cal R}$ where $S = {\cal R} - {\mathfrak m}$.
Let $\Pi_i$ denote the one projector that is not in maximal ideal
${\mathfrak m}_i$, then for all other projectors,
$(\Pi_{k \neq i})_{ {\mathfrak m}_i } = 0$, essentially because
$\Pi_i \in S$ and so can be used to project out ideals generated by
other projectors.
In physics terms, the restriction of $\Pi_j$ to the point
corresponding to ${\mathfrak m}_i$ vanishes if $j \neq i$; only for
the point corresponding to ${\mathfrak m}_i$ can the restriction
of $\Pi_i$ be nonzero.

We should mention that the fact that
all points are reduced, that the ring has no nilpotents,
(a mathematical consequence of the construction reviewed in
subsection~\ref{sect:projectors},)
is physically a reflection of cluster decomposition.
Specifically, we assume that each universe has a unique (up to scaling)
dimension-zero operator, and so the points appearing in the ring should
all be reduced.

In passing, since these rings of dimension-zero operators describe isolated
points (i.e. have finitely many maximal ideals), they are semi-local rings,
see e.g. \cite{hirano,lam,gks},
and since there are no nilpotents (reflecting cluster decomposition 
in each separate universe), the nilradical vanishes, hence since ${\cal R}/J({\cal R})$
is semisimple in semi-local rings, where $J({\cal R})$ denotes the Jacobson
radical, which coincides with the nilradical for polynomial rings,
the rings are semisimple.  From another perspective,
the Jacobson radical is the intersection of the maximal ideals, so for
a ring describing isolated points, $J({\cal R}) = 0$, and the ring is
semisimple.

\subsection{Wilson lines}
\label{sect:Wlines}

So far, we have focused on dimension-zero operators.
Next, we turn to Wilson lines in these theories.
To clarify, there are two types of Wilson lines that one might be
interested in:  bulk Wilson lines, associated with the (finite) gauge
theory, and boundary Wilson lines, also known as Chan-Paton factors in
open strings.
If there is no discrete torsion, then those Chan-Paton factors
are in an ordinary representation of the orbifold group.  If there is
discrete torsion, on the other hand, then they are in a projective
representation, determined by a cocycle $\omega$ representing an
element of group cohomology.

The two types of Wilson lines have somewhat different interpretations:
bulk Wilson lines act as bridges between universes, whereas boundary
Wilson lines 
can be associated with different universes in the
decomposition.  (Geometrically, in the latter context, 
K theory and sheaves on gerbes decompose
into K theory and sheaves on universes, as discussed in
\cite{Hellerman:2006zs}.)  We shall see that the projectors
$\Pi_R$ we computed in subsection~\ref{sect:projectors} project onto
Wilson lines associated with the universe corresponding to $R$.
We will set up the computational technology in this section, then see
that decomposition in examples later.

Formally, several fusion categories of topological line operators have been
defined in e.g. \cite{Brunner:2013ota,Brunner:2013xna,Brunner:2014lua,Bhardwaj:2017xup}.  
Briefly, the objects in one such category are topological line operators
(such as Wilson lines), 
and the morphisms are local operators at
junctions of the line operators.
In this paper, we will associate such line operators with
topological Wilson lines, associated to pairs $(C, \rho)$,
where $C$ is a curve and $\rho$ is a representation.
For simplicity we focus on open string diagrams with
a Wilson line on the boundary and local operator insertions in the bulk,
so we will usually omit the curve $C$, and focus on the representation reelvant
to any given Wilson line.

\subsubsection{Bulk Wilson lines: defects bridging universes}
\label{sect:overview:bulkW}

In most of this paper, for simplicity we will focus on boundary
Wilson lines (Chan-Paton factors).  In this section, we will discuss
how bulk Wilson lines act as defects spanning universes in a decomposition.

Before discussing orbifolds specifically, it may be useful to consider
a related example in a different theory.  Specifically, consider
$U(1)$ $BF$ theory at level $k$.  This theory is known to decompose
(see e.g.~\cite[appendix B]{Hellerman:2010fv}), 
and the local operators ${\cal O}_p$ and
Wilson lines $W_q$ obey clock-shift commutation relations 
(see e.g.~\cite[appendix B]{Hellerman:2010fv})
\begin{equation}
{\cal O}_p W_q \: = \: \xi^{pq} W_q {\cal O}_p,
\end{equation}
where $\xi = \exp(2 \pi i/k)$.  Now, the projectors are
\begin{equation}
\Pi_m \: = \: \frac{1}{k} \sum_{n=0}^{k-1} \xi^{nm} {\cal O}_n,
\end{equation}
and it is straightforward to check that the clock-shift commutation
relations are algebraically equivalent to
\begin{equation}
\Pi_m W_p \: = \: W_p \Pi_{m+p \mod k}.
\end{equation}
In other words, moving a projector through a Wilson line changes
the projector -- bulk Wilson lines in abelian $BF$ theory act as
defects linking different universes of the decomposition.

We shall see that
the same effect happens in orbifolds, 
later in section~\ref{sect:pairing}.

\subsubsection{Boundary Wilson lines:  Decomposition of bundles and sheaves}
\label{sect:decomp:sheaves}

Next we turn to boundary Wilson lines (Chan-Paton factors),
which most of this paper will focus on.
As mentioned previously, in principle sheaves and bundles on an
orbifold $[X/\Gamma]_{\omega}$, with trivially-acting
$K \subset \Gamma$, are equivalent to sheaves and bundles on the universes
of decomposition.  Briefly, the association is computed by
comparing the (projective) representation of $K$ associated to the bundle
or sheaf with that associated to each universe of decomposition.

Let us walk through that more carefully.  
A bundle or sheaf
on $[X/\Gamma]_{\omega}$ is a bundle or sheaf on $X$ with a
(projective) $\Gamma$-equivariant structure.  If we restrict to $K \subset
\Gamma$, then since $K$ acts trivially on $X$, the $\Gamma$-equivariant
structure determines a (projective) representation of $K$, call it 
$\rho$.  (In passing, if $\rho$ is a projective representation associated
to a nontrivial element of $H^2(\Gamma, U(1))$, then $\rho$ is necessarily
of dimension greater than one.)

Now, as described earlier in
section~\ref{sect:dictionary}, we can associate a representation $R$
of $K$ to each universe of the decomposition.  
That representation is of the form
\begin{equation}
R \: = \: R_1 \oplus \cdots \oplus R_{\ell},
\end{equation}
which each $R_i$ is an irreducible projective representation of $K$,
and the isomorphism classes $\{ [ R_i ] \}$ span the orbit of the
$G = \Gamma/K$ action on $\hat{K}_{\iota^* \omega}$.

In order to associate a given bundle or sheaf on $[X/\Gamma]_{\omega}$ to
a bundle or sheaf on one of the universes of decomposition, the idea then
is that
we take the (projective) representation $\rho$ of $K$, derived
from the (projective) $\Gamma$-equivariant structure, decompose into its
irreducible components, and compare to the representation $R$ associated
to each of the universes.  If a (sum of) component(s) match, that piece
lives in the corresponding universe.

We shall see this correspondence in examples.

The projectors, defined in section~\ref{sect:projectors}, also have the property
of projecting Wilson lines into the pertinent universes.
After describing general aspects of the relation between twist fields
and representations, we will discuss the pairing between the two
provided by characters, and determine the action of the projectors
on Wilson lines.

\subsubsection{Pairing between twist fields and representations}
\label{sect:pairing}

In general terms, for a fixed group $\Gamma$, although ($\omega$-regular)
conjugacy classes are in one-to-one correspondence with
(projective) representations, bijections are not\footnote{
In fact, unlike conjugacy classes,
for projective representations for 
a fixed $[\omega] \in H^2(G,U(1))$, there is no ring structure, as the
product does not preserve $\omega$:
if
$R$ is projective with respect to $\omega_1$ and $S$ is projective with
respect to $\omega_2$ then $R \otimes S$ is projective with respect to
$\omega_1 \otimes \omega_2$ \cite[section 7.1, prop. 1.2]{karpilovsky}.
} canonical.
Instead, the two are more nearly dual, with a canonical pairing defined by
characters.

Recall that the character 
$\chi_R(g)$ of a (possibly projective)
representation $R$ of a group $\Gamma$, evaluated on group element 
$g \in \Gamma$, is 
Tr $T^R_g$, where $T^R_g$ is a matrix representing $g$ in $R$.
In the case of ordinary representations, characters are constant on
conjugacy classes (more formally, they are class functions).
For projective representations, their properties 
are somewhat different, as explained in
e.g. \cite[section 7.2, prop. 2.2]{karpilovsky}.  For example:
\begin{itemize}
\item Characters no longer need be constant on conjugacy classes,
and so need not be class functions.
Instead,
\begin{equation} \label{eq:char-non-class}
\chi(g) \: = \: 
\frac{ \omega(g,h^{-1}) }{ \omega(h^{-1},hgh^{-1}) } \,
\chi(h g h^{-1}),
\end{equation}
\item If $g \in \Gamma$ is not $\omega$-regular\footnote{
For $g \in \Gamma$ to be $\omega$-regular means that $\omega(g,h) = 
\omega(h,g)$ for all $h$ commuting with $g$.
}
then $\chi(g) = 0$,
\item If for any $\omega$-regular element $g \in \Gamma$ and for any $h \in \Gamma$,
\begin{equation}
\omega(g,h) \: = \: \omega(h,h^{-1}gh),
\end{equation}
then the character is a class function.
\end{itemize}

The pairing between conjugacy classes and representations,
via characters, is reflected in physics.
Consider a Wilson loop in representation $R$ of $\Gamma$ encircling a twist
operator $\tau_g$ associated to $g \in \Gamma$.  The Wilson loop will cross a $g$ branch
cut, effectively inserting a $g$ inside the Wilson loop,
implemented by a matrix $T^{R}_g$ describing $g$ in representation
$R$.  Except in special cases\footnote{
For example, in the special case that
$R$ is a one-dimensional representation, then
the matrix $T^{R}_g$ is just the number $\chi_R(g)$, and in that case, 
$\tau_g W_{R} = \chi_{R}(g) W_{R}$.  That will not be the case
in general, however.
}, $T^{R}_g$ will not be proportional to
the identity, and so the effect of $\tau_g$ is not to merely
multiply $W_R$ by a phase -- $W_R$ need not be an eigenbrane of $\tau_g$
in general.

Consider on the other hand the effect of multiplying
$\sigma_{[g]}$ into $W_R$.  From the definition~(\ref{eq:sigma-defn}),
this should insert into the Wilson line $W_R$ the matrix
\begin{equation}
\frac{1}{|G|} \sum_{h \in G}
\frac{\omega(h,g) \omega(hg,h^{-1}) }{ \omega(h,h^{-1}) }
T^R_{hgh^{-1}}.
\end{equation}
It can be shown that the quantity above commutes with other matrices
(see appendix~\ref{app:casimirs}), just as $\sigma_{[g]}$ commutes with
other local operators, hence the sum above is 
proportional to the identity, and a short computation demonstrates
that
\begin{equation}
\frac{1}{|G|} \sum_{h \in G}
\frac{\omega(h,g) \omega(hg,h^{-1}) }{ \omega(h,h^{-1}) }
T^R_{hgh^{-1}} \: = \:
\frac{ \chi_R(g) }{ \dim R} I,
\end{equation}
(See appendix~\ref{app:casimirs} for further details.)

As an upshot of this, we then have that\footnote{
See also e.g.~\cite[footnote 10]{Delmastro:2021otj} 
for the special case of one-dimensional representations $R$ and
no discrete torsion, and also \cite[equ'n (2.2)]{Heidenreich:2021tna}.
}
\begin{equation}  \label{eq:W-twist:bulk}
\sigma_{[g]} W_R \: = \:  
 \frac{ \chi_R(g) }{\dim R}
W_R \sigma_{ [g] }
\end{equation}
for bulk Wilson lines,
an analogue of the clock-shift commutation relations of $BF$ theory
discussed in section~\ref{sect:overview:bulkW},
and
\begin{equation}  \label{eq:W-twist}
\sigma_{[g]} W_R \: = \: 
 \frac{ \chi_R(g) }{\dim R}
W_R,
\end{equation}
for boundary Wilson lines.
which we will confirm in examples.
In particular, in the boundary case,
although $W_R$ is not an eigenbrane of $\tau_g$,
it is an eigenbrane of $\sigma_{[g]}$.

Before going on, let us check the consistency of this expression.
Specifically, if there is nonzero discrete torsion, two of the factors are
ambiguous:  both $\sigma_{[g]}$ and $\chi_R(g)$ depend upon the choice of
representative of the conjugacy class, and different representatives 
give results that differ by phases.  Specifically, if 
$g' = h g h^{-1}$, then from equation~(\ref{eq:sigma-var}) and
\cite[section 3]{cheng},
\begin{equation}
\sigma_{[hgh^{-1}]} \: = \:
\frac{ \omega( gh^{-1}, h) }{ \omega(h, gh^{-1} ) }
\sigma_{[g]}.
\end{equation}
Similarly, if there is discrete torsion, so that the representation
$R$ is projective, the characters $\chi_R$ are not true class functions -- they
are not constant on conjugacy classes -- but
instead obey  \cite[section 7.2, prop. 2.2]{karpilovsky}
\begin{equation}
\chi_R(hgh^{-1}) \: = \:
\frac{ \omega(h^{-1}, hgh^{-1} ) }{ \omega(g,h^{-1}) } 
\chi_R(g).
\end{equation}
Indeed, from the product
\begin{equation}
(d \omega)(h^{-1},h,gh^{-1}) \, (d \omega)(g, h^{-1}, h) \: = \: 1,
\end{equation}
one finds
\begin{equation}
\frac{ \omega(h^{-1}, h g h^{-1}) }{ \omega(g, h^{-1}) }
\: = \:
\frac{ \omega(gh^{-1}, h) }{ \omega(h, gh^{-1}) },
\end{equation}
and so we see that 
\begin{equation}
\sigma_{[hgh^{-1}]} W_R \: = \: \frac{ \chi_R( hgh^{-1}) }{ \dim R} W_R,
\end{equation}
as needed for consistency.

The relations~(\ref{eq:W-twist:bulk}), (\ref{eq:W-twist})
are similar to the usual relation between 't Hooft loops
and Wilson lines, which is of the schematic form
(see e.g.~\cite[equ'n (3.8)]{tHooft:1977nqb})
\begin{equation}
T \cdot W \: = \: ({\rm phase}) W \cdot T
\end{equation}
in three and four dimensions.
As twist fields generate branch cuts, they are natural analogues of
't Hooft loops in higher dimensions, and so the analogy with 't Hooft-Wilson
line relations is precise.

\subsubsection{Action of projectors on Wilson lines}

Now, let us apply projectors $\Pi_R$ (defined in
subsection~\ref{sect:projectors}) to Wilson lines.
We will see that 
the projectors commute with bulk Wilson lines in such a way as to
describe defects bridging universes, and
the projectors project boundary Wilson lines into corresponding
universes, in the fashion outlined in subsection~\ref{sect:decomp:sheaves}.

The reader should note that
\begin{itemize}
\item $R$ is a representation of the trivially-acting subgroup $K$.  
For the projector $\Pi_R$ corresponding to any given universe,
$R$ is a sum of representatives of the orbit of $G = \Gamma/K$ in
$\hat{K}_{\iota^* \omega}$, and so is of the form
\begin{equation}
R \: = \: R_1 \oplus \cdots \oplus R_{\ell}.
\end{equation}
\item $S$, on the other hand, 
is a representation of $\Gamma$.  In evaluating the action of
$\Pi_R$ on $W_S$, we effectively restrict $S$ to $K$.  Even if $S$ itself
were irreducible, in general its restriction is reducible, and so $S|_K$
can also be written as a sum of irreducible (projective) representations of
$K$.  
\item We will see in appendix~\ref{app:induced}
that all representations 
of $K$ appear as summands in restrictions of representations of $\Gamma$;
however, we will see in examples in e.g. sections~\ref{sect:specex:d4}, 
\ref{sect:specex:d8}
that 
some irreducible representations of $K$ only appear as summands and never
as the entire restriction of an irreducible representation of $\Gamma$.
\end{itemize}

First, we consider bulk Wilson lines.
Here, for simplicity, we specialize for the moment 
to the special case of a $G = {\mathbb Z}_k$
orbifold, where all of $G$ acts trivially.
Let $\sigma_j$ denote the twist field associated to $g^j$, for $g$
the generator of $G$.  Projectors are of the form *** CITE ****
\begin{equation}
\Pi_i \: = \: \frac{1}{n} \sum_{j=0}^{n-1} \xi^{ij} \sigma_j
\end{equation}
for $\xi = \exp(2 \pi i/k)$. 
Applying~(\ref{eq:W-twist:bulk}), we find
\begin{eqnarray}
\Pi_i W_R & = &
\frac{1}{k} \sum_j \xi^{ij} \sigma_j W_R,
\\
& = &
\frac{1}{k} \sum_j \chi_R(g^j) W_R \sigma_j,
\end{eqnarray}
using the fact that for $G = {\mathbb Z}_k$, all irreducible representations
have dimension $1$.  Now, suppose $R$ is the $n$th
irreducible representation, meaning $\chi_R(g^j) = \xi^{jn}$, then
\begin{eqnarray}
\Pi_i W_R & = &
W_R \left( \frac{1}{k} \sum_j \xi^{j(i+n)} \sigma_j \right),
\\
& = & W_R \Pi_{i+n \mod k},
\end{eqnarray}
so that again, moving the projector through the Wilson line changes the
projector, just as in $BF$ theory as in section~\ref{sect:overview:bulkW}.
We will see closely analogous 
relations of this form in less trivial examples later in this paper.

Next, we consider boundary Wilson lines, returning to more general
orbifold groups $G$.
For simplicity, and without loss of significant generality,
let us restrict to a single irreducible (projective) representation
$R$ of $K$, and let us denote by $S$ a single summand in the restriction of
a representation of $\Gamma$ to $K$.

Then, for boundary Wilson lines,
using the pairing~(\ref{eq:W-twist}) and the
identities~(\ref{eq:chi-ginv}),
(\ref{eq:chi-orthog}),
we see that
\begin{eqnarray}
\Pi_R W_S & = & 
\frac{\dim R}{|K|} \sum_{k \in K} \frac{ \chi_R(k^{-1}) }{ \omega(k,k^{-1}) }
 \frac{\chi_S(k)}{\dim S}
W_S,
\\
& = & \frac{\dim R}{\dim S} \frac{1}{|K|}
\sum_{k \in K} \overline{\chi_R(k)} \chi_S(k) W_S,
\\
& = & \delta_{R,S} W_S.
\end{eqnarray}
In effect, the $W_S$ are eigenbranes of the projectors, with eigenvalues
$0$, $1$.

Walking through the details of the prototypical computation above,
the effect of acting on $W_S$ with $\Pi_R$ is to restrict the
$\Gamma$ representation $S$ to $K$, and then project into universes based
on comparing that projection with the $K$ representations associated with
each universe.  This is a computational realization of the
correspondence described in subsection~\ref{sect:decomp:sheaves}
and \cite{Hellerman:2006zs}
between bundles and sheaves on gerbes and on universes -- 
the projectors implement the mathematical statement that bundles and
sheaves on gerbes decompose into bundles and sheaves on constituent
universes, by picking out the Wilson lines that exist on each
universe.  We will see this explicitly in examples later.

Next, we will discuss symmetries of these theories.

\subsection{Symmetries}
\label{sect:symm}

Now that we have thoroughly set up the mechanics of dimension-zero
twist fields, projectors, Wilson lines, and corresponding geometries,
let us take a moment to describe the (possibly noninvertible)
one-form 
symmetries present in theories obeying decomposition.
We will propose that invertibility of the 1-form symmetries is correlated
to properties of the representations characterizing the universes:
one-form symmetries go hand-in-hand with universes corresponding to
one-dimensional representations, and propose that
noninvertible (one-form) symmetries
are related to universes corresponding to higher-dimensional representations.

It is sometimes said that a $q$-form symmetry in $d$ dimensions is
generated by an operator
\begin{equation}
U_g(M_{d-q-1}),
\end{equation}
supported on a codimension-$(q+1)$ submanifold $M$, acting as a 't Hooft
operator.  It is invertible if it obeys an
(abelian) group law 
\begin{equation}
U_q U_{q'} \: =  \: U_{q q'},
\end{equation}
while an analogous noninvertible symmetry has a generator on a submanifold
of the same codimension, but obeying a fusion algebra instead of a group
law:
\begin{equation}
U_q U_{q'} \: = \: \sum_p N^p_{q q'} U_p.
\end{equation}
We are focused on two-dimensional theories, 
hence our symmetry generators are supported at points.
Given that they generate branch cuts, it is natural to identify the 
$U_g$ with twist fields (see e.g. \cite[section III]{Nguyen:2021naa}), 
which in two-dimensional orbifolds play a role
analogous to Gukov-Witten operators \cite{Gukov:2006jk,Gukov:2008sn}
in higher dimensions.
Specifically, we are interested in twist fields associated to conjugacy
classes whose elements lie in the trivially-acting subgroup $K$.

If the trivially-acting subgroup $K$ is a subgroup of the center of the
orbifold group $\Gamma$, and there is no discrete
torsion -- corresponding to a banded abelian gerbe -- then each conjugacy
class has one element, and the conjugacy classes obey a group law:
\begin{equation}
\sigma_g \sigma_{g'} \: = \: \sigma_{g g'}.
\end{equation}
In this case, the twist fields $\sigma_g$ are invertible,
and the universes of decomposition are all identical
(modulo theta angle / discrete torsion phases).
As a consequence, there is an ordinary one-form
symmetry, no noninvertible symmetry.

In more general cases, if $K$ is not in the center,
twist fields are typically associated to conjugacy classes containing
multiple elements of $K$.  In such cases, as we have computed in
section~\ref{app:computation} (and will see in examples later), the 
twist fields obey a fusion algebra structure:
\begin{equation}
\sigma_{[g]} \sigma_{[g']} \: = \:
\sum_h N^h_{g g'} \sigma_{[h]}
\end{equation}
in general.  Mathematically, at least in the absence of discrete torsion,
these correspond to nonbanded abelian and
nonabelian gerbes.
In these cases, the universes of decomposition are not all identical,
and at least some of the twist fields
$\sigma_{[g]}$ are noninvertible.  These would appear to correspond to
noninvertible symmetries.

All that said, there is an important subtlety:  abstractly
in the vector space of dimension-zero operators spanned by the
twist fields, in general there is not to our knowledge
a natural mathematical way to distinguish
twist fields associated to conjugacy classes from general linear combinations
of twist fields (spanning the center of the twisted group algebra).
(In fact, some of the group structure is also washed out by the
group algebra.  For example, ${\mathbb C}[{\mathbb Z}_4] \cong
{\mathbb C}[{\mathbb Z}_2\times{\mathbb Z}_2]$ even though the groups
${\mathbb Z}_4$, ${\mathbb Z}_2 \times {\mathbb Z}_2$ are not isomorphic.)
In special cases, there may be a pertinent construction,
a means to detect ``group-like'' elements.
For example, in any group ring ${\mathbb C}[G]$ for finite $G$,
there is a comultiplication $\Delta$ that maps $g \mapsto g \otimes g$
for $g \in G$, and is extended linearly to other elements of the
group ring.  
(This action of the comultiplication is one of the defining
properties of group-like elements of a Hopf algebra.)
One can define group-like elements to be those elements $\sigma$
for which $\Delta(\sigma) = \sigma \otimes \sigma$.
However, the $\sigma$'s only correspond to $g$'s in
a group ring in the special cases of an abelian orbifold group without
discrete torsion, and so in general this cannot be used to identify
twist fields associated with conjugacy classes in a generic
ring of dimension-zero operators.
Similarly, in the symmetry categories discussed in e.g.
\cite{Bhardwaj:2017xup}, one can define `simple' objects
(namely, objects $a$ such that Hom$(a,a) \cong {\mathbb C}$); however,
we do not know of an analogous construction here.

Furthermore,
as we have seen, the vector space of dimension-zero operators
always include noninvertible
operators, regardless of whether $K$ is central.  Put another way, even
if there are no noninvertible symmetries, even if all the higher symmetries are
(invertible) one-form symmetries, then nevertheless the theory still contains
noninvertible dimension-zero
operators, noninvertible linear combinations of twist fields,
that form a codimension-one subspace of the space of all linear combinations
of twist fields (the center of the twisted group algebra).

To be clear, the vector space of dimension-zero operators does have a natural
basis of twist fields, so one can try to use those elements.
For banded abelian gerbes, this appears to work well, as outlined above.
However, for nonbanded
gerbes, this approach runs into subtleties.
For example, for the nonbanded
${\mathbb Z}_4$ gerbe we will discuss in section~\ref{sect:ex:nonband:z4},
the projection operators onto the two identical universes in the
decomposition (which should be related by a $B {\mathbb Z}_2$ symmetry)
mix invertible and noninvertible twist fields,
and the projection operator onto the third, distinct, universe,
involves only invertible twist fields.

This suggests that
a more invariant method to identify noninvertible 1-form
symmetries may be desirable.
To that end, in the spirit of \cite{Bhardwaj:2017xup,Brunner:2013ota,Brunner:2013xna,Brunner:2014lua}, 
we propose instead to associate symmetries with
representations
\footnote{This is in the spirit of topological defect lines.
In this language, intersection vertices of topological defect lines carry
complex vector spaces of operators -- the twist fields $\sigma_{[g]}$.
The operators $\tau_g$ are naturally associated with endpoints of 
topological defect lines.
} classified by 
representations.  One-dimensional representations obey a group-like
multiplication.
Higher-dimensional representations 
obey a more general algebra, and so
in general are not invertible.  For our purposes, we will see
in examples that it is more useful
to characterize noninvertible symmetries in terms of dimensions of
corresponding representations.

In terms of that dictionary, we can make a very explicit connection to
decomposition.
As we noted in section~\ref{sect:projectors}, associated to each universe
is a (projective)
representation $R$ of $K$
(a direct sum of representatives of the elements of a $G$ orbit on
$\hat{K}_{\iota^* \omega}$), which corresponds to a projector $\Pi_R$.  
Universes for which $R$ is a one-dimensional
representation are, in general, identical to one another (modulo
theta angle / discrete torsion shifts), and so are related by 
ordinary one-form
symmetries.  Universes for which $R$ has higher dimension need not be the
same, and are related by noninvertible symmetries.
Thus, properties of decomposition naturally tie into this proposed
characterization of invertibility of
higher-form symmetries:  in general terms, copies of universes are associated
with (invertible) one-form symmetries, whereas distinct universes
are associated with noninvertible symmetries.

\section{Example:  Sigma model on a disjoint union}
\label{sect:ex:disjoint}

Before studying orbifolds with trivially-acting subgroups,
let us first consider sigma models whose targets are disjoint unions,
to clarify expectations in an example which, by virtue of decomposition,
will often be equivalent to the orbifolds we will study later.

Consider first a sigma model on a disjoint union of $n$ copies of
the space $X$.  For simplicity, we assume $X$ is Calabi-Yau, so that the
sigma model is a CFT.  As discussed in e.g. \cite{Sharpe:2019ddn},
this theory has a ${\mathbb Z}_n^{(1)} = B {\mathbb Z}_n$ (one-form)
symmetry.  The spectrum of this theory contains
\begin{itemize}
\item an $n$-dimensional space of dimension-zero operators, with basis
given by the identity operators on each component,
\item a set of $n$ projection operators $\Pi_k$, each dimension zero,
obeying
\begin{equation}
\Pi_k \Pi_{\ell} \: = \: \delta_{k,\ell} \Pi_k, \: \: \:
\sum_k \Pi_k \: = \: 1,
\end{equation}
and each corresponding to a (projective) representation of $K$.
\end{itemize}

We can recast this theory in the form of an orbifold of $X$ by
a trivially-acting ${\mathbb Z}_n$, as follows.  Let $\xi$ generate
$k$th roots of unity, and define
\begin{equation}
y \: = \: \sum_{k=0}^{n-1} \xi^k \Pi_k.
\end{equation}
Then, it is straightforward to check that
\begin{equation}
y^p \: = \: \sum_{k=0}^{n-1} \xi^{pk} \Pi_k,
\end{equation}
and $y^p y^q = y^{p+q}$, with identity given by
\begin{equation}
y^0 \: = \: 1 \: = \: \sum_{k=0}^{n-1} \Pi_k,
\end{equation}
and relation $y^n = 1$.

The ring of dimension-zero operators is then given by
${\mathbb C}[y] / (y^n - 1)$.
There is one summand in the disjoint union, one universe,
for every solution of $y^n = 1$.
In addition to the projection operators, which are clearly noninvertible,
this ring has a codimension-one space of noninvertible operators.
We will explore the structure of this ring in greater detail
in section~\ref{sect:bandedex}, 
when we find the same ring of dimension-zero operators
in orbifolds describing banded ${\mathbb Z}_n$ gerbes (i.e.,
orbifolds with trivially-acting central ${\mathbb Z}_n$).
(Indeed, from decomposition, such orbifolds are physically equivalent
to sigma models on disjoint unions, so the same strucure is expected.)

As discussed in detail in \cite{Sharpe:2019ddn},
a sigma model on a disjoint union of $n$ identical copies of a space
has a $B {\mathbb Z}_n$ (one-form) symmetry.  In the present language,
that one-form symmetry acts by phases:
\begin{equation}
y \: \mapsto \: \xi y,
\end{equation}
for $\xi$ an $n$th root of unity.  This is a symmetry of the ring, leaving
the relation $y^n = 1$ invariant.  It does, however, interchange the
vevs of $y$ at which the ring has support, the values of $y$ for
which $y^n = 1$.  This structure is remniscent of spontaneous symmetry
breaking, in that the different vacua are interchanged by the action
of the symmetry, but unlike spontaneous symmetry breaking, here we have
a decomposition and not mere superselection sectors, in this case
by explicit construction of the disjoint union.

Next, consider a sigma model with target a disjoint union of two
different Calabi-Yau spaces, say $X$ and $Y$.  This theory has
a $B {\mathbb Z}_2$ action, but not a symmetry.
The ring of dimension-zero operators
is two-dimensional, from linear combinations of the pair of
identity operators, and has two projectors $\Pi_X$, $\Pi_Y$:
\begin{equation}
\Pi_X^2 \: = \: \Pi_X,  \: \: \:
\Pi_Y^2 \: = \: \Pi_Y, \: \: \:
\Pi_X \Pi_Y \: = \: 0 \: = \: \Pi_Y \Pi_X,
\: \: \:
1 \: = \: \Pi_X + \Pi_Y.
\end{equation}

In principle, we could define the same structure on the ring that we
have discussed previously, defining
\begin{equation}
z \: = \: \Pi_X - \Pi_Y,
\end{equation}
so that $z^2 = 1$, but the ring would then have a symmetry $z \mapsto -z$
which is not reflected in the physics.  A little more generally, we could
instead define
\begin{equation}
z \: = \: \Pi_X  - \alpha \Pi_Y,
\end{equation}
for some number $\alpha \neq 0, -1$,
so that
\begin{equation}
\Pi_X \: = \: \frac{1}{2} \left( 1 + \alpha^{-1} z \right),
\: \: \:
\Pi_Y \: = \: \frac{1}{1 + \alpha} \left( 1 - z \right),
\end{equation}
and
\begin{equation}
z^2 \: = \: \left( \frac{1 + \alpha + 2 \alpha^2 }{ 2(1 + \alpha) } \right)
\: + \:
\left( \frac{1 + \alpha - 2 \alpha^3}{2 \alpha (1 + \alpha) } \right) z,
\end{equation}
in other words a deformation of the ring relation $z^2 = 1$.
We will see an orbifold that gives a ring of this form in
section~\ref{sect:ex:nonbanded:z2z2}.

Finally, let us comment on the boundary Wilson lines.
If $W_i$ is a brane corresponding in the $i$th
universe, then
\begin{equation}
\Pi_i W_j \: = \: \delta_{ij} W_i.
\end{equation}
We will see the same structure when discussing Wilson lines in orbifolds.

\section{Examples in orbifolds} 
\label{sect:ex:orbs}

Next we will describe these structures in orbifold examples.
In each case, we will explicitly describe the ring of dimension zero
operators (twist fields, fusion algebra, projectors, support loci,
and loci of noninvertible operators), as well as the corresponding
structure of Wilson lines.  We will also discuss symmetries in each
example, describing both ordinary 
one-form and noninvertible symmetries, and their
correspondence to the structure of the universes.

Many of these examples have appeared previously in the literature,
where projectors were often computed, on an ad-hoc basis.
We will see that the general formula for projectors in
section~\ref{sect:projectors} correctly predicts the projectors worked
out previously.

\subsection{Banded ${\mathbb Z}_n$ gerbe}
\label{sect:bandedex}

Suppose our banded ${\mathbb Z}_n$ gerbe is an orbifold
$[X/\Gamma]$, where ${\mathbb Z}_n$ is a (trivially-acting)
subgroup of the center of
$\Gamma$.  Let $G = \Gamma/K$, then in general terms decomposition predicts
\begin{equation}
{\rm QFT}\left( [X/\Gamma] \right) \: = \:
{\rm QFT}\left( \coprod_{ \hat{K} } [X/G]_{\hat{\omega}} \right)
\end{equation}
for choices of discrete torsion $\hat{\omega}$. 

In this section we begin by considering general aspects of
banded ${\mathbb Z}_n$ gerbes.  We inclue some specific concrete
examples.

\subsubsection{Ring of dimension-zero operators}

Let $g$ denote a generator of ${\mathbb Z}_n$.  In a sigma model on
a banded ${\mathbb Z}_n$ gerbe, meaning an orbifold or a gauge theory
with a trivially-acting central ${\mathbb Z}_n$, there exist dimension-zero
twist fields corresponding to the elements of ${\mathbb Z}_n$, namely
\begin{equation}
\sigma_1 \: = \: 1, \: \: \:
\sigma_g, \: \: \:
\sigma_{g^2} \: = \: \sigma_g^2, \: \: \:
\cdots \: \: \:
\sigma_{g^{n-1}} \: = \: \sigma_g^{n-1},
\end{equation}
where $\sigma_g^n = 1$.

This ring can be described as ${\mathbb C}[y]/(y^n-1)$,
where $y$ is identified with $\sigma_g$.

The reader should note that
\begin{equation}
{\rm Spec}\, {\mathbb C}[y]/(y^n-1) \: = \: \mbox{ $n$ points }
\: = \: \coprod_n {\rm pt},
\end{equation}
which corresponds to the fact that there are $n$ universes in this example.
In fact, the locations of the points, $y$ such that $y^n = 1$,
correspond to expectation values of the physical order parameter
which distinguish the various universes.

Next, let us compute the projectors.
There are $n$ irreducible representations of $K = {\mathbb Z}_n$,
all one-dimensional and invertible, and in the banded case,
because $K$ is central, the representations are invariant under
$G = \Gamma/K$, so there is
a one-to-one correspondence between irreducible representations of $K$
and universes.
Label those irreducible representations / universes as
 $\rho_i$,
so that $\rho_0$ is the identity operator, and $i$ is counted mod $n$,
then their products are simply
\begin{equation}
\rho_i \otimes \rho_j \: = \: \rho_{i+j}.
\end{equation}
If we let $g$ denote the generator of ${\mathbb Z}_n$, $\xi = \exp(2 \pi i/n)$,
and let representation $i$ denote the representation with character
\begin{equation}
\chi_i(g^k) \: = \: \xi^{-ik},
\end{equation}
then the general expression for projectors~(\ref{eq:genl-def-proj})
reduces to
\begin{equation}
\Pi_k \: = \: \frac{1}{n} \sum_{i=0}^{n-1} \xi^{ik} \sigma_g^k
 \: = \: \frac{1}{n} \sum_{i=0}^{n-1} \xi^{ik} y^i,
\end{equation}
in the notation above.  These projectors -- whose derivation we have
outlined from~(\ref{eq:genl-def-proj}) -- are precisely of the
expected form.
It is straightforward to check
that
\begin{equation}
\Pi_k \Pi_{\ell} \: = \: \delta_{k,\ell} \Pi_k,
\: \: \:
\sum_k \Pi_k \: = \: 1.
\end{equation}
Let $p_i$ be the point $y = \xi^{-i}$, then it is straightforward to
check that
\begin{equation}
\Pi_k |_{p_i} \: = \: \delta_{k,i}.
\end{equation}

Now, let us compute the noninvertible elements of these rings.
The noninvertible elements will include operators proportional to projectors,
but can also include other elements as well.

First, consider the case $n=2$.
Here, it is straightforward to compute that
\begin{equation}
\left( a + b y \right)^{-1} \: = \: \Delta^{-1} \left( a - b y \right),
\end{equation}
for
\begin{equation}
\Delta \: = \: a^2 - b^2.
\end{equation}
Thus, the dimension-zero operator $a + b y$ is noninvertible precisely when
$a^2 = b^2$.  In this simple case
the
projectors onto universes are proportional to $1 \pm y$, so we see immediately
that for $n=2$, the non-invertible elements are proportional to projectors.
(For higher $n$, only some non-invertible elements will be proportional to
projectors.)

One efficient way to do this computation is to identify $y$ with the
matrix
\begin{equation}
\left[ \begin{array}{cc} 0 & 1 \\ 1 & 0 \end{array} \right],
\end{equation}
which encodes the fact that $y^2 = 1$.  Then, 
\begin{equation}
a + b y \: \sim \: \left[ \begin{array}{cc} a & b \\ b & a \end{array} \right],
\end{equation}
and the inverse above can be read off immediately.

Next, consider the case $n=3$.  Here, we identify
\begin{equation}
y \: \sim \: \left[ \begin{array}{ccc} 0 & 0 & 1 \\ 1 & 0 & 0 \\ 0 & 1 & 0
\end{array} \right],
\: \: \:
y^2 \: \sim \: \left[ \begin{array}{ccc} 0 & 1 & 0 \\ 0 & 0 & 1 \\ 1 & 0 & 0
\end{array} \right],
\end{equation}
and proceeding as before one finds 
\begin{equation}
\left( a_0 + a_1 y + a_2 y^2 \right)^{-1} \: = \:
\Delta^{-1} \left(
a_0^2 - a_1 a_2,
a_2^2 - a_0 a_1,
a_1^2 - a_0 a_2 \right),
\end{equation}
where
\begin{eqnarray}
\Delta & = & a_0^3 + a_1^3 + a_2^3 - 3 a_0 a_1 a_2 ,
\\
& = &
(a_0 + a_1 + a_2) 
(a_0^2 - a_0 a_1 + a_1^2 - a_0 a_2 - a_1 a_2 + a_2^2).
\end{eqnarray}
Thus, the noninvertible operators lie along the locus $\{ \Delta = 0 \}$.
This locus include operators proportional to the projectors,
but as those lie along one-dimensional lines, whereas 
$\{ \Delta = 0 \}$ has dimension two.
For example, the operator $(1 + y - 2 y^2)$ is not invertible, but is also
not proportional to a projector.

While considering the case $n=3$, let us also look in more detail at the
projectors and maximal ideals.
The three points at which the order parameters have nonzero vevs are
$y = \xi^i$, for $\xi$ a third root of unity, so the corresponding
maximal ideals are
\begin{equation}
{\mathfrak m}_0 \: = \: (y - 1), \: \: \:
{\mathfrak m}_1 \: = \: (y - \xi^{-1}), \: \: \:
{\mathfrak m}_2 \: = \: (y - \xi^{-2}),
\end{equation}
and the corresponding projectors are 
\begin{equation}
\Pi_0 \: = \: \frac{1}{3} \left(1 + y + y^2 \right), \: \: \:
\Pi_1 \: = \: \frac{1}{3}\left(1 + \xi y + \xi^2 y^2 \right), \: \: \:
\Pi_2 \: = \: \frac{1}{3}\left( 1 + \xi^2 y + \xi y^2 \right).
\end{equation}
It is straightforward to verify that
\begin{equation}
\Pi_0 |_{y=1} \: = \: 1 \: = \: \Pi_1|_{y = \xi^{-1}} \: = \:
\Pi_2 |_{y = \xi^{-2}},
\end{equation}
with other restrictions vanishing, and furthermore
\begin{eqnarray}
(y - \xi^{-1}) (y - \xi^{-2}) & = & 1 + y + y^2 \: \propto \: \Pi_0,
\\
\xi^2 (y - 1) ( y - \xi^{-2}) & = & 1 + \xi y + \xi^2 y^2 \: \propto \: 
\Pi_1,
\\
\xi (y - 1)(y - \xi^{-1}) & = & 1 + \xi^2 y + \xi y^2 \: \propto \: 
\Pi_2,
\end{eqnarray}
hence
\begin{equation}
\Pi_0 \: \in \: {\mathfrak m}_1 \cap {\mathfrak m}_2, \: \: \:
\Pi_1 \: \in \: {\mathfrak m}_0 \cap {\mathfrak m}_2, \: \: \:
\Pi_2 \: \in \: {\mathfrak m}_0 \cap {\mathfrak m}_1.
\end{equation}
Similarly,
\begin{equation}
\{ \Pi_0 = 0 \} \cap \{ \Pi_1 = 0 \} \: = \: \{ y = \xi^{-2} \},
\end{equation}
and so forth.

Finally, for the case $n=4$, one can similarly demonstrate that amongst the
operators $a_0 + a_1 y + a_2 y^2 + a_3 y^3$, the noninvertible
operators lie along
the locus $\{ \Delta = 0 \}$ for
\begin{eqnarray}
\Delta & = &
a_0^4 - a_1^4 + 4 a_0 a_1^2 a_2 - 2 a_0^2 a_2^2 + a_2^4 
- 4 a_0^2 a_1 a_3 - 4 a_1 a_2^2 a_3 + 2 a_1^2 a_3^2 + 4 a_0 a_2 a_3^2
- a_3^4,
\nonumber \\
& = &
( a_0 + a_1 + a_2 + a_3)
( a_0 - a_1 + a_2 - a_3)
( a_0^2 + a_1^2 - 2 a_0 a_2 + a_2^2 - 2 a_1 a_3 + a_3^2 ).
\end{eqnarray}

For completeness, the maximal ideals describing order parameters at which
the universes have support are
\begin{equation}
{\mathfrak m}_0 \: = \: (y-1), \: \: \:
{\mathfrak m}_1 \: = \: (y - \xi^{-1}), \: \: \:
{\mathfrak m}_2 \: = \: (y - \xi^{-2}), \: \: \:
{\mathfrak m}_3 \: = \: (y - \xi^{-3}),
\end{equation}
where $\xi$ generates fourth roots of unity, and the projectors are
\begin{equation}
\Pi_0 \: = \: \frac{1}{4} \left( 1 + y + y^2 + y^3 \right), \: \: \:
\Pi_1 \: = \: \frac{1}{4} \left( 1 + \xi y + \xi^2 y^2 + \xi^3 y^3 \right),
\end{equation}
\begin{equation}
\Pi_2 \: = \: \frac{1}{4} \left( 1 + \xi^2 y + y^2 + \xi^2 y^3 \right), \: \: \:
\Pi_3 \: = \: \frac{1}{4} \left( 1 + \xi^3 y + \xi^2 y^2 + \xi y^3 \right).
\end{equation}
When restricted to points, the restriction of $\Pi_i$ is nonzero only at
$y = \xi^{-i}$, and furthermore, for example,
\begin{equation}
(y - \xi^{-1}) (y - \xi^{-2}) (y - \xi^{-3}) \: = \:
1 + y + y^2 + y^3 \: \propto \: \Pi_0,
\end{equation}
hence
\begin{equation}
\Pi_0 \: \in \: {\mathfrak m}_1 \cap {\mathfrak m}_2 \cap {\mathfrak m}_3,
\end{equation}
and so forth.

\subsubsection{Wilson lines}

Let us return momentarily to the presentation of banded
${\mathbb Z}_k$ gerbes as orbifolds $[X/\Gamma]$ where a central
subgroup ${\mathbb Z}_k \subset \Gamma$ acts trivially.
As reviewed earlier in section~\ref{sect:Wlines}, bulk Wilson lines act
as defect operators linking universes, and boundary Wilson lines are
stratified by choices of universe -- sheaves and bundles on
$[X/\Gamma]$ are equivalent to sheaves and bundles on the
constituent universes $[X/G]_{\hat{\omega}}$, for $G = \Gamma/{\mathbb Z}_k$.

In fact, since ${\mathbb Z}_k$ is central, we can make a slightly
stronger statement:  representations of $\Gamma$ decompose into
projective representations of $G$,
with projectivity determined by the restriction of $\Gamma$ to $K$, which
determines the
discrete torsion $\hat{\omega}$.

In principle, the projectors $\Pi_k$ described in the previous subsection
should project onto the representations associated with various
universes in the decomposition, which we will confirm in examples.

\subsubsection{Symmetries}

Let us briefly make some general comments on one-form symmetries in
banded ${\mathbb Z}_n$ gerbes.  Here, since $K = {\mathbb Z}_n$ acts
trivially, and is both abelian and central, the twist fields all
correspond to single elements of $K$:  $\sigma_g$ rather than merely
$\sigma_{[g]}$.  As a result, the twist fields obey a group-like
multiplication:
\begin{equation}
\sigma_g \sigma_{g'} \: = \: \sigma_{g g'},
\end{equation}
and so define invertible symmetries.

Hand-in-hand, the representations $R$ associated to the universes are
all one-dimensional, because irreducible representations of $K$ are
one-dimensional (since $K$ is abelian), and because the action of $G$
is trivial, so orbits of the $G$ action consist of single elements of
$\hat{K}$.

We emphasize that even in these examples, there exist non-invertible
dimension-zero twist fields, defined by the locus $\{ \Delta = 0 \}$
computed earlier, even though this theory contains only invertible
symmetries.

\subsubsection{Specific example: $\Gamma = D_4$, $K = {\mathbb Z}_2$}
\label{sect:specex:d4}

Next, let us consider a specific concrete example of a banded
${\mathbb Z}_2$ gerbe, namely an orbifold $[X/D_4]$, where
$D_4$ is the eight-element dihedral group, and the center
${\mathbb Z}_2$ acts trivially.
(This example has been previously discussed in e.g.
\cite[section 2.0.1]{Pantev:2005rh}, \cite[section 5.2]{Hellerman:2006zs}.)

We denote the elements of $D_4$ by
\begin{equation}
\{ 1, z, a, b, az, bz, ab, ba = abz \},
\end{equation}
where $z$ generates the center, $a^2 = z^2 = 1$, and $b^2 = z$.

Since the trivially-acting subgroup is (in) the center, this is a banded
gerbe, and this theory admits a $B {\mathbb Z}_2$ symmetry.

Most of the analysis of this example proceeds as in the rest of this
subsection.  For example, there are two irreducible representations of
${\mathbb Z}_2$, call them $\pm$, which are invariant under
$D_4/{\mathbb Z}_2 = {\mathbb Z}_2 \times {\mathbb Z}_2$ as $K$
is central.  From the general formula~(\ref{eq:genl-def-proj})
one quickly computes that the projectors are
\begin{equation}
\Pi_{\pm} \: = \: \frac{1}{2}\left( 1 \pm \tau_z \right),
\end{equation}
where $z$ denotes the generator of the ${\mathbb Z}_2$ center.

Comparing to Wilson lines is more interesting in this example.
The group $D_4$ has five irreducible representations, four one-dimensional,
and one two-dimensional.  The two-dimensional
representation can be given explicitly as
\begin{equation}
a \: = \: \left[ \begin{array}{rr} 1 & 0 \\ 0 & -1 \end{array} \right], \: \: \:
b \: = \: \left[ \begin{array}{rr} 0 & -1 \\ 1 & 0 \end{array} \right], 
\: \: \:
z \: = \: \left[ \begin{array}{rr} -1 & 0 \\ 0 & -1 \end{array} \right],
\end{equation}
and the character table is
\begin{center}
\begin{tabular}{c|crrrr}
& $\{1\}$ & $\{z\}$ & $\{a, az\}$ & $\{b, bz\}$ & $\{ab, ba\}$ \\ \hline
$1$ & $1$ & $1$ & $1$ & $1$ & $1$ \\
$1_a$ & $1$ & $1$ & $1$ & $-1$ & $-1$ \\
$1_b$ & $1$ & $1$ & $-1$ & $1$ & $-1$ \\
$1_{ab}$ & $1$ & $1$ & $-1$ & $-1$ & $1$ \\
$2$ & $2$ & $-2$ & $0$ & $0$ & $0$
\end{tabular}
\end{center}

From the character table, we can see explicitly that each of the
one-dimensional representations $1$, $1_a$, $1_b$, $1_{ab}$ restricts
to the trivial one-dimensional representation of 
${\mathbb Z}_2$.  The nontrivial one-dimensional representation of 
${\mathbb Z}_2$ is not the restriction of any one-dimensional representation
of $D_4$.  However, the two-dimensional representation restricts to the sum
of two copies of the nontrivial one-dimensional representation of
${\mathbb Z}_2$.

It is straightforward to check that
bulk Wilson lines obey
\begin{equation}
\Pi_+ W_{1, 1_a, 1_b, 1_{ab}} \: = \: 
W_{1, 1_a, 1_b, 1_{ab}} \Pi_+, 
\: \: \:
\Pi_+ W_2 \: = \: W_2 \Pi_-,
\end{equation}
\begin{equation}
\Pi_- W_{1, 1_a, 1_b, 1_{ab}} \: = \: W_{1, 1_a, 1_b, 1_{ab}} \Pi_-,
\: \: \:
\Pi_- W_2 \: = \: W_2 \Pi_+,
\end{equation}
so that $W_2$ acts as a defect linking the two universes,
and that boundary Wilson lines obey
\begin{equation}
\Pi_+ W_{1, 1_a, 1_b, 1_{ab}} \: = \: W_{1, 1_a, 1_b, 1_{ab}}, \: \: \:
\Pi_+ W_2 \: = \: 0,
\end{equation}
\begin{equation}
\Pi_- W_{1, 1_a, 1_b, 1_{ab}} \: = \: 0, 
\: \: \:
\Pi_- W_2 \: = \: W_2.
\end{equation}

As expected, the
four one-dimensional representations of $D_4$ correspond to the
four one-dimensional ordinary representations of
${\mathbb Z}_2 \times {\mathbb Z}_2$, and the 
two-dimensional irreducible
representation of $D_4$ corresponds to the single irreducible
(two-dimensional) projective representation of ${\mathbb Z}_2 \times
{\mathbb Z}_2$ \cite[section 3.7]{karpilovsky}.

Not only does the counting match, but indeed so do the representations
themselves.  Since the four one-dimensional representations of $D_4$
are trivial
on the center generator $z$, they descend to four honest representations
of $D_4/{\mathbb Z}_2 = {\mathbb Z}_2 \times {\mathbb Z}_2$, which are easily checked from
the character table to be distinct.  Similarly, the fact that the
two-dimensional representation of $D_4$ is nontrivial on $z$ means that
it descends to a projective representation of $D_4/{\mathbb Z}_2$.
Explicitly, if we let the generators of ${\mathbb Z}_2 \times
{\mathbb Z}_2$ be denoted $\overline{a}$, $\overline{b}$, and define
a representation $\rho$ by
\begin{equation}
\rho(\overline{a}) \: = \: \left[ \begin{array}{rr} -1 & 0  \\ 0 & 1 \end{array}
\right], \: \: \:
\rho(\overline{b}) \: = \: \left[ \begin{array}{rr} 0 & -1 \\ 1 & 0 \end{array}
\right], \: \: \:
\rho( \overline{a}\overline{b}) \: = \: \left[
\begin{array}{cc} 0 & 1 \\ 1 & 0 \end{array} \right]
\end{equation}
(the images of $az$, $b$, $abz$ under the two-dimensional $D_4$ representation),
then we find that $\rho$ is a projective representation of 
${\mathbb Z}_2 \times {\mathbb Z}_2$, coinciding with that given in
\cite[section 3.7]{karpilovsky}.

Finally, we observe that the projectors $\Pi_{\pm}$ correctly select
out the representations associated with each universe.
Specifically, $\Pi_+$ projects onto the four one-dimensional representations
of $D_4$, which we have seen descend to
honest representations of ${\mathbb Z}_2 \times {\mathbb Z}_2$,
and $\Pi_-$ projects onto the irreducible
two-dimensional representation of $D_4$, which descends to a projective
representation of ${\mathbb Z}_2 \times {\mathbb Z}_2$.
Thus, the projectors are correctly projecting onto boundary Wilson lines associated
with the two universes.

If instead we had used $\Gamma = {\mathbb H}$, the eight-element group of
unit quaternions, and picked the trivially-acting subgroup to be the
center $K = {\mathbb Z}_2$, we would have identical results:
the projectors have the same form, ${\mathbb H}$ has five irreducible
representations, four of which are one-dimensional, one of which is
two-dimensional, and the character table is the same as for $D_4$,
so that the four one-dimensional representations all restrict to the
trivial representation of $K = {\mathbb Z}_2$, and the two-dimensional
representation restricts to two copies of the nontrivial representation
of ${\mathbb Z}_2$.

The banded gerbe $[X/{\mathbb H}]$ has the same decomposition as for $D_4$:
\begin{equation}
{\rm QFT}\left( [X/{\mathbb H}] \right) \: = \:
{\rm QFT}\left( [X/{\mathbb Z}_2 \times {\mathbb Z}_2] \, \coprod \,
[X / {\mathbb Z}_2 \times {\mathbb Z}_2]_{\rm d.t.} \right).
\end{equation}
The five irreducible representations of ${\mathbb H}$ also
naturally decompose into four ordinary irreducible representations of
${\mathbb Z}_2 \times {\mathbb Z}_2$ plus one projective (two-dimensional)
irreducible representation of ${\mathbb Z}_2 \times {\mathbb Z}_2$,
which are selected by the corresponding projectors, as expected.

\subsubsection{Specific example:  $\Gamma = D_8$, $K = {\mathbb Z}_2$}
\label{sect:specex:d8}

In this section we consider one more example of a banded gerbe,
namely, $[X/D_8]$ with the central $K = {\mathbb Z}_2 \subset D_8$
acting trivially.

The group $D_8$ is the sixteen-element dihedral group generated by
$\tilde{a}$, $\tilde{b}$, subject to the relations
\begin{equation}
\tilde{a}^2 \: = \: 1, \: \: \:
\tilde{b}^{8} \: = \: 1, \: \: \:
\tilde{a} \tilde{b} \tilde{a} \: = \: \tilde{b}^{-1} \: = \: \tilde{b}^7.
\end{equation}
The center is generated by $\tilde{b}^4$.

Computing e.g. the genus-one partition function, one can quickly verify the
decomposition
\begin{equation}
{\rm QFT}\left( [X/D_8] \right) \: = \:
{\rm QFT}\left( [X/D_4] \, \coprod \,
[X/D_4]_{\rm d.t.} \right),
\end{equation}
where $H^2(D_4,U(1)) = {\mathbb Z}_2$, with phases given in
\cite[appendix D.3]{Robbins:2021ylj}, \cite[section 3.7]{karpilovsky}.

If we let $z$ denote the central element of $D_8$,
then the projectors can be computed in exactly the same fashion as in
the previous example, and one finds from the 
general formula~(\ref{eq:genl-def-proj}) that, for the irreducible
representations $\pm 1$ of ${\mathbb Z}_2$,
\begin{equation}
\Pi_{\pm 1} \: = \: \frac{1}{2} \left( 1 \pm \tau_z \right).
\end{equation}

The group $D_8$ has seven irreducible representations, four of dimension one
and three of dimension two.  The conjugacy classes of $D_8$ are
\begin{equation}
\{ 1 \}, \: \: \:
\{ \tilde{z} \}, \: \: \:
\{ \tilde{b}, \tilde{b}^7 = \tilde{b}^3 \tilde{z} \}, \: \: \:
\{ \tilde{b}^2, \tilde{b}^6 = \tilde{b}^2 \tilde{z} \}, \: \: \:
\{ \tilde{b}^3, \tilde{b}^5 = \tilde{b} \tilde{z} \}, \: \: \:
\{ \tilde{a}, \tilde{a}\tilde{z}, \tilde{a}\tilde{b}^2, 
\tilde{a}\tilde{b}^2\tilde{z} \}, \: \: \:
\{ \tilde{b}\tilde{a} = \tilde{a}\tilde{b}^7, \tilde{a}\tilde{b}, 
\tilde{a}\tilde{b}^5, \tilde{a}\tilde{b}^3 \}.
\end{equation}
for $\tilde{z} = \tilde{b}^4$.

The character table of ordinary irreducible representations of $D_8$ is
\begin{center}
\begin{tabular}{c|ccccccc}
 & $1$ & $\tilde{z}$ & $\{\tilde{b}, \tilde{b}^7\}$ &
$\{ \tilde{b}^2, \tilde{b}^6\}$ &
$\{\tilde{b}^3, \tilde{b}^5\}$ &
$\{ \tilde{a}, \cdots \}$ &
$\{ \tilde{b} \tilde{a}, \cdots \}$ \\ \hline
$1$ & $1$ & $1$ & $+1$ & $1$ & $+1$ & $+1$ & $+1$ \\
$1_b$ & $1$ & $1$ & $+1$ & $1$ & $+1$ & $-1$ & $-1$ \\
$1_c$ & $1$ & $1$ & $-1$ & $1$ & $-1$ & $+1$ & $-1$ \\
$1_d$ & $1$ & $1$ & $-1$ & $1$ & $-1$ & $-1$ & $+1$ \\
$2_1$ & $2$ & $-2$ & $2 \cos( \pi/4)$ & $2\cos( \pi/2)$ & $2 \cos(3 \pi/4)$ 
& $0$ & $0$ \\
$2_2$ & $2$ & $+2$ & $2 \cos(\pi/2)$ & $-2$ & $2 \cos(3 \pi/2)$ 
& $0$ & $0$ \\
$2_3$ & $2$ & $-2$ & $2 \cos(3 \pi / 4)$ & $2 \cos(3 \pi / 2)$ & $2\cos(\pi/4)$
& $0$ & $0$
\end{tabular}
\end{center}
The three ordinary two-dimensional irreducible representations are defined by
\begin{equation}
\rho_r(b^i a^j) \: = \: B_r^i A_r^j
\end{equation}
for 
\begin{equation}
A_r \: = \: \left[ \begin{array}{cc} 0 & 1 \\ 1 & 0 \end{array} \right],
\: \: \:
B_r \: = \: \left[ \begin{array}{cc} \xi^r & 0 \\ 0 & \xi^{-r} \end{array}
\right],
\end{equation}
for $r \in \{1, 2, 3 \}$, $\xi = \exp(2 \pi i/8)$.

Applying the projectors $\Pi_{\pm 1}$ to bulk Wilson lines, we find
\begin{equation}
\Pi_{\pm 1} W_{1, 1_b, 1_c, 1_d} \: = \: W_{1, 1_b, 1_c, 1_d} \Pi_{\pm 1},
\end{equation}
\begin{equation}
\Pi_{\pm 1} W_{2_1} \: = \: W_{2_1} \Pi_{\mp 1},
\: \: \:
\Pi_{\pm 1} W_{2_2} \: = \: W_{2_2} \Pi_{\pm 1},
\: \: \:
\Pi_{\pm 1} W_{2_3} \: = \: W_{2_3} \Pi_{\mp 1},
\end{equation}
so that $W_{2_1, 2_3}$ act as defects linking different universes,
and for boundary Wilson lines, we find
\begin{equation}
\Pi_{+1} W_{1, 1_b, 1_c, 1_d} \: = \: W_{1, 1_b, 1_c, 1_d},
\: \: \:
\Pi_{+1} W_{2_1, 2_3} \: = \: 0, \: \: \:
\Pi_{+1} W_{2_2} \: = \: W_{2_2},
\end{equation}
\begin{equation}
\Pi_{-1} W_{1, 1_b, 1_c, 1_d} \: = \: 0, \: \: \:
\Pi_{-1} W_{2_1, 2_3} \: = \: W_{2_1, 2_3}, \: \: \:
\Pi_{-1} W_{2_2} \: = \: 0.
\end{equation}
Much as in the last example, the nontrivial representation of $K = 
{\mathbb Z}_2$ only appears in the restriction of (two of the)
two-dimensional representations, and not as the restriction of a 
one-dimensional representation.

By comparison, the group $D_4$ has five ordinary irreducible representations,
four one-dimensional and one two-dimensional,
and two irreducible projective representations, both two-dimensional
\cite[section 3.7]{karpilovsky}.  This matches the prediction of
decomposition:  
the $D_8$ representations
are a union of $D_4$ representations with and without discrete torsion.
The projectors $\Pi_{\pm 1}$ above correctly identify which representations
of $D_8$ descend to the twisted and
untwisted representations of $D_4$:  $\Pi_{+1}$ projects onto
ordinary $D_4$ representations, and $\Pi_{-1}$ projects onto projective
$D_4$ representations.  Indeed, the representations selected out by
$\Pi_{+1}$ are invariant under $\tilde{z}$, and so descend to honest 
representations of $D_4$, whereas the representations selected out by
$\Pi_{-1}$ are not invariant under $\tilde{z}$, and so descend to
projective representations of $D_4$, in essentially the same fashion as
we observed in the previous subsection for the example there.

\subsection{Nonbanded ${\mathbb Z}_4$ gerbe}
\label{sect:ex:nonband:z4}

Next, consider a nonbanded ${\mathbb Z}_4$ gerbe, described as the
orbifold $[X/{\mathbb H}]$, where ${\mathbb H}$ is the eight-element group
of unit quaternions, and ${\mathbb Z}_4 \cong \langle i \rangle$ acts trivially.
This example decomposes into three universes as
\begin{equation}
{\rm QFT}\left( [X/{\mathbb H}] \right)
\: = \:
{\rm QFT}\left( X \, \coprod \, [X/{\mathbb Z}_2] \, \coprod \,
[X/{\mathbb Z}_2] \right).
\end{equation}
(This example is discussed in \cite[section 2.0.4]{Pantev:2005rh} as well as
\cite[section 5.4]{Hellerman:2006zs},
where its decomposition was checked via its spectrum of operators and
in multiloop partition functions.)

\subsubsection{Ring of dimension-zero operators}

Here, as discussed in section~\ref{app:computation},
the dimension-zero twist fields are
\begin{equation}
\sigma_{[+1]}, \: \: \:
\sigma_{[-1]}, \: \: \:
\sigma_{[i]},
\end{equation}
which obey
\begin{equation}
\sigma_{[+1]}^2 \: = \: \sigma_{[+1]},
\: \: \:
\sigma_{[+1]} \sigma_{[-1]} \: = \: \sigma_{[-1]},
\: \: \:
\sigma_{[+1]} \sigma_{[i]} \: = \: \sigma_{[i]},
\end{equation}
\begin{equation}
\sigma_{[-1]}^2 \: = \: \sigma_{[+1]},
\: \: \:
\sigma_{[-1]} \sigma_{[i]} \: = \: \sigma_{[i]},
\: \: \:
\sigma_{[i]}^2 \: = \: (1/2) \left( \sigma_{[+1]} + \sigma_{[-1]} \right).
\end{equation}
Clearly we can identify $\sigma_{[+1]}$ with the identity,
and $\sigma_{[-1]}$ generates a ${\mathbb Z}_2$.  These are consistent
with the fact that, although this is a ${\mathbb Z}_4$ gerbe,
it only has a $B {\mathbb Z}_2$ (one-form) symmetry, as only
${\mathbb Z}_2 \subset {\mathbb H}$ is central.

Identifying $x$ with $\sigma_{[-1]}$ and $y$ with $\sigma_{[i]}$,
we can write the ring of dimension-zero operators more efficiently as
\begin{equation}
{\mathbb C}[x,y] / \left( x^2 - 1, xy - y, y^2 - (1/2) (1+x) \right).
\end{equation}
Geometrically, this describes a complete intersection
of three quadrics in ${\mathbb C}^2$, and it is straightforward
to see the only solutions are
\begin{equation}
(x,y) \: = \: \left\{ (+1,+1), \: \: \:
 (+1,-1), \: \: \:
 (-1,0) \right\},
\end{equation}
corresponding to three points.

Physically, as discussed in \cite[section 5.4]{Hellerman:2006zs}, 
this orbifold decomposes into
three pieces:
\begin{equation}
{\rm QFT}\left( [X/{\mathbb H}] \right) \: = \:
{\rm QFT}\left( X \coprod [X/{\mathbb Z}_2] \coprod [X/{\mathbb Z}_2] \right).
\end{equation}
The fact that two of the universes match reflects a
$B {\mathbb Z}_2$ (one-form) symmetry in the decomposition,
matching the (center) symmetry of the ${\mathbb H}$ orbifold.
The fact that the third does not, reflects a noninvertible symmetry.

In terms of the ring of dimension-zero operators,
this $B {\mathbb Z}_2$ acts as
\begin{eqnarray}
y & \mapsto & - y, \\
x & \mapsto & x \mbox{ (invariant)}.
\end{eqnarray}
This symmetry leaves the relations invariant, but interchanges
two of the universes (with nonzero $y$ vevs), as one would expect
from the physical decomposition.
(As remarked elsewhere, this structure is reminiscent of spontaneous
symmetry breaking, though the details differ here, as the theory
exhibits decomposition and not superselection sectors.)

In passing, let us describe another way to arrive at the ring above.
Mathematically, the $[X/{\mathbb H}]$ gerbe is a nonbanded
${\mathbb Z}_4$ gerbe, which means it looks like a fiber bundle
with fibers $B {\mathbb Z}_4$, but the transition functions include
a nontrivial bundle of outer automorphisms, and as a result, the ring
in this case can be derived by taking the ${\mathbb Z}_2$-invariants
of the ring of dimension-zero operators of a banded ${\mathbb Z}_4$
gerbe.  In terms of the quaternions ${\mathbb H}$ and
${\mathbb Z}_4 \cong \langle i \rangle \subset {\mathbb H}$,
we can lift the ${\mathbb Z}_2 = {\mathbb H}/\langle i \rangle$ via a 
section $s$ which we take to be $s(+1) = 1$,
$s(-1) = j$, so that $s(-1)^{-1} = -j$.
The action on any element $g \in {\mathbb Z}_4$ is
$g \mapsto s^{-1} g s$, for which we find $\pm 1 \in \langle i \rangle$
are invariant but $i \mapsto -i$.  In the ring ${\mathbb C}[y]/(y^4-1)$
of the banded ${\mathbb Z}_4$ gerbe, this maps $y \mapsto -y$, leaving
$y^2$ invariant.  Define $\tilde{x} = y^2$,
$\tilde{y} = (1/2)(y + y^3)$, and then the ${\mathbb Z}_2$
invariants are
\begin{equation}
\left( {\mathbb C}[y]/(y^4 - 1) \right)^{ {\mathbb Z}_2 } \: = \:
{\mathbb C}[\tilde{x},\tilde{y}]/( \tilde{x}^2 - 1, \tilde{x} \tilde{y} - 
\tilde{y}, \tilde{y}^2 - (1/2)(1 + \tilde{x}) ),
\end{equation}
recovering the ring of the nonbanded ${\mathbb Z}_4$ gerbe,
as desired.

Next, let us compute the noninvertible operators.
It is straightforward to show that
\begin{eqnarray}
\lefteqn{
\left(a + b x + cy \right)^{-1}
}  \\
& = &
\frac{1}{2}\left( \frac{1}{a-b} - \frac{ (a+b) }{  c^2 - (a+b)^2 }
\right) \: - \:
 \frac{x}{2} \left( \frac{1}{a-b} + \frac{ (a+b) }{  c^2 - (a+b)^2 } \right)
\: + \: y \frac{c}{c^2 - (a+b)^2},  \nonumber
\end{eqnarray}
hence we see the operator is noninvertible 
if it lies along the locus $\{\Delta = 0\}$ where
\begin{equation}
\Delta \: = \: (a-b) \left( c^2 - (a+b)^2 \right).
\end{equation}

Now, let us compute the projectors onto universes.
Such projectors are already listed in \cite[section 5.4]{Hellerman:2006zs};
let us instead compute them from the general expression~(\ref{eq:genl-def-proj})
and then compare to the results in \cite{Hellerman:2006zs}.
The (one-dimensional) irreducible representations of $\langle i \rangle 
\cong {\mathbb Z}_4$ can be characterized by their values on $i$:
$\rho_{\pm 1}(i) = \pm 1$, $\rho_{\pm i}(i) = \pm i$.
From the discussion above, we see that the action of 
${\mathbb H}/\langle i \rangle = {\mathbb Z}_2$ on the
irreducible representations leaves $\rho_{\pm 1}$ invariant but
exchanges $\rho_i \leftrightarrow \rho_{-i}$.
Therefore, the universes of decomposition correspond to the representations
$\rho_{+1}$, $\rho_{-1}$, and $\rho_{+i} \oplus \rho_{-i}$ of 
${\mathbb Z}_4$.

Letting $g$ denote the generator of ${\mathbb Z}_4$,
we have the ${\mathbb Z}_4$ characters
\begin{equation}
\chi_{\pm 1}(g) \: = \: \pm 1, \: \: \:
\chi_{\pm i}(g) \: = \: \pm i,
\end{equation}
from which
we find from the general formula~(\ref{eq:genl-def-proj}) that
\begin{eqnarray}
\Pi_{R=1} & = & \frac{1}{| {\mathbb Z}_4 |} \sum_{k=0}^3
\chi_1(g^{-k}) \tau_g^k,
\\
& = & \frac{1}{4} \left( 1 + x + \tau_i + \tau_{-i} \right),
\\
& = & \frac{1}{4} \left( 1 + x + 2y \right),
\\
\Pi_{R = -1} & = & \frac{1}{|{\mathbb Z}_4|} \sum_{k=0}^3
\chi_{-1}(g^{-k}) \tau_g^k,
\\
& = & \frac{1}{4} \left( 1 + x - 2y \right),
\end{eqnarray}
Finally, from
\begin{eqnarray}
\Pi_{\rho_i} & = & \frac{1}{| {\mathbb Z}_4|}  \sum_{k=0}^3
\chi_i(g^{-k}) \tau_g^k, 
\\
& = & \frac{1}{4} \left( 1 - i \tau_i - \tau_{-1} + i \tau_{-i} \right),
\\
\Pi_{\rho_{-i}}  & = & \frac{1}{| {\mathbb Z}_4|}  \sum_{k=0}^3
\chi_{-i}(g^{-k}) \tau_g^k, 
\\
& = & \frac{1}{4} \left( 1 + i \tau_i - \tau_{-1} - i \tau_{-i} \right),
\end{eqnarray}
we find
\begin{eqnarray}
\Pi_{R=[2]} & = & \Pi_{\rho_i} + \Pi_{\rho_{-i}}
\: = \:
\frac{1}{4}\left(2 - 2 \tau_{-1} \right)
\: = \: \frac{1}{2}\left( 1 - x \right).
\end{eqnarray}

The projectors computed above match the projectors worked out on an ad-hoc
basis in \cite[section 5.4]{Hellerman:2006zs}, that project operators
onto each of the three universes:
\begin{eqnarray}
\Pi_1 & = & \frac{1}{4} \left( 1 + x + 2y \right),
\\
\Pi_{-1} & = & \frac{1}{4} \left( 1 + x - 2y \right),
\\
\Pi_2 & = & \frac{1}{2} \left( 1 - x \right),
\end{eqnarray}
which are easily checked to obey
\begin{equation}
\Pi_i \Pi_j \: = \: \delta_{i,j} \Pi_i, \: \: \:
\Pi_1 + \Pi_{-1} + \Pi_2 \: = \: 1.
\end{equation}
As observed in \cite[section 5.4]{Hellerman:2006zs},
from looking at twisted sector states,
$\Pi_{\pm 1}$ project onto the universes $[X/{\mathbb Z}_2]$,
and $\Pi_2$ projects onto universe $X$.
It is also straightforward to check that each of these projectors
lies along the locus $\{ \Delta = 0 \}$, as expected as they are not
invertible.

In passing, note that the projectors above have the property that
when restricted to the possible vacua, the complete intersection of
quadrics, each projector is nonzero on precisely one point:
\begin{equation}
\Pi_1|_{(x,y) = (+1,+1)} \: = \: 1,
\: \: \:
\Pi_{-1}|_{(x,y) = (+1,-1)} \: = \: 1,
\: \: \:
\Pi_2|_{(x,y) = (-1,0)} \: = \: 1,
\end{equation}
with all other restrictions zero.  Phrased another way,
the point $(x,y) = (+1,+1)$ is the locus $\Pi_{-1} = 0 = \Pi_2$, 
and so forth.

These statements correspond more formally to statements in
commutative algebra.  First, the points $(x,y) = (+1,+1),
(+1,-1), (-1,0)$ correspond to the maximal ideals
\begin{equation}
{\mathfrak m}_1 \: = \: (x-1,y-1), \: \: \:
{\mathfrak m}_{-1} \: = \: (x-1,y+1), \: \: \:
{\mathfrak m}_2 \: = \: (x+1,y).
\end{equation}
Each projector is in all of the maximal ideals save one.
For example,
\begin{equation}
(x-1) + 2(y+1) \: = \: 1 + x + 2y \: \propto \: \Pi_1,
\end{equation}
hence $\Pi_1 \in {\mathfrak m}_{-1}$, and similarly,
\begin{equation}
(1+y)(x+1) \: = \: 1 + x + 2y \: \propto \: \Pi_1,
\end{equation}
hence $\Pi_1 \in {\mathfrak m}_2$ also.
However, $\Pi_1 \not\in {\mathfrak m}_1$, as its restriction to the
corresponding point is nonzero.
Similarly, one can show
\begin{equation}
\Pi_{-1} \: \in \: {\mathfrak m}_1 \cap {\mathfrak m}_2, \: \: \:
\Pi_2 \: \in \: {\mathfrak m}_1 \cap {\mathfrak m}_{-1}.
\end{equation}
As a result, for example,
\begin{equation}
( \Pi_{-1} )_{ {\mathfrak m}_1 } \: = \: 0 \: = \:
( \Pi_2 )_{ {\mathfrak m}_1 }
\end{equation}
since $\Pi_1 \not\in {\mathfrak m}_1$.

\subsubsection{Wilson lines}

In this section we will consider bulk and boundary
Wilson lines on $[X/{\mathbb H}]$, the latter via
representation of ${\mathbb H}$, and how they correspond to defects between
and sheaves on
univeres of the decomposition of $[X/{\mathbb H}]$.

The group ${\mathbb H}$ has five irreducible representations,
four one-dimensional, and one two-dimensional.
The two-dimensional representation can be given explicitly as
\begin{equation}
\rho_2(1) \: = \: \left[ \begin{array}{cc} 1 & 0 \\ 0 & 1 \end{array} \right],
\: \: \:
\rho_2(-1) \: = \: \left[ \begin{array}{rr} -1 & 0 \\ 0 & -1 \end{array} \right],
\end{equation}
\begin{equation}
\rho_2(i) \: = \: \left[ \begin{array}{rr} i & 0 \\ 0 & -i \end{array} \right],
\: \: \:
\rho_2(j) \: = \: \left[ \begin{array}{rr} 0 & -1 \\ 1 & 0 \end{array} \right],
\: \: \:
\rho_2(k) \: = \: \left[ \begin{array}{rr} 0 & -i \\ -i & 0 \end{array} \right].
\end{equation}
It will be convenient to refer to the character table,
which we give below:
\begin{center}
\begin{tabular}{c|crrrr}
 & $+1$ & $-1$ & $\{ \pm i \}$ & $\{ \pm j \}$ & $\{ \pm k \}$ \\ \hline
$1$ (trivial) & $1$ & $1$ & $1$ & $1$ & $1$ \\
$1_i$ & $1$ & $1$ & $1$ & $-1$ & $-1$ \\
$1_j$ & $1$ & $1$ & $-1$ & $1$ & $-1$ \\
$1_k$ & $1$ & $1$ & $-1$ & $-1$ & $1$ \\
$2$ & $2$ & $-2$ & $0$ & $0$ & $0$
\end{tabular}
\end{center}
Using the character table and also \cite{ty}, one can show
\begin{eqnarray}
1_i^2 & = & 1, \\
1_i \otimes 1_j & = & 1_k \mbox{ and cyclically,} \\
1_{i,j,k} \otimes 2 & = & 2, \\
2 \otimes 2 & = & 1 + 1_i + 1_j + 1_k.
\end{eqnarray}

Now, consider the restriction to $\langle i \rangle \cong {\mathbb Z}_4$.
We characterize the one-dimensional
irreducible representations of $\langle i \rangle$
by their values on $i$:
$\rho_{\pm 1}(i) = \pm 1$, $\rho_{\pm i}(i) = \pm i$.
\begin{center}
\begin{tabular}{cc}
Rep' of ${\mathbb H}$ & Restriction \\ \hline
$1$ & $\rho_{+1}$ \\
$1_i$ & $\rho_{+1}$ \\
$1_j$ & $\rho_{-1}$ \\
$1_k$ & $\rho_{-1}$ \\
$2$ & $\rho_{i} \oplus \rho_{-i}$
\end{tabular}
\end{center}
Thus, we see that three representations appear in the restriction:
$\rho_{+1}$, $\rho_{-1}$, and $[2] \equiv \rho_{i} \oplus \rho_{-i}$.
It is straightforward
to check that multiplication in ${\mathbb H}$ induces
\begin{equation}
\rho_{+1}  \otimes \rho_{-1}  \: = \: \rho_{-1}, \: \: \:
\rho_{-1} \otimes \rho_{-1} \: = \: \rho_{+1}, 
\end{equation}
\begin{equation}
\rho_{-1} \otimes [2] \: = \: [2], \: \: \:
[2] \otimes [2] \: = \: \rho_{+1} \oplus \rho_{+1} \oplus 
\rho_{-1} \oplus \rho_{-1},
\end{equation}
consistent with the restrictions.
Clearly, $\rho_{-1}$ is invertible, but $[2]$ is not.

Now, let us compute the representations associated with the universes in
the decomposition
\begin{equation}
\left[ \frac{ X \times \hat{K} }{ G } \right]_{\hat{\omega}}
\: = \:
X \, \coprod \, [X/{\mathbb Z}_2] \, \coprod \, [X/{\mathbb Z}_2],
\end{equation}
for $G = {\mathbb H}/\langle i \rangle
= {\mathbb Z}_2$.
In the decomposition of $[X/{\mathbb H}]$, $G$
acts trivially on $\rho_{\pm 1}$ but exchanges
$\rho_{+i} \leftrightarrow \rho_{-i}$, so the representations associated
with the three universes are $\rho_{+1}$, $\rho_{-1}$, and
$\rho_{+i} \oplus \rho_{-i}$, which happen to match the representations 
appearing as restrictions of the irreducible representations of
${\mathbb H}$.  Of these, $\rho_{+1}$ and $\rho_{-1}$ each correspond to
an $[X/{\mathbb Z}_2]$, whereas $\rho_{[2]}$ corresponds to
\begin{equation}
\left[ \frac{ X \times \hat{\mathbb Z}_2 }{ {\mathbb Z}_2 } \right]
\: = \: X.
\end{equation}

From the character table and the restrictions to $\langle i \rangle \subset 
{\mathbb H}$, 
\begin{itemize}
\item the representations $1$, $1_i$ transform as $\rho_{+1}$ under $K = 
\langle i \rangle$
and so correspond to one copy of $[X/{\mathbb Z}_2]$,
\item the representations $1_j$, $1_k$ transform as $\rho_{-1}$ under
$K$ and so correspond to another copy of $[X/{\mathbb Z}_2]$,
\item the representation $2$ transforms as $[2]$ under
$K$ and so corresponds to $X$.
\end{itemize}

Next, we will consider the action on bulk and boundary Wilson lines,
and in the latter, recover that same classification above from the projectors.

For bulk Wilson lines, 
if we label Wilson lines by the restrictions of representations of
${\mathbb H}$ to $\langle i \rangle$,
we have
\begin{equation}
\sigma_{[-1]} W_1 \: = \: W_1 \sigma_{[-1]}, \: \: \:
\sigma_{[-1]} W_{-1} \: = \: W_{-1} \sigma_{[-1]},
\end{equation}
\begin{equation}
\sigma_{[i]} W_1 \: = \: W_1 \sigma_{[i]}, \: \: \:
\sigma_{[i]} W_{-1} \: = \: - W_{-1} \sigma_{[i]}.
\end{equation}
Since $W_{[2]}$ involves a higher-dimension representation, we proceed
more carefully.  Note
\begin{equation}
\sigma_{[-1]} \rho_i \: = \: - \rho_i \sigma_{[-1]}, \: \: \:
\sigma_{[-1]} \rho_{-i} \: = \: - \rho_{-i} \sigma_{[-1]},
\end{equation}
\begin{equation}
\sigma_{[i]} \rho_{\pm i} \: = \: \pm i \rho_{\pm i} \sigma_{[i]}.
\end{equation}
This implies
\begin{equation}
\sigma_{[-1]} W_{[2]} \: = \: - W_{[2]} \sigma_{[-1]}, \: \: \:
\sigma_{[i]} W_{[2]} \: = \: 0.
\end{equation}

For the projection operators
\begin{equation}
\Pi_{\pm 1} \: = \: \frac{1}{4} \left( 1 + \sigma_{[-1]} \pm 2 \sigma_{[i]} 
\right), \: \: \:
\Pi_2 \: = \: \frac{1}{2} \left( 1 - \sigma_{[-1]} \right),
\end{equation}
now labelling them by representations of ${\mathbb H}$,
we find
\begin{equation}
\Pi_{\pm 1} W_{1, 1_i} \: = \: W_{1, 1_i} \Pi_{\pm 1}, 
\: \: \:
\Pi_2 W_{1, 1_i} \: = \: W_{1, 1_i} \Pi_2,
\end{equation}
\begin{equation}
\Pi_{1} W_{1_j, 1_k} \: = \: W_{1_j, 1_k} \Pi_{-1},
\: \: \: 
\Pi_{-1} W_{1_j, 1_k} \: = \: W_{1_j, 1_k} \Pi_{+1},
\: \: \:
\Pi_2 W_{-1} \: = \: W_{-1} \Pi_2,
\end{equation}
\begin{equation}
\Pi_1 W_2 \: = \: \frac{1}{2} W_2 \Pi_2,
\: \: \:
\Pi_{-1} W_2 \: = \: \frac{1}{2} W_2 \Pi_2,
\: \: \:
\Pi_2 W_2 \: = \: W_2 \left( \Pi_{+1} + \Pi_{-1} \right).
\end{equation}
Thus, we see that $W_{1_j, 1_k}$ acts as a defect linking universes
$\pm 1$ on either side, and $W_2$ acts as a defect linking universe 2
to either of universes $\pm 1$.

Next, we repeat this analysis for boundary Wilson lines.
From the pairing $\tau_g W_{\rho} = \chi_{\rho}(g) W_{\rho}$,
if we label Wilson lines by the restrictions of representations of
${\mathbb H}$ to $\langle i \rangle$,
we have
\begin{equation}
\sigma_{[-1]} W_1 \: = \: W_1, \: \: \:
\sigma_{[-1]} W_{-1} \: = \: W_{-1}, 
\end{equation}
\begin{equation}
\sigma_{[i]} W_1 \: = \: W_1, \: \: \:
\sigma_{[i]} W_{-1} \: = \: - W_{-1}.
\end{equation}
Since $W_{[2]}$ involves a higher-dimension representation, we should
be more careful.  First, note
\begin{equation}
\sigma_{[-1]} \rho_i \: = \: - \rho_i, \: \: \:
\sigma_{[-1]} \rho_{-i} \: = \: - \rho_{-i},
\end{equation}
\begin{equation}
\sigma_{[i]} \rho_i \: = \: 0 \: = \: \sigma_{[i]} \rho_{-i}.
\end{equation}
Then, the effect of $\sigma_{[-1]}$ is to insert in $W_{[2]}$ the
diagonal matrix $-I$, so that
\begin{equation}
\sigma_{[-1]} W_{[2]} \: = \: - W_{[2]},
\end{equation}
and the effect of $\sigma_{[i]}$ is to insert in $W_{[2]}$ the
diagonal matrix $0$, so that
\begin{equation}
\sigma_{[i]} W_{[2]} \: = \: 0.
\end{equation}
Note that in both cases, the eigenvalue corresponding to $W_{[2]}$ is
precisely $\chi_{[2]}(g)/\dim [2]$, as expected.

For the projection operators
\begin{equation}
\Pi_{\pm 1} \: = \: \frac{1}{4} \left( 1 + \sigma_{[-1]} \pm 2 \sigma_{[i]} 
\right), \: \: \:
\Pi_2 \: = \: \frac{1}{2} \left( 1 - \sigma_{[-1]} \right),
\end{equation}
now labelling them by representations of ${\mathbb H}$,
we find
\begin{equation}
\Pi_{1} W_{1, 1_i} \: = \: W_{1, 1_i}, \: \: \:
\Pi_{-1} W_{1, 1_i} \: = \: 0 \: = \: \Pi_2 W_{1, 1_i},
\end{equation}
\begin{equation}
\Pi_{1} W_{1_j, 1_k} \: = \: 0, \: \: \:
\Pi_{-1} W_{1_j, 1_k} \: = \: W_{1_j, 1_k}, 
\: \: \:
\Pi_2 W_{1_j, 1_k} \: = \: 0,
\end{equation}
\begin{equation}
\Pi_{\pm 1} W_{2} \: = \: 0, \: \: \:
\Pi_2 W_{2} \: = \: W_{2}.
\end{equation}
As expected, the projectors project onto Wilson lines associated with
the corresponding universes.

\subsubsection{Symmetries}

Now, let us turn to the symmetries of the $[X/{\mathbb H}]$ orbifold.
There is a $B {\mathbb Z}_2$ symmetry, corresponding to the fact that
a ${\mathbb Z}_2$ subgroup of the trivially-acting $\langle i \rangle$
is the center of ${\mathbb H}$.  In addition, there is a noninvertible
symmetry, corresponding to the fact that one of the three universes is not
like the others.

The theory has twist fields
\begin{equation}
x \: = \: \sigma_{[-1]}, \: \: \:
y \: = \: \sigma_{[i]}.
\end{equation}
The twist field $x$ obeys $x^2 = 1$, and is invertible: 
$x^{-1} = x$.  The twist field $y$ obeys a fusion rule
$y^2 = (1/2)(1+x)$, and lies along the locus $\{\Delta = 0\}$,
so it is not invertible.

As mentioned earlier, the one-form symmetry $B {\mathbb Z}_2$ acts on
the twist fields as $y \mapsto - y$, leaving $x$ invariant.
Note that although $y$ is not invertible, $B {\mathbb Z}_2$ acts on it
nontrivially.

The three universes correspond to the three representations $R = \pm 1,
2$.  Of these, $R = \pm 1$ correspond to the two copies of the universe
$[X/{\mathbb Z}_2]$, while $R = 2$ corresponds to the universe $X$. 
Note that the projectors $\Pi_{R = \pm 1}$, onto the two universes
$[X/{\mathbb Z}_2]$ exchanged by the one-form symmetry, both involve
$x$ and $y$, even though $y$ is not invertible, whereas the
projector $\Pi_{R=2}$ onto the distinct universe $X$ only involves the
invertible twist field $x$.  Thus, the relation between invertibility of
twist fields and universes of decomposition is nontrivial -- as stated
elsewhere, conjugacy classes and representations are merely dual,
and not canonically isomorphic, as this example illustrates.

In terms of the representations, the two identical universes $[X/{\mathbb Z}_2]$
both correspond to one-dimensional representations, as expected for the
one-form symmetry, while the distinct universe $X$ is associated to a
two-dimensional representation, as appropriate for a noninvertible symmetry.

\subsection{Nonbanded ${\mathbb Z}_2 \times {\mathbb Z}_2$ gerbe}
\label{sect:ex:nonbanded:z2z2}

In this section we consider the orbifold $[X/A_4]$,
where $A_4$ is the group of alternating permutations of four elements,
and $K = {\mathbb Z}_2 \times {\mathbb Z}_2 \subset A_4$ acts
trivially.  This example exhibits decomposition, but has no one-form
symmetry at all.

In this case, $A_4/K = {\mathbb Z}_3$, and it was argued in
\cite[section 5.5]{Hellerman:2006zs} that
\begin{equation}
{\rm QFT}\left( [X/A_4] \right) \: = \: 
{\rm QFT}\left( X \, \coprod \, [X/{\mathbb Z}_3] \right).
\end{equation}

\subsubsection{Ring of dimension-zero operators}

As elements of $A_4$, the elements of $K$ form two conjugacy classes:
\begin{equation}
\{ 1 \}, \: \: \:
\{ (12)(34), (13)(24), (14)(23) \},
\end{equation}
hence there are corresponding dimension-zero twist fields
\begin{equation}
\sigma_1 \: = \: 1, \: \: \:
\sigma \: = \:
\frac{1}{3} \left( \tau_{(12)(34)} + \tau_{(13)(24)} + \tau_{(14)(23)} \right),
\end{equation}
which obey
\begin{equation}
\sigma^2 \: = \: \frac{1}{3} \: + \: \frac{2}{3} \sigma.
\end{equation}

The ring of dimension-zero operators is then
\begin{equation}
{\mathbb C}[\sigma] / (\sigma^2 - (1/3) - (2/3) \sigma),
\end{equation}
so the corresponding universes have support at
\begin{equation}
\langle \sigma \rangle \: = \:
1, \: -1/3.
\end{equation}

The fact this has support at two points is consistent with the
statement of \cite[section 5.5]{Hellerman:2006zs} that this example
decomposes into two universes.

Next, let us compute the noninvertible operators.  It is straightforward to
show that
\begin{equation}
(a + b \sigma)^{-1} \: = \:
\Delta^{-1} \left( a + \frac{2}{3} b - b \sigma \right),
\end{equation}
where
\begin{equation}
\Delta \: = \: a^2 \: + \: \frac{2}{3} a b \: - \: \frac{b^2}{3},
\end{equation}
hence we see the operator $(a + b \sigma)$ is noninvertible if it lies
along the locus $\{ \Delta = 0 \}$.

In this theory, $\sigma$ itself is invertible, with inverse
\begin{equation}
\sigma^{-1} \: = \: -2 \: + \: 3 \sigma.
\end{equation}
Thus, we see that even in this case, where all of the non-identity
elements of ${\mathbb Z}_2 \times {\mathbb Z}_2$ are collected
into a single operator $\sigma$, $\sigma$ is invertible,
so, roughly speaking, nonbanded does not necessarily imply that the
twist fields associated to conjugacy classes are noninvertible.

Next, let us compute the projectors onto the two universes
from equation~(\ref{eq:genl-def-proj}).
Write $K = {\mathbb Z}_2 \times {\mathbb Z}_2 = \langle a, b \rangle$,
where $a^2 = 1 = b^2$, and label the four one-dimensional
irreducible representations of $K$ as
$1$, $\rho_a$, $\rho_b$, $\rho_{ab}$, with character table
\begin{center}
\begin{tabular}{c|rrrr} 
 & $1$ & $a$ & $b$ & $ab$ \\ \hline
$1$ & $1$ & $1$ & $1$ & $1$ \\
$\rho_a$ & $1$ & $1$ & $-1$ & $-1$ \\
$\rho_b$ & $1$ & $-1$ & $1$ & $-1$ \\
$\rho_{ab}$ & $1$ & $-1$ & $-1$ & $1$
\end{tabular}
\end{center}
It is straightforward to compute that
the action of $A_4/K = {\mathbb Z}_3$ on the representations is to
permute $\rho_a$, $\rho_b$, $\rho_{ab}$, while leaving $1$ invariant.
Thus, the universes correspond to the representations
\begin{equation}
1, \: \: \: 
\rho_a \oplus \rho_b \oplus \rho_{ab}
\end{equation}
of $K$.  Applying the definition~(\ref{eq:genl-def-proj}), we have
\begin{eqnarray}
\Pi_{R=1} & = & \frac{1}{4} \left( 1 + \tau_{(12)(34)} + \tau_{(13)(24)}
+ \tau_{(14)(23)} \right),
\\
& = & \frac{1}{4} \left( 1 + 3 \sigma \right),
\end{eqnarray}
and from
\begin{eqnarray}
\Pi_a & = & \frac{1}{| {\mathbb Z}_4|} \sum_{g \in {\mathbb Z}_2 \times
{\mathbb Z}_2} \chi_{\rho_a} \left( g^{-1} \right) \tau_g,
\\
& = &
\frac{1}{4} \left( 1 + \tau_a - \tau_b - \tau_{ab} \right),
\\
\Pi_b & = & \frac{1}{| {\mathbb Z}_4|} \sum_{g \in {\mathbb Z}_2 \times
{\mathbb Z}_2} \chi_{\rho_b} \left( g^{-1} \right) \tau_g,
\\
& = & \frac{1}{4} \left( 1 - \tau_a + \tau_b - \tau_{ab} \right),
\\
\Pi_{ab} & = & \frac{1}{| {\mathbb Z}_4|} \sum_{g \in {\mathbb Z}_2 \times
{\mathbb Z}_2} \chi_{\rho_{ab}} \left( g^{-1} \right) \tau_g,
\\
& = & \frac{1}{4} \left( 1 - \tau_a - \tau_b + \tau_{ab} \right),
\end{eqnarray}
we find
\begin{eqnarray}
\Pi_{R=a \oplus b \oplus ab} & = &
\Pi_a + \Pi_b + \Pi_{ab} \: = \: 
\frac{1}{4} \left( 3 - \tau_{(12)(34)} - \tau_{(13)(24)} - \tau_{(14)(23)}
\right),
\\
& = & \frac{3}{4}\left( 1 - \sigma \right).
\end{eqnarray}
Thus, the projectors onto the two universes are
\begin{equation}
\Pi_1 \: \equiv \:
\Pi_{R=1} \: = \: \frac{1}{4} \left( 1 + 3 \sigma \right), \: \: \:
\Pi_2 \: \equiv \:
\Pi_{R=a\oplus b \oplus ab} \: = \: \frac{3}{4} \left( 1 - \sigma \right).
\end{equation}
Specifically, $\Pi_1$ projects onto $[X/{\mathbb Z}_3]$,
and $\Pi_2$ projects onto $X$.
It is straightforward to check that
\begin{equation}
\Pi_i \Pi_j \: = \: \delta_{i,j} \Pi_i, \: \: \:
\Pi_1 + \Pi_2 \: = \: 1,
\end{equation}
and that they lie along the non-invertible locus $\{ \Delta = 0 \}$.

Note also that
\begin{equation}
\Pi_1 |_{\sigma = 1} \: = \: 1, \: \: \:
\Pi_1 |_{\sigma=-1/3} \: = \: 0,
\end{equation}
\begin{equation}
\Pi_1 |_{\sigma = 1} \: = \: 1, \: \: \:
\Pi_2 |_{\sigma = -1/3} \: = \: 0,
\end{equation}
so we see that $\Pi_1$ projects onto states associated with the
universe at $\sigma=1$, and $\Pi_2$ projects onto states associated with the
universe at $\sigma = -1/3$.

\subsubsection{Wilson lines}

Now, let us turn to Wilson lines, and see how bulk Wilson lines provide
defects linking universes, and from boundary Wilson lines, how
bundles and sheaves on
$[X/A_4]$ decompose into bundles and sheaves on the universes
$X$, $[X/{\mathbb Z}_3]$.

The group $A_4$ has four conjugacy classes, namely
\begin{equation}
\{ 1 \}, \\
\{ (12)(34), (13)(24), (14)(23) \}, \\
\{ (123), (421), (243), (341) \}, \\
\{ (132), (412), (234), (314) \},
\end{equation}
(of which the first two form $K$),
hence four irreducible representations.
From \cite[section 2.3]{fh}, three of those representations are one-dimensional,
and the other is three-dimensional, with character table
\begin{center}
\begin{tabular}{c|cccc}
& $1$ & $(12)(34)$ & $(123)$ & $(132)$ \\ \hline
$1$ & $1$ & $1$ & $1$ & $1$ \\
$1_a$ & $1$ & $1$ & $\omega$ & $\omega^2$ \\
$1_b$ & $1$ & $1$ & $\omega^2$ & $\omega$ \\
$3$ & $3$ & $-1$ & $0$ & $0$
\end{tabular}
\end{center}
where $\omega = \exp(2 \pi i/3)$.

From the character table, we can read off the products
\begin{equation}
1_a^2 \: = \: 1_b, \: \: \:
1_b^2 \: = \: 1_a, \: \: \:
1_a 1_b \: = \: 1,
\end{equation}
\begin{equation}
1_{a,b} \otimes 3 \: = \: 3, \: \: \:
3 \otimes 3 \: = \: 3 \oplus 3 \oplus 1 \oplus 1_a \oplus 1_b.
\end{equation}
Clearly, the one-dimensional representations are invertible, but not the
three-dimensional representation.

Now, let us look at restrictions of these representations.
In particular, mathematically we expect that any bundle or sheaf
on $[X/A_4]$ such that the restriction of the $A_4$-equivariant structure
to $K$ is the representation $1$, descends to universe $[X/{\mathbb Z}_3]$,
whereas those that restrict to $\rho_a \oplus \rho_b \oplus \rho_{ab}$,
descend to universe $S$.

From the character table, it is straightforward to see that
the restrictions of the irreducible representations of
$A_4$ to $K$ are as follows:
\begin{center}
\begin{tabular}{cc}
Representation of $A_4$ & Restriction to $K$ \\ \hline
$1$ & $1$ \\
$1_a$ & $1$ \\
$1_b$ & $1$ \\
$3$ & $\rho_a \oplus \rho_b \oplus \rho_{ab}$
\end{tabular}
\end{center}
Thus, in the restriction, only two representations of $K$ appear,
corresponding to the two universes in the decomposition.

Using those restrictions,
let us compute the action of the projectors.
We begin with bulk Wilson lines.
Trivially,
\begin{equation}
\sigma W_1 \: = \: W_1 \sigma,
\end{equation}
and the effect of  $\sigma$ on $W_{a+b+ab}$ is to insert a matrix
\begin{equation}
\frac{1}{3} {\rm diag}\left( 1-1-1, -1+1-1, -1-1+1 \right) \: = \:
- \frac{1}{3} I,
\end{equation}
so we see that
\begin{equation}
\sigma W_{a+b+ab} \: = \: - \frac{1}{3} W_{a+b+ab} \sigma.
\end{equation}
Then, in terms of Wilson lines in representations of $A_4$, we compute
\begin{equation}
\Pi_1 W_{1, 1_a, 1_b} \: = \: W_{1, 1_a, 1_b} \Pi_1,
\: \: \:
\Pi_2 W_{1, 1_a, 1_b} \: = \: W_1 \Pi_2,
\end{equation}
\begin{equation}
\Pi_1 W_3 \: = \: \frac{1}{3} W_3 \Pi_2,
\: \: \:
\Pi_2 W_3 \: = \: W_3 \left( \Pi_1 + \frac{2}{3} \Pi_2 \right).
\end{equation}
In particular, $W_3$ acts as a defect bridging universes.

Next, we turn to boundary Wilson lines (Chan-Paton factors).
For the moment, we label the Wilson lines by representations of $K$
(obtained as restrictions of representations of $A_4$).
Trivially,
\begin{equation}
\sigma W_1 \: = \: W_1.
\end{equation}
As before,
the effect of $\sigma$ on $W_{a+b+ab}$ is to insert a matrix
\begin{equation}
\frac{1}{3} {\rm diag}\left( 1-1-1, -1+1-1, -1-1+1 \right) \: = \:
- \frac{1}{3} I,
\end{equation}
so we see that
\begin{equation}
\sigma W_{a+b+ab} \: = \: - \frac{1}{3} W_{a+b+ab}.
\end{equation}
Note that the eigenvalue corresponding to
$W_{a+b+ab}$ is $\chi_3(g) / \dim 3$, as expected.

Then, in terms of boundary Wilson lines in representations of $A_4$,
we compute
\begin{equation}
\Pi_1 W_{1, 1_a, 1_b} \: = \: W_{1, 1_a, 1_b}, \: \: \:
\Pi_2 W_{1, 1_a, 1_b}1 \: = \: 0,
\end{equation}
\begin{equation}
\Pi_1 W_{3} \: = \: 0, \: \: \:
\Pi_2 W_{3} \: = \: W_{3},
\end{equation}
so that each projector selects out Wilson lines in the fashion predicted
mathematically, as expected.

\subsubsection{Symmetries}

In this example, since the two constituent universes are distinct,
we expect no ordinary one-form symmetries, only noninvertible symmetries.

This is reflected at several levels.  At a group theoretic level,
the alternating group $A_4$ has no center (beyond the identity),
so no one-form symmetry is expected there.  Similarly, the ring of 
dimension-zero operators does not have any multiplicative symmetries,
again in accord with a lack of one-form symmetries.

Perhaps surprisingly in this case, the single twist field
$\sigma$ is invertible -- but as we have seen, twist fields are merely dual
to representations, not canonically isomorphic, and there is no canonical
method to distinguish a single twist field from a general linear combination,
which always contains noninvertible elements.  
In terms of representations,
the universe $[X/{\mathbb Z}_3]$ is associated with a one-dimensional
representation but the other universe ($X$) is associated with a 
three-dimensional representation of ${\mathbb Z}_2 \times {\mathbb Z}_2$.
Since only one universe is associated with a one-dimensional representation,
there is no reason to expect an invertible one-form symmetry, 
and since there exists a universe associated with a higher-dimension 
representation, one does expect a noninvertible symmetry.

\subsection{Nonabelian $D_4$ gerbe}
\label{sect:ex:nonabelian:d4}

In this section we consider the orbifold $[{\rm point}/D_4]$,
where all of the eight-element dihedral group
$D_4$ acts trivially.

Geometrically, this is a $D_4$-gerbe (over a point).
Since $D_4$ is nonabelian, only its center (${\mathbb Z}_2$) defines
a one-form symmetry ($B{\mathbb Z}_2$).  From decomposition 
\cite{Hellerman:2006zs,Robbins:2021ylj},
\begin{eqnarray}
{\rm QFT}\left( [{\rm point}/D_4] \right)
& = &
{\rm QFT}\left( \coprod_5 {\rm point} \right),
\\
& = &
{\rm QFT}\left(
[{\rm point}/{\mathbb Z}_2 \times {\mathbb Z}_2] \, \coprod \,
[{\rm point}/{\mathbb Z}_2 \times {\mathbb Z}_2]_{\rm d.t.} \right).
\end{eqnarray}

For use here, we present the 
elements of $D_4$ as
\begin{equation}
D_4 \: = \: \{1, z, a, b, az, bz, ab, ba \},
\end{equation}
with the relations
\begin{equation}
a^2 \: = \: 1 \: = \: b^4, \: \: \:
b^2 \: = \: z, \: \: \:
ba \: = \: abz,
\end{equation}
in the same notation as \cite{Robbins:2021ylj},
where $z$ generates the center.

\subsubsection{Ring of dimension-zero operators}

First, let us consider the twist fields, corresponding to conjugacy classes
of group elements.  
Following appendix~\ref{app:computation},
the conjugation-invariant dimension-zero twist fields are
\begin{equation}
\sigma_{[+1]} \: = \: 1, \: \: \:
\sigma_{[-1]} \: = \: \tau_z, 
\end{equation}
\begin{equation}
\sigma_{[a]} \: = \: (1/2)\left( \tau_a + \tau_{az} \right), \: \: \:
\sigma_{[b]} \: = \: (1/2)\left( \tau_b + \tau_{bz} \right), \: \: \:
\sigma_{[ab]} \: = \: (1/2)\left( \tau_{ab} + \tau_{ba} \right).
\end{equation}
Using the multiplication rule
\begin{equation}
\tau_g \tau_h \: = \: \tau_{gh}
\end{equation}
as in section~\ref{app:computation},
one quickly finds
\begin{equation}
\sigma_{[+1]}^2 \: = \: \sigma_{[+1]},
\: \: \:
\sigma_{[+1]} \sigma_{[-1]} \: = \: \sigma_{[-1]},
\: \: \:
\sigma_{[+1]} \sigma_{[a,b,ab]} \: = \: \sigma_{[a,b,ab]},
\end{equation}
\begin{equation}
\sigma_{[-1]}^2 \: = \: \sigma_{[+1]},
\: \: \:
\sigma_{[-1]} \sigma_{[a,b,ab]} \: = \: \sigma_{[a,b,ab]},
\: \: \:
\sigma_{[a,b,ab]}^2 \: = \: (1/2) \left( \sigma_{[+1]} + \sigma_{[-1]} \right).
\end{equation}
\begin{equation}
\sigma_{[a]} \sigma_{[b]} \: = \:
\sigma_{[ab]},
\: \: \:
\sigma_{[b]} \sigma_{[ab]} \: = \: \sigma_{[a]},
\: \: \:
\sigma_{[ab]} \sigma_{[a]} \: = \: \sigma_{[b]}.
\end{equation}

Identifying $x$ with $\sigma_{[-1]}$ and $y_{1,2,3}$ with
$\sigma_{[a,b,ab]}$, we can write the 
ring of dimension-zero operators more efficiently as
\begin{equation}
{\mathbb C}[x,y_1,y_2,y_3] / \left(
x^2 - 1, x y_i - y_i, y_i^2 - (1/2)(1+x), y_1 y_2 - y_3, y_2 y_3 - y_1,
y_3 y_1 - y_2 \right).
\end{equation}
Clearly, $x$ is invertible 
but none of $\{y_1, y_2, y_3\}$ are invertible.

Geometrically, this ring describes five points in ${\mathbb C}^4$, 
as many points as the number of components of the decomposition,
at the
locations
\begin{eqnarray}
\lefteqn{
(x,y_1,y_2,y_3)
} \nonumber \\
& = & \left\{
(+1, +1, +1, +1), \: \: \:
(+1, +1, -1, -1), \: \: \:
(+1, -1, +1, -1), \: \: \:
\right. \nonumber \\
& & \hspace*{0.5in} \left.
(+1, -1, -1, +1), \: \: \:
(-1, 0, 0, 0) \right\}.
\end{eqnarray}

Physically,
\begin{equation}
{\rm QFT}\left( [{\rm point}/D_4] \right) \: = \: 
{\rm QFT}\left( \coprod_5 {\rm point} \right),
\end{equation}
a disjoint union of five points, corresponding to the number of conjugacy
classes.  We can equivalently decompose this theory using the fact that
the center of $D_4$ is ${\mathbb Z}_2$, with quotient
$D_4/{\mathbb Z}_2 = {\mathbb Z}_2 \times {\mathbb Z}_2$, 
and write the decomposition as
\begin{equation}
{\rm QFT}\left( [{\rm point}/D_4] \right) \: = \:
{\rm QFT}\left( [{\rm point}/{\mathbb Z}_2 \times {\mathbb Z}_2]
\, \coprod \, [ {\rm point}/{\mathbb Z}_2 \times {\mathbb Z}_2]_{\rm dt}
\right).
\end{equation}
As explained in \cite[section 4.1]{Robbins:2021ylj},
\begin{eqnarray}
{\rm QFT}\left(  [{\rm point}/{\mathbb Z}_2 \times {\mathbb Z}_2]
\right) & = & {\rm QFT}\left( \mbox{four points} \right),
\\
{\rm QFT}\left(  [{\rm point}/{\mathbb Z}_2 \times {\mathbb Z}_2]_{\rm dt}
\right) & = & {\rm QFT}\left( \mbox{one point} \right),
\end{eqnarray}
so again we get a total of five points, but this alternative description
may make the role of the central ${\mathbb Z}_2$ more clear.
Note that this structure is reflected in the order parameters,
the points on ${\mathbb C}^4$ where the decomposition has support:
\begin{itemize}
\item the single point at $x=-1$ corresponds to 
the ${\mathbb Z}_2 \times {\mathbb Z}_2$ orbifold with discrete torsion,
\item the four points at $x=+1$ correspond to the
${\mathbb Z}_2 \times {\mathbb Z}_2$ orbifold without discrete torsion,
\end{itemize}
as will be confirmed from the representations associated to the universes.

Now, let us compute the projectors from~(\ref{eq:genl-def-proj}).
For this we need some facts about the representation theory of $D_4$.
This group has five irreducible representations, correspodning to
the five conjugacy classes
\begin{equation}
\{1 \}, \{ z \}, \{a, az\}, \{b, bz\}, \{ab, ba=abz\}.
\end{equation}
It has five irreducible representations: four one-dimensional,
and one two-dimensional.

The two-dimensional representation can be given explicitly as
\begin{equation}
a \: = \: \left[ \begin{array}{rr} 1 & 0 \\ 0 & -1 \end{array} \right], \: \: \:
b \: = \: \left[ \begin{array}{rr} 0 & -1 \\ 1 & 0 \end{array} \right], 
\: \: \:
z \: = \: \left[ \begin{array}{rr} -1 & 0 \\ 0 & -1 \end{array} \right],
\end{equation}
and the group has character table
\begin{center}
\begin{tabular}{c|crrrr}  
& $\{1\}$ & $\{z\}$ & $\{a, az\}$ & $\{b, bz\}$ & $\{ab, ba\}$ \\ \hline
$1$ & $1$ & $1$ & $1$ & $1$ & $1$ \\
$1_a$ & $1$ & $1$ & $1$ & $-1$ & $-1$ \\
$1_b$ & $1$ & $1$ & $-1$ & $1$ & $-1$ \\
$1_{ab}$ & $1$ & $1$ & $-1$ & $-1$ & $1$ \\
$2$ & $2$ & $-2$ & $0$ & $0$ & $0$
\end{tabular}
\end{center}

Here, since $K=\Gamma$, there is no quotient group to interchange the
representations, and so there is one universe for each representation.
Applying~(\ref{eq:genl-def-proj}), we find
\begin{eqnarray}
\Pi_{R=1} & = &
\frac{1}{|G|} \sum_{g \in G} \chi_1\left( g^{-1} \right) \tau_g,
\\
& = &
\frac{1}{8} \left( 1 + \tau_z + \tau_a + \tau_{az} + \tau_b + \tau_{bz}
+ \tau_{ab} + \tau_{ba} \right),
\\
& = & \frac{1}{8} \left( 1 + \sigma_{[-1]} + 2(\sigma_{[a]} + 
\sigma_{[b]} + \sigma_{[ab]}) \right),
\\
\Pi_{R=1_a} & = &
\frac{1}{|G|} \sum_{g \in G} \chi_a\left( g^{-1} \right) \tau_g,
\\
& = &
\frac{1}{8} \left( 1 + \sigma_{[-1]} + 2(\sigma_{[a]} - \sigma_{[b]}
- \sigma_{[ab]} ) \right),
\\
\Pi_{R=1_b} & = &
\frac{1}{|G|} \sum_{g \in G} \chi_b\left( g^{-1} \right) \tau_g,
\\
& = &
\frac{1}{8} \left( 1 + \sigma_{[-1]} + 2( -\sigma_{[a]} + \sigma_{[b]}
- \sigma_{[ab]} ) \right),
\\
\Pi_{R=1_{ab}} & = &
\frac{1}{|G|} \sum_{g \in G} \chi_{ab}\left( g^{-1} \right) \tau_g,
\\
& = &
\frac{1}{8} \left( 1 + \sigma_{[-1]} + 2(- \sigma_{[a]} - \sigma_{[b]}
+ \sigma_{[ab]} ) \right),
\\
\Pi_{R=2} & = &
\frac{2}{|G|}  \sum_{g \in G} \chi_{2}\left( g^{-1} \right) \tau_g,
\\
& = &
\frac{2}{8}\left( 2 - 2 \sigma_{[-1]} \right) \: = \: \frac{1}{2} \left( 1 -
\sigma_{[-1]} \right).
\end{eqnarray} 
Thus, the projectors are
\begin{eqnarray}
\Pi_{+++} \: \equiv \: \Pi_{R=1} & = & \frac{1}{8}\left( 1 + x + 2 (y_1 +  y_2 +  y_3) \right),
\\
\Pi_{+--} \: \equiv \:\Pi_{R=1_a} & = & \frac{1}{8}\left( 1 + x + 2 (y_1 - y_2 - y_3) \right),
\\
\Pi_{--+} \: \equiv \: \Pi_{R=1_{ab}} & = & \frac{1}{8}\left( 1 + x + 2( -y_1 - y_2 + y_3) \right),
\\
\Pi_{-+-} \: \equiv \: \Pi_{R=1_b} & = & \frac{1}{8} \left( 1 + x + 2(-y_1 + y_2 - y_3) \right),
\\
\Pi_5 \: \equiv \: \Pi_{R=2} & = & \frac{1}{2} (1 - x ).
\end{eqnarray}
It is straightforward to check that
\begin{equation}
\Pi_i \Pi_j \: = \: \delta_{i,j} \Pi_i, \: \: \:
\sum_i \Pi_i \: = \: 1.
\end{equation}
It is also straightforward to check that each projector is nonzero at
exactly one of the points where the order parameter vevs are nonzero:
\begin{equation}
\Pi_{+++}|_{(+1,+1,+1,+1)} \: = \: 1 \: = \:
\Pi_{+--}|_{(+1,+1,-1,-1)} \: = \:
\Pi_{--+}|_{(+1,-1,-1,+1)} \: = \:
\Pi_{-+-}|_{(+1,-1,+1,-1)},
\end{equation}
\begin{equation}
\Pi_5|_{(-1,0,0,0)} \: = \: 1,
\end{equation}
with other restrictions of projectors to points above vanishing.

As the center of $D_4$ is ${\mathbb Z}_2$, one would expect that
the orbifold $[{\rm point}/D_4]$ would have a $B {\mathbb Z}_2$
(one-form) symmetry.  However, in fact, this algebra is actually
consistent with a $B({\mathbb Z}_2 \times {\mathbb Z}_2)$ symmetry,
generated by
\begin{equation}
\left\{ \begin{array}{ccl}
y_{1,2} & \mapsto & - y_{1,2}, \\
x, y_3 & \mapsto & x, y_3 \mbox{ (invariant)},
\end{array} \right.
\: \: \:
\left\{ \begin{array}{ccl}
y_{1,3} & \mapsto & - y_{1,3}, \\
x, y_2 & \mapsto & x, y_2 \mbox{ (invariant).}
\end{array}
\right.
\end{equation}
This permutes the points at which the order parameters have
vevs, which is reminiscent of spontaneous symmetry breaking,
though again we observe that decomposition is a stronger statement.
Existence of a $B({\mathbb Z}_2 \times {\mathbb Z}_2)$
is consistent with the decomposition under the $B {\mathbb Z}_2$
into a pair of ${\mathbb Z}_2 \times {\mathbb Z}_2$ orbifolds of
points discussed above.

In passing, the orbifold $[{\rm point}/{\mathbb H}]$ is nearly
identical:  the ring of dimension-zero fields is the same, with
$y_{1,2,3}$ corresponding to
\begin{equation}
\sigma_{[i]} \: = \: (1/2)\left( \tau_i + \tau_{-i} \right),
\: \: \:
\sigma_{[j]} \: = \: (1/2)\left( \tau_j + \tau_{-j} \right),
\: \: \:
\sigma_{[k]} \: = \: (1/2) \left( \tau_k + \tau_{-k} \right),
\end{equation}
and it also decomposes into five points:
\begin{equation}
{\rm QFT}\left( [ {\rm point}/{\mathbb H}] \right)
\: = \:
{\rm QFT}\left( [{\rm point}/{\mathbb Z}_2 \times {\mathbb Z}_2]
\, \coprod \,
[{\rm point}/{\mathbb Z}_2 \times {\mathbb Z}_2]_{\rm dt} \right)
\: = \:
{\rm QFT}\left( \mbox{5 points} \right).
\end{equation}
One difference is that the group ${\mathbb H}$ does not admit
discrete torsion: $H^2({\mathbb H},U(1)) = 0$, unlike $D_4$ for which
$H^2(D_4,U(1)) = {\mathbb Z}_2$ \cite[appendix D.3]{Robbins:2021ylj}.

\subsubsection{Wilson lines}

As described in the previous section,
the group $D_4$ has five irreducible representations:
four one-dimensional,
and one two-dimensional, which we label as $1, 1_a, 1_b, 1_{ab}, 2$.

From the character table (see also \cite{ty}), we can read off
\begin{equation}
1_a \otimes 1_b \: = \: 1_{ab}, \: \: \:
1_a^2 \: = \: 1_b^2 \: = \: 1_{ab}^2 \: = \: 1,
\end{equation}
\begin{equation}
1_{a,b,ab} \otimes 2 \: = \: 2, \: \: \:
2 \otimes 2 \: = \: 1 + 1_a + 1_b + 1_{ab}.
\end{equation}

We begin with bulk Wilson lines.
Letting for example $W_a$ denote a Wilson
line associated to representation $1_a$,
it is straightforward to compute
\begin{equation}
\Pi_{\pm \pm \pm} W_1 \: = \: W_1 \Pi_{\pm \pm \pm},
\: \: \:
\Pi_5 W_1 \: = \: W_1 \Pi_5,
\end{equation}
\begin{equation}
\Pi_{+++} W_a \: = \: W_a \Pi_{+--},
\: \: \:
\Pi_{+--} W_a \: = \: W_a \Pi_{+++},
\end{equation}
\begin{equation}
\Pi_{--+} W_a \: = \: W_a \Pi_{-+-},
\: \: \:
\Pi_{-+-} W_a \: = \: W_a \Pi_{--+},
\: \: \:
\Pi_5 W_a \: = \:  W_a \Pi_5,
\end{equation}
\begin{equation}
\Pi_{+++} W_b \: = \: W_b \Pi_{-+-},
\: \: \:
\Pi_{+--} W_b \: = \: W_b \Pi_{--+},
\end{equation}
\begin{equation}
\Pi_{--+} W_b \: = \: W_b \Pi_{+--},
\: \: \:
\Pi_{-+-} W_b \: = \: W_b \Pi_{+++},
\: \: \:
\Pi_5 W_b \: = \: W_b \Pi_5,
\end{equation}
\begin{equation}
\Pi_{+++} W_{ab} \: = \: W_{ab} \Pi_{--+},
\: \: \:
\Pi_{+--} W_{ab} \: = \: W_{ab} \Pi_{-+-},
\end{equation}
\begin{equation}
\Pi_{--+} W_{ab} \: = \: W_{ab} \Pi_{+++},
\: \: \:
\Pi_{-+-} W_{ab} \: = \: W_{ab} \Pi_{+--},
\: \: \:
\Pi_5 W_{ab} \: = \: W_{ab} \Pi_5,
\end{equation}
\begin{equation}
\Pi_{\pm \pm \pm} W_2 \: = \: \frac{1}{4} W_2 \Pi_5,
\: \: \:
\Pi_5 W_2 \: = \: W_2 \left( \Pi_{+++} + \Pi_{+--} +
\Pi_{--+} + \Pi_{-+-} \right).
\end{equation}
We see that the $W_{a,b,ab}$ act as defects bridging the $\pm \pm \pm$
universes, and $W_2$ acts as a defect bridging the universe $2$ with the
$\pm \pm \pm$ universes.

Next, we turn to boundary Wilson lines (Chan-Paton factors).
As before, sheaves and bundles on $[{\rm point}/D_4]$ 
are sheaves and bundles on the
universes of the decomposition.  Since all of $D_4$ acts trivially,
we do not restrict to a subgroup, and consider the action of
projectors on the Wilson lines, which are both associated to
representations of $D_4$.
Thus, $1$, $1_a$, $1_b$, $1_{ab}$ correspond to sheaves on
$[{\rm point}/{\mathbb Z}_2 \times {\mathbb Z}_2]$
(which itself decomposes into four points, one for each of those
one-dimensional representations),
since their restriction to the center is trivial,
and $2$ corresponds to a sheaf on
$[{\rm point}/{\mathbb Z}_2 \times {\mathbb Z}_2]_{\rm d.t.}$,
since its restriction to the center is nontrivial.

Letting for example $W_a$ denote a Wilson
line associated to representation $1_a$, 
it is straightforward to compute
\begin{equation}
\Pi_{+++} W_1 \: = \: W_1, \: \: \:
\Pi_{+--} W_a \: = \: W_a, \: \: \:
\Pi_{-+-} W_b \: = \: W_b, \: \: \:
\Pi_{--+} W_{ab} \: = \: W_{ab},
\end{equation}
\begin{equation}
\Pi_5 W_{2} \: = \: W_2,
\end{equation}
with other projectors annihilating other Wilson lines.
This is consistent with the identification above of projectors with
irreducible representations of $D_4$.

\subsubsection{Symmetries}

Since all of the orbifold group acts trivially,
this is a rather degenerate special case, and as such,
there are unexpected symmetries.

From the group theory alone, as the center of $D_4$ is only
${\mathbb Z}_2$, one would expect in general only a $B {\mathbb Z}_2$
symmetry.  On the other hand, since the decomposition is into five points
(naturally grouped into two sets, one of four points and the other of one),
one could reasonably expect a larger one-form symmetry.

Amongst the twist fields associated to conjugacy
classes, only one twist field ($x$) is invertible, while the others
($y_{1,2,3}$) are noninvertible; however, as noted elsewhere, that is somewhat
ambiguous.

Abstractly, the ring of dimension-zero operators has three
${\mathbb Z}_2$ symmetries, under which $x$ and one of the $y_i$ is
invariant while flipping the signs of the other two $y$'s.

In terms of the representations associated to the universes,
four of the universes are associated to one-dimensional representations of
$D_4$, while one is associated to the irreducible two-dimensional
representation of $D_4$.  The four one-dimensional representations
form the group ${\mathbb Z}_2 \times {\mathbb Z}_2$, so this is
consistent with a $B({\mathbb Z}_2 \times {\mathbb Z}_2)$ symmetry,
plus a noninvertible symmetry relating to the fifth universe.

Of the projectors corresponding to these universes, the projectors onto
the four universes associated to one-dimensional representations involve
both $x$ and $y$'s; only the projectors onto
the universe associated to a higher-dimensional
representation involves only the invertible twist field $x$.  None of the
projectors is itself invertible, and as remarked elsewhere, twist fields and
representations are merely dual, not canonically bijective.

\subsection{Nonabelian $D_4$ gerbe with discrete torsion}
\label{sect:ex:nonabelian:d4dt}

Next, consider the orbifold $[{\rm point}/D_4]$ with discrete torsion.
(Here, we use the fact that $H^2(D_4,U(1)) = {\mathbb Z}_2$.)
In general terms, we will see the same phenomena as in previous examples.

It will be handy in this section to have an explicit cocycle representing
the nontrivial element of $H^2(D_4,U(1))$.  
Following the conventions of \cite[section 3.7]{karpilovsky},
we take the discrete torsion cocycle $\omega$ to be 
\begin{center}
\begin{tabular}{c|cccccccc}
& $1$ & $b$ & $z$ & $bz$ & $a$ & $ba$ & $az$ & $ab$ \\ \hline
$1$ & $1$ & $1$ & $1$ & $1$ & $1$ & $1$ & $1$ & $1$ \\
$b$ & $1$ & $1$ & $1$ & $1$ & $1$ & $1$ & $1$ & $1$ \\
$z$ & $1$ & $1$ & $1$ & $1$ & $1$ & $1$ & $1$ & $1$ \\
$bz$ & $1$ & $1$ & $1$ & $1$ & $1$ & $1$ & $1$ & $1$ \\
$a$ & $1$ & $\xi$ & $\xi^2$ & $\xi^3$ & $1$ & $\xi$ & $\xi^2$ & $\xi^3$ \\
$ba$ & $1$ & $\xi$ & $\xi^2$ & $\xi^3$ & $1$ & $\xi$ & $\xi^2$ & $\xi^3$ \\
$az$ & $1$ & $\xi$ & $\xi^2$ & $\xi^3$ & $1$ & $\xi$ & $\xi^2$ & $\xi^3$ \\
$ab$ & $1$ & $\xi$ & $\xi^2$ & $\xi^3$ & $1$ & $\xi$ & $\xi^2$ & $\xi^3$ 
\end{tabular}
\end{center} 
where $\xi = \exp(2 \pi i / 4) = i$, so that $\xi^4 = 1$.
(As a consistency test, the invariant ratios $\omega(g,h)/\omega(h,g)$ match
those in e.g. \cite[table D.4]{Robbins:2021ylj}.)
The cocycles can also be described as \cite[section 3.7]{karpilovsky}
\begin{equation}
\omega(b^i, b^j a^k) \: = \: 1,  \: \: \:
\omega(b^i a, b^j a^k) \: = \: \xi^j.
\end{equation}

It is straightforward to check that $D_4$ has two irreducible projective
representations, hence decomposition implies 
\cite{Hellerman:2006zs,Robbins:2021ylj}
\begin{equation}
{\rm QFT}\left( [{\rm point}/D_4]_{\omega} \right)
\: = \: {\rm QFT}\left( {\rm point} \, \coprod \, {\rm point} \right).
\end{equation}

\subsubsection{Ring of dimension-zero operators}

Using the methods of section~\ref{app:computation}
and the cocycle for the nontrivial element of $H^2(D_4,U(1))$
given above
we find that the only conjugation-invariant dimension-zero twist fields
are
\begin{equation}
\sigma_1 \: = \: 1, \: \: \:
\sigma_{[b]} \: = \: (1/2) \left( \tau_b + i \tau_{bz} \right).
\end{equation}
Consistent with expectations, these are the only two conjugacy classes that
correspond to irreducible projective representations of $D_4$
\cite[appendix D.3]{Robbins:2021ylj}.
Since the cocycle is trivial on $\langle b \rangle \subset D_4$,
we use the fact that
\begin{equation}
\tau_b \tau_b \: = \: \tau_z, \: \: \:
\tau_{bz} \tau_{bz} \: = \: \tau_z, \: \: \:
\tau_b \tau_{bz} \: = \: 1 \: = \: \tau_{bz} \tau_b
\end{equation}
to derive that
\begin{equation}
\sigma_{[b]}^2 \: = \:  i / 2.
\end{equation}
Identifying $\sigma_{[b]}$ with $y$, the ring of dimension-zero opeators
is then given by
\begin{equation}
{\mathbb C}[y] / (y^2 - (i/2)  ),
\end{equation}
which corresponds to a pair of points, supported at
\begin{equation}
y_{\pm} \: = \: \pm \frac{1}{\sqrt{2}} \exp(+ \pi i / 4) \: = \:
\pm \frac{1}{2}\left( 1 + i \right).
\end{equation}
Each point corresponds to one of the two universes appearing in the
decomposition of $[X/D_4]_{\omega}$.

The reader should note that in this ring, $y \sim \sigma_{[b]}$ is invertible:
\begin{equation}
y^{-1} \: = \: - 2 i y.
\end{equation}
Before turning on discrete torsion, in the $D_4$ orbifold we discussed
in the previous section, the twist field built from the conjugacy class
$\{b, bz\}$ was not invertible.

For completeness,
\begin{equation}
(a + b y)^{-1} \: = \: \Delta^{-1} \left( a - b y \right),
\end{equation}
for
\begin{equation}
\Delta \: = \: a^2 - (i/2)  b^2,
\end{equation}
so we see that noninvertible ring elements lie along the locus
$\{ \Delta = 0 \}$.

Next, we compute the projectors from~(\ref{eq:genl-def-proj}),
for which we need the projective representations of $D_4$.
As discussed in e.g. \cite[appendix D.3]{Robbins:2021ylj},
\cite[example 3.12]{cheng}, \cite[section 3.7]{karpilovsky},
there are two irreducible projective representations of $D_4$,
and they both have dimension two.  In the conventions of
\cite[section 3.7]{karpilovsky}, they are given by
\begin{equation}
\rho_r(b^i a^j) \: = \: B_r^i A_r^j,
\end{equation}
for 
\begin{equation}
A_r \: = \: \left[ \begin{array}{cc} 0 & 1 \\ 1 & 0 \end{array} \right],
\: \: \:
B_r \: = \: \left[ \begin{array}{cc} \xi^r & 0 \\ 0 & \xi^{1-r} \end{array}
\right],
\end{equation}
for $r \in \{1, 2\}$ indexing the two representations.
For example, it is straightforward to check that
\begin{equation}
\rho_r(a) \, \rho_r(b) \: = \: \omega(a,b) \, \rho_r(ab).
\end{equation}

Since there is no quotient group $G = \Gamma/K$, as $\Gamma=K$,
there is nothing to interchange the two representations, so there are
two universes, corresponding to each of those two irreducible representations.

In order to apply~(\ref{eq:genl-def-proj}), we need the character table,
which we compute next.
To that end, we note that of the five conjugacy classes of $D_4$,
namely
\begin{equation}
\{ 1 \}, \: \: \:
\{ z \}, \: \: \:
\{ a, az \}, \: \: \:
\{ b, bz \}, \: \: \:
\{ ab, ba \},
\end{equation}
only two of them ($\{1\}$, $\{b, bz\}$) are $\omega$-regular,
meaning that for all elements $x$ of the conjugacy class,
$\omega(x,g) = \omega(g,x)$ for all $g$ commuting with $x$.
For example, $\omega(z,a) \neq \omega(a,z)$, hence neither $\{ z\}$
nor $\{a, az\}$ can be $\omega$-regular, and similarly 
since $\omega(ab,z) \neq \omega(z,ab)$, $\{ab, ba\}$ also cannot be
$\omega$-regular.

The fact that there are two $\omega$-regular conjugacy classes correctly
matches the number of irreducible projective representations.

From the explicit form of the representations above, 
given the definition $\chi_r(g) = {\rm Tr}
\rho_r(g)$, we find that the characters are given by
\begin{center}
\begin{tabular}{c|cccccccc}
$\chi$ & $1$ & $z$ & $a$ & $az$ & $b$ & $bz$ & $ab$ & $ba$ \\ \hline
$r=1$ & $2$ & $0$ & $0$ & $0$ & $1+i$ & $1-i$ & $0$ & $0$ \\
$r=2$ & $2$ & $0$ & $0$ & $0$ & $-1-i$ & $-1+i$ & $0$ & $0$
\end{tabular}
\end{center}
As expected, characters of elements of non-$\omega$-regular conjugacy classes
vanish.  Also as expected, the characters are not class functions,
but instead obey~(\ref{eq:char-non-class}).  For example,
using the fact that $bz = a b a^{-1}$ and
\begin{equation}
\frac{ \omega(b, a^{-1}) }{ \omega(a^{-1}, a b a^{-1}) }
\: = \:
\frac{ \omega(b,a) }{ \omega(a,bz) } \: = \:
\xi \: = \: i,
\end{equation}
we can confirm
\begin{eqnarray}
\chi_1(b) & = & 1 + i \: = \: i (1-i) \: = \: i \chi_1(bz),
\\
\chi_2(b) & = & -1-i \: = \: i (-1+i) \: = \: i \chi_2(bz),
\end{eqnarray}
as predicted by~(\ref{eq:char-non-class}).

Now, we can compute projectors.
From equation~(\ref{eq:genl-def-proj}),
we have that
\begin{equation}
\Pi_r \: = \: \frac{2}{|D_4|} \sum_{g \in D_4}
\frac{ \chi_r(g^{-1}) }{ \omega(g,g^{-1}) } \tau_g,
\end{equation}
from which we find
\begin{eqnarray}
\Pi_{r=1} & = & \frac{1}{2} \left[
1 + \frac{1}{2}(1-i) \tau_b + \frac{1}{2}(1+i) \tau_{bz} \right],
\\
& = & \frac{1}{2} \left[ 1 + \sqrt{2}\exp(-\pi i/4) \sigma_{[b]} \right],
\\
\Pi_{r=2} & = & \frac{1}{2} \left[
1 - \frac{1}{2}(1-i) \tau_b - \frac{1}{2}(1+i) \tau_{bz} \right],
\\
& = &
\frac{1}{2} \left[ 1 - \sqrt{2}\exp(-\pi i/4) \sigma_{[b]} \right].
\end{eqnarray}
It is straightforward to check that they obey
\begin{equation}
\Pi_r \Pi_s \: = \: \delta_{r,s} \Pi_r, \: \: \:
\Pi_1 + \Pi_2 \: = \: 1,
\end{equation}
and are easily checked to lie along the locus $\{ \Delta = 0 \}$.

\subsubsection{Wilson lines}

For simplicity, in this section, we will only
compute the action of the projectors on the boundary Wilson lines.
Since the entire orbifold group acts trivially, both the projectors
and the Wilson lines are associated to projective representations of $D_4$.

Using the relation
\begin{equation}
\sigma_{[g]} W_R \: = \: \frac{ \chi_R(g) }{ \dim R} \, W_R,
\end{equation}
it is straightforward to compute that
\begin{equation}
\Pi_r W_s \: = \: \delta_{r,s} W_r.
\end{equation}
This is precisely as expected -- since all of $D_4$ acts trivially,
and each universe is associated to an irreducible projective representation
of $D_4$, the Wilson lines obey the same decomposition.

\subsubsection{Symmetries}

Much as in the last example, since this is an orbifold of a point,
it is a rather degenerate special case, and so can have unexpected
symmetries.

Here, there are two ($\omega$-regular) conjugacy classes, including
$\{ 1 \}$.  The twist field associated to the nontrivial conjugacy
class is invertible.

Since the theory decomposes into two copies of a point, one expects
a $B {\mathbb Z}_2$ symmetry.  

On the other hand, the representations associated to each universe
are both two-dimensional, suggesting that this theory only has
noninvertible symmetries (which in this case exchange identical copies).

\section{Examples in supersymmetric gauge theories}
\label{sect:ex:gauge}

The rings of dimension-zero operators that we have described
are also visible in two-dimensional supersymmetric gauge theories,
as we shall now describe.  In these examples, the one-form symmetry
will always be a center symmetry, corresponding to a banded abelian
gerbe, so the structure we derive will coincide with that of the
banded abelian examples discussed previously.

First, consider a $U(1)$ gauge theories with nonminimal charges.
Theories of this form were first discussed in 
\cite{Pantev:2005rh,Pantev:2005wj,Pantev:2005zs}, and include
variations of the supersymmetric ${\mathbb P}^n$ model.
Here, let us briefly consider a family of variations of the
${\mathbb P}^n$ model, discussed in e.g. \cite{Pantev:2005zs},
corresponding to sigma models on the ${\mathbb Z}_k$ gerbes over
${\mathbb P}^n$.  These are described by $U(1)^2$ gauge theories
with chiral superfields $x_{0,\cdots,n}$, $z$ of charges
\begin{center}
\begin{tabular}{cc}
  $x_i$ & $z$ \\ \hline
$1$ & $-m$ \\
$0$ & $k$.
\end{tabular}
\end{center}
The $x_i$ act as homogeneous coordinates on 
${\mathbb P}^n$, and $z$ is a coordinate along the fibers of a 
${\mathbb C}^{\times}$ bundle
over ${\mathbb P}^n$.  The second $U(1)$ 
`overgauges' the ${\mathbb C}^{\times}$, so that the fibers
become $B {\mathbb Z}_k$, hence a ${\mathbb Z}_k$ gerbe
(with a one-form translation symmetry along the fibers).
The characteristic class of the gerbe is $-m \mod k$.
The case $m=1$ is equivalent to a ${\mathbb P}^n$ model in which
charges of the fields are multiplied by $k$.

The quantum cohomology ring one derives from the Coulomb branch of this
gauged linear sigma model is \cite{Pantev:2005zs}
\begin{equation}
{\mathbb C}[x,y] / (x^k -1, y^{n+1} - x^m q).
\end{equation}

The dimension-zero operators are encoded in
\begin{equation}
{\mathbb C}[x] / (x^k - 1),
\end{equation}
the same structure we have previously seen in sigma models on disjoint
unions of $k$ copies of a space, and in banded ${\mathbb Z}_k$ gerbes.

As discussed in \cite{Pantev:2005zs}, this structure also arises in the
mirror.  Computing the mirror ala \cite{Hori:2000kt}, one finds
Landau-Ginzburg models with superpotentials of the form
\begin{equation}
W \: = \: \exp(-X_1) \: + \: \cdots \: + \:
\exp(-X_n) \: + \: q \Upsilon^{-m} \exp\left(+ X_1 + X_2 + \cdots + X_n \right),
\end{equation}
where $\Upsilon$ is a field valued in $k$th roots of unity.
Here, $\Upsilon$ is a dimension-zero field, the mirror of the
``$x$'' appearing in the quantum cohomology ring, obeying the same
relation ($\Upsilon^k = 1$) that we have already discussed.
(This description of the mirror implicitly encodes decomposition:
the path integral's sum over values of $\Upsilon$ is a finite sum
that can be pulled out of the path integral, making it clear that the
theory is equivalent to a disjoint union of ordinary Landau-Ginzburg models
with staggered complex structures).

Next, let us turn to an $SU(2)$ gauge theory with matter
invariant under the central ${\mathbb Z}_2$, so that the
theory has a $B {\mathbb Z}_2$ (one-form) symmetry.  
Here, decomposition predicts \cite{Sharpe:2014tca}, schematically,
\begin{equation}
SU(2) \: = \: SO(3)_+ \: + \: SO(3)_-,
\end{equation}
where the $\pm$ indicate discrete theta angles.

Mirrors to
such theories and their generalizations were discussed in
\cite{Gu:2018fpm,Chen:2018wep,Gu:2019zkw,Gu:2019byn,Gu:2020ivl,Gu:2020ana}.
For example, the mirror to the pure $SU(2)$ theory is described
by the superpotential
\cite[equ'n (12.3)]{Gu:2018fpm}
\begin{equation}
W \: = \: \sigma \ln \left( \frac{ X_{12} }{ X_{21} } \right)^2
\: + \: X_{12} \: + \: X_{21}.
\end{equation}
Taking the square root, this could be equivalently described
as a theory with a ${\mathbb Z}_2$-valued field $\Upsilon$
and superpotential
\begin{equation}  \label{eq:su2mirror}
W \: = \: \sigma \ln \left( \Upsilon \frac{ X_{12} }{ X_{21} } \right)
\: + \: X_{12} \: + \: X_{21}.
\end{equation}
Such a theory is equivalent to a disjoint union of two Landau-Ginzburg
models, with either value of $\Upsilon = \pm 1$.

For comparison, the mirror to the pure $SO(3)_+$ theory is described by
the superpotential \cite[equ'n (12.9)]{Gu:2018fpm}
\begin{equation}
W \: = \: \sigma \ln \left( \frac{ X_{12} }{ X_{21} } \right)
\: + \: X_{12} \: + \: X_{21},
\end{equation}
which has no vacua (corresponding to the fact that the $SO(3)_+$ theory
dynamically breaks supersymmetry), and the mirror to the pure $SO(3)_-$
theory is described by the superpotential
\cite[equ'n (12.14)]{Gu:2018fpm}
\begin{equation}
W \: = \: \sigma \ln \left( - \frac{ X_{12} }{ X_{21} } \right)
\: + \: X_{12} \: + \: X_{21},
\end{equation}
which does have supersymmetric vacua.

In any event, we now see that the two constituent theories of
the $SU(2)$ mirror, at either value of $\Upsilon = \pm 1$,
are precisely the $SO(3)_{\pm}$ mirrors, recovering the decomposition statement
\begin{equation}
SU(2) \: = \: SO(3)_+ \: + \: SO(3)_-.
\end{equation}
Furthermore, $\Upsilon$ acts as a dimension-zero field, with a ring
relation $\Upsilon^2 = 1$, the same structure we have seen previously
in sigma models on disjoint unions and in other banded abelian gerbes.
Similar structures arise in mirrors to other two-dimensional
gauge theories with center-invariant matter, see for example
\cite{Gu:2018fpm,Chen:2018wep,Gu:2019zkw,Gu:2019byn,Gu:2020ivl}.
As the stories are closely related, we will not describe them explicitly
here.

\section{Four-dimensional analogues}
\label{sect:4d}

The paper \cite{Tanizaki:2019rbk} considered four-dimensional versions of
decomposition.  Specifically, beginning with ordinary bosonic Yang-Mills
theory in four dimensions, with action
\begin{equation}
S \: = \: \frac{1}{2g^2} \int {\rm Tr}\, F \wedge *F,
\end{equation}
they construct a modified theory with a restriction on instanton numbers,
with action of the form
\begin{eqnarray}
S & = & 
\frac{1}{2g^2} \int {\rm Tr}\, F \wedge *F
\nonumber \\
& & 
\: + \:
i \int B  \left( \frac{1}{8 \pi^2} {\rm Tr}\, F \wedge F
\: - \: \frac{k}{2 \pi} d C^{(3)} \right)
\: + \:
\frac{i \hat{\theta}}{2\pi} \int d C^{(3)},
\label{eq:analogues:ym3form}
\end{eqnarray}
where $k$ is an integer, $B$ is a scalar field of periodicity $2 \pi$,
and $C^{(3)}$ is a three-form gauge field.  As they discuss, the
equations of motion for $B$ are
\begin{equation} \label{eq:decomp4d:ceom}
 \frac{1}{8 \pi^2} {\rm Tr}\, F \wedge F
\: = \:
 \frac{k}{2 \pi} d C^{(3)},
\end{equation}
which implies that this theory
has a restriction on instantons (to instanton numbers divisible by $k$),
and exhibits a decomposition.
Rather than study Gukov-Witten operators, we will look at a different
class of (nonlocal) operators in this section.

Briefly, hand-in-hand with existence of a decomposition,
this theory has projection operators, of the form
\begin{equation}
\Pi_n \: = \: \frac{1}{k} \sum_{m=0}^{k-1} \xi^{nm} \exp\left[ i m
\int_{M_3} \left( C^{(3)} - \frac{1}{k} \omega_{\rm CS}(A) \right) \right],
\end{equation}
where
\begin{equation}
d \omega_{\rm CS}(A) \: = \: \frac{1}{8 \pi^2} {\rm Tr}\, F \wedge F.
\end{equation}
and $\xi$ is a $k$th root of unity.
($C$ absorbs the lack of gauge-invariance of $(1/k) \omega_{\rm CS}$.)
These operators are volume operators, defined on three-manifolds $M_3$,
and obey
\begin{equation}
\Pi_n \Pi_m \: = \: \delta_{n,m} \Pi_n,
\end{equation}
using the equations of motion~(\ref{eq:decomp4d:ceom}) in the form
\begin{equation}
        \exp\left[ i k \int_{M_3} \left( C^{(3)} - \frac{1}{k} \omega_{\rm CS}(A) \right) \right] 
        \: = \: 1.
\end{equation}
These operators
project onto particular universes in the decomposition.

In passing, since $B$ can take on only finitely many values,
inserting $\Pi_n$ into the path integral is equivalent to coupling 
the original Yang-Mills theory to the 
TFT to restrict instanton numbers, as the path integral over $B$ is equivalent
to the sum over $m$, and the $\xi^{nm}$ term corresponds to 
$\hat{\theta} = 2 \pi n$.

Physically, these projection operators form domain walls that are
also projectors:  ``end-of-the-world projectors'' that separate distinct
universes.

In any event, given a set of projection operators, we can proceed precisely
as we have previously in this paper, for e.g. sigma models on disjoint
unions and banded ${\mathbb Z}_k$ gerbes, and take linear combinations
of projectors so that for any
three-submanifold $M_3$, we have a ring of operators given by
${\mathbb C}[x]/(x^k - 1)$, which includes the projectors as a subset,
and which describes a set of $k$ points (one for each universe in the
decomposition).

\section{Conclusions}

In this paper we have discussed dimension-zero operators and Wilson
lines in two-dimensional 
theories, especially orbifolds, exhibiting decomposition.
We have described the computation of fusion algebras of twist fields,
given a systematic construction of projectors onto universes,
and used those tools to verify that the projectors project Wilson lines
onto universes in the fashion predicted in \cite{Hellerman:2006zs}.

We have also discussed the geometries underlying these computations.
These algebras of twist fields are commutative algebras, and the methods
of commutative ring theory give one a perspective on their properties --
for example, the rings describe a set of points, as many as universes
in decomposition.

We have also discussed the symmetries of these theories -- both ordinary
one-form
and noninvertible symmetries -- and their descriptions in
terms of twist fields (Gukov-Witten operators).

\section{Acknowledgements}

We would like to thank S.~Katz, D.~Robbins,
S.~Seifnashri, Y.~Tanizaki, M.~\"Unsal, and T.~Vandermeulen for
useful conversations.  In addition,
we would particularly
like to thank Z.~Komargodski for asking the questions that
led to this paper, and T.~Pantev for discussions of Wedderburn's theorem
and many years of collaboration on decomposition generally.
E.S. was partially supported by NSF grant
PHY-2014086.

\appendix

\section{Casimirs}
\label{app:casimirs}

In this section we will justify the fact that the action of
the twist fields $\sigma_{[g]}$ on Wilson lines is to multiply by a factor
proportional to a character.

Briefly, each $\sigma_{[g]}$ is a linear combination of operators
$\tau_g$, and the effect of any $\tau_g$ on a Wilson line $W_R$ in
representation $R$ is to
insert the matrix $T^R_g$ into the Wilson line.  We will demonstrate that
the linear combination of $T^R_g$'s provided by $\sigma_{[g]}$ commutes
with all other matrices in that representation, and so (by e.g. Schur's
lemma) is proportional to the identity.  This is equivalent to multiplying
the Wilson line by a numerical factor.

We first discuss the pertinent claim for ordinary representations,
and then discuss the more complicated case of projective representations
subsequently.
Let $G$ be a finite group, $R$ an (ordinary) irreducible
representation of $G$,
and let $T^R_g$ denote a matrix representing $g \in G$ in representation $R$.
Let $T_{[g]}$ be the matrix inserted in Wilson lines $W_R$ by the branch
cut induced by the twist field $\sigma_{[g]}$.  Specifically, 
in this case (\ref{eq:sigma-defn-nodt})
\begin{equation}
\sigma_{[g]} \: = \: \frac{1}{|[g]|} \sum_{x \in [g]} \tau_x,
\end{equation}
we have 
\begin{equation}
T_{[g]} \: = \: \frac{1}{|[g]|} \sum_{x \in [g]} T^R_x,
\end{equation}
where $[g]$ is a conjugacy class of $G$ containing $g$, 
$|[g]|$ is the number of elements in that conjugacy class.

We first demonstrate that
\begin{equation}
T_{[g]} \: = \:
\frac{ \chi_R(g) }{ \dim R} I,
\end{equation}
where $I$ is the identity matrix.
This is a standard result, but we pause to prove it en route
to establishing an analogue for projective representations.
The key result is that 
for any $y \in G$, $T_{[g]} T^R_y =
T^R_y T_{[g]}$, which we verify as follows:
\begin{equation}
T_{[g]} T^R_y \: = \: 
\frac{1}{|[g]|}
\sum_{x \in [g]} T^R_x T^R_y \: = \: 
\frac{1}{|[g]|}
\sum_{x \in [g]} T^R_y T^R_{y^{-1} x y}
\: = \: T^R_y T_{[g]}.
\end{equation}
Since $T_{[g]}$ commutes with all the $T^R_y$,
Schur's lemma then implies that $T_{[g]}$ must be proportional to the
identity, and the normalization can be easily checked.  The desired
result follows.

Next, we consider the analogue of this statement for projective
representations.  To that end, let $\omega \in H^2(G,U(1))$
be a normalized cocycle, meaning
\begin{equation}
\omega(1,g) \: = \: 1 \: = \: \omega(g,1)
\end{equation}
for all $g \in G$,
and consider an irreducible projective representation $R$,
representing elements of $G$ by matrices
$T^R_g$ such that
\begin{equation}
T^R_g T^R_h \: = \: \omega(g,h) T^R_{gh}.
\end{equation}
From equation~(\ref{eq:sigma-defn}), 
\begin{equation}
\sigma_{[g]} \: = \: \frac{1}{|G|} \sum_{h \in G}
\frac{\omega(h,g) \omega(hg,h^{-1}) }{ \omega(h,h^{-1}) }
\tau_{h g h^{-1}},
\end{equation}
so the effect of $\sigma_{[g]}$ is to insert into a Wilson line $W_R$ the
matrix
\begin{equation}
T_{[g]} \: = \: \frac{1}{|G|} \sum_{h \in G}
\frac{\omega(h,g) \omega(hg,h^{-1}) }{ \omega(h,h^{-1}) }
T^R_{h g h^{-1}}.
\end{equation}

We now check that
$T_{[g]} T^R_y \: = \: T^R_y T_{[g]}$:
\begin{eqnarray}
T_{[g]} T^R_y & = & \frac{1}{|G|}
\sum_{h \in G} \frac{ \omega(h,g) \omega(hg,h^{-1}) }{ \omega(h,h^{-1}) }
T^R_{hgh^{-1}} T^R_y,
\\
& = & \frac{1}{|G|}
\sum_{h \in G}
\frac{ \omega(h,g) \omega(hg,h^{-1}) \omega(hgh^{-1},y) }{ \omega(h,h^{-1}) }
T^R_{hgh^{-1}y},
\\
& = & \frac{1}{|G|}
\sum_{h \in G}
\frac{
\omega(h,g) \omega(hg,h^{-1}) \omega(hgh^{-1},y)
}{ \omega(h,h^{-1})
\omega(y,y^{-1}hgh^{-1}y)
}
T^R_y T^R_{y^{-1}hgh^{-1}y}.
\end{eqnarray}
It can be shown that
\begin{equation}
\frac{
\omega(h,g) \omega(hg,h^{-1}) \omega(hgh^{-1},y)
}{ \omega(h,h^{-1})
\omega(y,y^{-1}hgh^{-1}y)
}
\: = \:
\frac{ \omega(y^{-1} h, g) \omega(y^{-1} h g, h^{-1} y) }{
\omega(y^{-1} h, h^{-1} y) },
\end{equation}
by multiplying in 
\begin{equation}
\frac{
(d\omega)(y,y^{-1},hgh^{-1}y) (d\omega)(hg, h^{-1}, y)
(d\omega)(h^{-1},y,y^{-1}h)
}{
(d\omega)(y^{-1},hg,h^{-1}y) (d\omega)(y^{-1},h,g)
(d\omega)(y,y^{-1},h)
} \: = \: 1,
\end{equation}
and using the fact that for normalized cocycles, 
$\omega(h,h^{-1}) = \omega(h^{-1},h)$
for all $h$.
As a result, the effect of multiplying in $y$ is to (potentially)
interchange elements of the sum, and so we have
\begin{equation}
T_{[g]} T^R_y \: = \:
T^R_y T_{[g]},
\end{equation}
confirming that $T_{[g]}$ is central.

Now, we claim that $T_{[g]}$ is proportional to the identity matrix.
We have shown that $T_{[g]}$ commutes with other representation matrices,
and for honest non-projective representations, Schur's lemma implies the
desired result.

Now, in principle, Schur's lemma only applies to honest representations,
not projective representations, so we need to recast this as a question
about honest representations of a central extension $\Gamma$ of $G$:
\begin{equation}
1 \: \longrightarrow \: U(1) \: \longrightarrow \: \Gamma \:
\longrightarrow \: G \: \longrightarrow \: 1.
\end{equation}
To that end, we pick a splitting of $\Gamma$, and write $\gamma \in \Gamma$
as pairs $\gamma = (x,\lambda)$ for $x \in G$, $\lambda \in U(1)$,
with multiplication
\begin{equation}
(x,\lambda) \cdot (y,\mu) \: = \: (xy, \lambda \mu \omega(x,y)).
\end{equation}
Then, given $g \in G$, we lift to $(g,1) \in \Gamma$.
Conjugating $[(g,1)]$ in $\Gamma$ by $(h,\lambda)$ gives a pair
of the form
\begin{equation}
(hgh^{-1}, \omega(h,g) \omega(hg,h^{-1}) ).
\end{equation}
This pair acts on a representation as
\begin{equation}
\omega(h,g) \omega(hg,h^{-1}) T^R_{hgh^{-1}},
\end{equation}
so we see that the combination
\begin{equation}
\sum_{hgh^{-1} \in [g]} \omega(h,g) \omega(hg,h^{-1}) T^R_{hgh^{-1}}
\end{equation}
is a sum over lifts to $\Gamma$, and hence involves honest
representations, so we can apply Schur's lemma again.

Doing so, and checking the normalization, one finds
\begin{equation}
T_{[g]} \: = \: \frac{ \chi_R(g) }{ \dim R } I.
\end{equation}

In any event, given that the action of $\sigma_{[g]}$ on a Wilson line
$W_R$ is to insert a matrix proportional to the identity, it is now
clear that
\begin{equation}
\sigma_{[g]} W_R \: = \: \frac{ \chi_R(g) }{ \dim R } W_R.
\end{equation}

In passing, in the presence of discrete torsion, changing the representative
of the conjugacy class $[g]$ multiplies both sides of the expression above
by a phase.  We check in section~\ref{sect:pairing} 
that the phases are identical, so
that the expression above is consistent.

\section{Character identities}
\label{app:char}

We collect here some character identities for finite groups $G$, to make this
paper self-contained.  See e.g. \cite[section 2]{serrerep},
\cite[section 7.3]{karpilovsky},
\cite[chapter V]{cr}, \cite[chapter 2.1]{collins}, \cite[section 1.12]{georgi}.

Let $\omega$ denote a cocycle representing an element of $H^2(G,U(1))$,
corresponding to twisting, and we assume that all representations
are unitary (projective) representations.
The identities we will use are sometimes written in terms of
complex conjugates of characters, which for an irreducible representation
$R$ twisted by $\omega$, are related by
\cite[section 2]{cheng}
\begin{equation}
\chi_R(g^{-1}) \: = \: \omega(g,g^{-1}) \overline{ \chi_R(g) },
\end{equation}
which for trivial $\omega$ specializes to
\begin{equation}  \label{eq:chi-ginv}
\chi_R(g^{-1}) \: = \: \overline{ \chi_R(g) }.
\end{equation}

Let $T^R(g)$ denote a matrix representing $g \in G$ in irreducible
representation $R$,
which obeys
\begin{equation}
T^R(g) T^R(h) \: = \: \omega(g,h) T^R(gh).
\end{equation}
Then, the source of the identities we will primarily use is\footnote{
Experts should note that the sum is over all elements of $G$,
not just the $\omega$-regular elements (meaning, elements $g \in G$
such that $\omega(g,h) = \omega(h,g)$ for all $h$ commuting with $g$).
Sums over the latter sometimes arise in character identities for
projective representations, simply because characters of non-regular
elements vanish.  It is straightforward to check in examples that
the identity~(\ref{eq:char-master}) above is only valid when one sums
over all $g \in G$, not just $\omega$-regular elements of $G$.
}
\begin{equation} \label{eq:char-master}
\frac{1}{|G|} \sum_{g \in G} 
\frac{ T^R(g)_{ju} T^S(g^{-1})_{ik} }{ \omega(g,g^{-1}) }
\: = \:
\frac{ \delta_{R,S} }{ \dim R}
\delta_{jk} \delta_{ui}.
\end{equation}
(See e.g. \cite[section 31.1]{cr}, \cite[exercise 2.1.1, p. 58]{collins},
\cite[section 1.12]{georgi}
for a version without discrete
torsion.)

Let us quickly outline a proof of the assertion above, as this identity
may seem obscure.  Following \cite[section 31]{cr}, for any $(\dim R) \times 
(\dim S)$ matrix $C$, define
\begin{equation}
T \: = \: \sum_{g \in G} \frac{ T_R(g) C T_S(g^{-1}) }{ \omega(g,g^{-1}) },
\end{equation}
then one can show that $T$ intertwines the representations $R$, $S$,
meaning for any $h \in G$,
\begin{equation}
T_R(h) T \: = \: T T_S(h).
\end{equation}
To establish that, the key step is the identity
\begin{equation}
\frac{ \omega(h, h^{-1} g) }{ \omega(h^{-1} g, g^{-1} h)}
\: = \:
\frac{ \omega(g^{-1}, h) }{ \omega(g, g^{-1}) },
\end{equation}
which can be established by multiplying by the coboundaries
\begin{equation}
\frac{
(d \omega)(h, h^{-1}, g) (d \omega)(g^{-1}, h, h^{-1})
(d \omega)(h^{-1}, g, g^{-1} h) (d \omega)(g, g^{-1} h, h^{-1})
}{
\omega(h, h^{-1}, h)
} \: = \: 1,
\end{equation}
and using the normalization condition.
Then, given that intertwining, $T \propto \delta_{R,S} I$, and the 
normalization is straightforward to compute.  Taking $C$ to be the identity,
the result~(\ref{eq:char-master}) follows.

As a consistency test, note that if we contract $u$ and $i$,
and set $S=R$,
this identity reduces to
\begin{equation}
\frac{1}{|G|} \sum_{G \in G} T^R(1)_{jk} \: = \: \delta_{jk},
\end{equation}
which is trivially correct.

As another consistency test, if we contract the pair $(u,j)$ and the
pair $(i,k)$, this identity reduces to\footnote{
Because characters of non-$\omega$-regular elements vanish,
these identities are sometimes equivalently written as a sum over only
$\omega$-regular elements, see e.g. \cite[section 7.3]{karpilovsky},
instead of all elements of $G$.  For simplicity, we have chosen to 
write these in terms of sums over all elements of $G$.
}
\begin{equation}
\frac{1}{|G|} \sum_{g \in G} \frac{ \chi_R(g) \chi_S(g^{-1}) }{ \omega(g,g^{-1})
}
\: = \:
\delta_{R,S},
\end{equation}
or more simply,
\begin{equation}  \label{eq:chi-orthog}
\frac{1}{|G|} \sum_{g \in G} \chi_R(g) \overline{ \chi_S(g) }
\: = \: \delta_{R,S},
\end{equation}
which can be found in e.g.
\cite[section 2.3]{serrerep}, \cite[section 2]{cheng},
\cite[chapter 7.3]{karpilovsky}.

If we multiply factors of $T^R(a)$, $T^S(b)$ into 
expression~(\ref{eq:char-master}),
we find
\begin{equation}
\frac{1}{|G|} \sum_{g \in G} 
\frac{ \omega(a,g) \omega(g^{-1},b) }{ \omega(g,g^{-1}) } \,
T^R(ag)_{\ell u} \, T^S(g^{-1} b)_{im} \,
\: = \:
\frac{ \delta_{R,S} }{ \dim R}
\, \omega(a,b) \, T^R(ab)_{\ell m} \, \delta_{ui},
\end{equation}
whose traces gives
\begin{equation} \label{eq:char-om1}
\frac{1}{|G|} \sum_{g \in G} 
\frac{ \omega(a,g) \omega(g^{-1},b) }{ \omega(g,g^{-1}) } \,
\chi_R(ag) \, \chi_S(g^{-1} b)
\: = \:
\frac{ \delta_{R,S} }{\dim R} \, \omega(a,b) \, \chi_R(ab).
\end{equation}

Similarly, if we take $S=R$ and multiply factors of $T^R(a)$, $T^R(b)$
into expression~(\ref{eq:char-master}),
we find
\begin{equation}
\frac{1}{|G|} \sum_{g \in G} \frac{ \omega(g,a) \, \omega(g^{-1},b) \,
\omega(ga, g^{-1} b) }{ \omega(g,g^{-1}) } \,
T^R(g a g^{-1} b)_{jm}
\: = \:
\frac{1}{\dim R} \, \chi_R(a) \, T^R(b)_{jm},
\end{equation}
whose trace gives
\begin{equation} \label{eq:char-om2}
\frac{1}{|G|} \sum_{g \in G} \frac{ \omega(g,a) \, \omega(g^{-1},b) \,
\omega(ga, g^{-1} b) }{ \omega(g,g^{-1}) } \,
\chi_R(g a g^{-1} b) \: = \:
\frac{1}{\dim R} \, \chi_R(a) \, \chi_R(b).
\end{equation}

The expressions~(\ref{eq:char-om1}), (\ref{eq:char-om2})
may seem somewhat exotic, but 
in another context they may look more familiar.  Specifically,
if we specialize to the case of vanishing discrete torsion,
they reduce to
\begin{equation}  \label{eq:A}
\frac{1}{|G|} \sum_{g \in G} \chi_R(ag) \chi_S(g^{-1}b) \: = \:
\delta_{R,S} \frac{\chi_R(ab)}{\dim R},
\end{equation}
\begin{equation}
\frac{1}{|G|} \sum_{g \in G} \chi_R(g a g^{-1} b) \: = \:
\frac{\chi_R(a) \chi_R(b)}{\dim R}.
\end{equation}
which are simply finite-group analogues of perhaps more familiar
versions from Lie groups of nonzero dimension:
\begin{equation}
\int dV \chi_R(XV) \chi_S(V^{\dag} Y) \: = \:
\frac{ \delta_{RS} }{ \dim R} \chi_R(XY),
\end{equation}
\begin{equation}
\int dU \chi_R(AUBU^{\dag}) \: = \: \frac{1}{\dim R} \chi_R(A) \chi_R(B),
\end{equation}
(see e.g. \cite{Gross:1992tu}).

Another useful orthogonality relation arises from summing over all
the irreducible (projective) representations of a group
\cite[prop. 2.7]{serrerep},
\cite[section 7.3, theorem 3.2]{karpilovsky}:
\begin{equation}   \label{eq:orthog2}
\sum_R \frac{ \chi_R(g) \chi_R(h^{-1}) }{ \omega(g,g^{-1}) } \: = \:
\left\{ \begin{array}{cl}
0 & g, h \mbox{ not conjugate}, \\
\frac{ |G|}{|[g]|}  & g, h \mbox{ conjugate},
\end{array} \right.
\end{equation}
where $|[g]|$ denotes the number of elements in a conjugacy class
containing $g$.  For applications, it may be helpful to note that
the number of elements in the centralizer of $g$ (the set of elements that
commute with $g$) equals $|G|/|[g]|$.

\section{Induced representations}
\label{app:induced}

In this section we wil argue that the irreducible components of
restrictions of representations of $G$ to a normal subgroup $K$
span all irreducible representations of $K$.  The argument is a short
exercise in induced representations, which we review.

Let $K$ be a normal subgroup of a finite group $G$,
and let $\rho$ be a representation of $K$,
acting on a vector space $V$.
Briefly, $\rho$ induces a representation $\tilde{\rho}$ of $G$,
which can be described as the vector bundle over $G/K$
associated to the principal $K$ bundle $G \rightarrow G/K$.
(Note that as a vector space, this is $|G/K|$ copies of $V$.)
This has a natural $G$-equivariant structure, and so gives a 
representation of $G$.

In more detail, let $\{ g_i \}$ be a set of representatives of 
$G/K$, then the vector space $\tilde{V}$ on which the induced
representation is the sum
\begin{equation}
\tilde{V} \: = \: \oplus_i g_i V.
\end{equation}
Any element $g \in G$ acts as follows.  For each $i$, write
\begin{equation}
g g_i \: = \: g_{j(i)} k
\end{equation}
for some $k \in K$, whose action is given by $\rho$.

It will be helpful to consider an example.
Let $G = D_4$, the eight-element finite group,
and let $K = {\mathbb Z}_2$, its center.
We present $D_4$ as
\begin{equation}
D_4 \: = \: \{ 1, a, b, z, az, bz, ab, ba = abz \},
\end{equation}
where $a^2 = 1 = z^2$, $b^2 = z$, and $z$ generates the center.
Let the representatives of the coset $G/K$ be $\{1, a, b, ab \}$,
so that, for example,
\begin{equation}
a a \: = \: 1, \: \: \:
a b \: = \: ab, \: \: \:
a ab \: = \: b,
\end{equation}
\begin{equation}
b a \: = \: abz, \: \: \:
b b \: = \: z, \: \: \:
b ab \: = \: a,
\end{equation}
\begin{equation}
ab a \: = \: b z, \: \: \:
ab b \: = \: az, \: \: \:
ab ab \: = \: 1.
\end{equation}
Let $\rho$ be the nontrivial one-dimensional representation of $K = 
{\mathbb Z}_2$.  The induced representation $\tilde{\rho}$ has vector space
\begin{equation}
\tilde{V} \: = \: \oplus_4 V \: = \: V \oplus a V \oplus b V \oplus ab V.
\end{equation}
Let $E \equiv (x_1, x_a, x_b, x_{ab}) \in \tilde{V}$, 
then from the multiplication
rules above, we have
\begin{eqnarray}
z \cdot E & = & (-x_1, -x_a, -x_b, -x_{ab}),
\\
a \cdot E & = & (x_a, x_1, x_{ab}, x_b),
\\
b \cdot E & = & (-x_b, x_{ab}, x_1, -x_a),
\\
(ab) \cdot E & = & (x_{ab}, -x_b, -x_a, x_1),
\end{eqnarray}
so that
\begin{equation}
a \cdot (b \cdot E) \: = \: (x_{ab}, -x_b, -x_a, x_1) \: = \: (ab) \cdot E,
\end{equation}
\begin{equation}
b \cdot (a \cdot E) \: = \: (-x_{ab}, x_b, x_a, -x_1) \: = \: (ba = abz) \cdot E,
\end{equation}
as expected

If we restrict $\tilde{\rho}$ to $K$, we get four copies of $\rho$:
\begin{equation}
\tilde{\rho}|_K \: = \: \bigoplus_4 \rho.
\end{equation}

For another example, suppose $G = {\mathbb H}$, the eight-element group
of unit quaternions, and $K = \langle i \rangle \cong {\mathbb Z}_4$.
Let $\rho$ be an irreducible representation of ${\mathbb Z}_4$
that maps the generator of ${\mathbb Z}_4$ to multiplication by $\xi$,
for $\xi$ some fourth root of unity.
Let $\{1, j \}$ represent cosets in $G/K = {\mathbb Z}_2$, so that
\begin{equation}
\tilde{V} \: = \: V \oplus j V
\end{equation}
for $V \cong {\mathbb C}$.
Then, from the multiplications
\begin{equation}
i j \: = \: k \: = \: j (-i), \: \: \:
j j \: = \: -1,  \: \: \:
k j \: = \: -i,
\end{equation}
for $E = (x_1, x_j) \in \tilde{V}$, we have
\begin{equation}
i \cdot E \: = \: (\xi x_1, \xi^3 x_j), \: \: \:
j \cdot E \: = \: (\xi^2 x_j, x_1), \: \: \:
k \cdot E \: = \: (\xi^3 x_j, \xi^3 x_1),
\end{equation}
so that, for example,
\begin{equation}
i \cdot (j \cdot E) \: = \: (\xi^3 x_j, \xi^3 x_1) \: = \: (k = ij) \cdot E,
\end{equation}
as expected.

In this case, if we restrict the induced representation
$\tilde{\rho}$ to $K$, we find
\begin{equation}
\tilde{\rho}|_K \: = \: \rho \oplus \rho^3.
\end{equation}

In general, given any irreducible representation $\rho$ of $K$,
the restriction of the induced representation will be a sum of
irreducible representations of $K$, as many as elements of $G/K$,
with at least one copy (over the identity coset) equal to $\rho$.
If $K$ is central, then all the irreducible representations appearing
in the restriction will be copies of $\rho$.

In particular, every irreducible representation of $K$ appears
as a summand in the decomposition of the restriction of 
representations of $G$ to $K$:  for any representation $\rho$ of $K$,
at least one summand in the
restriction of the induced representation will be a copy of $\rho$.

\section{Miscellaneous group cohomology results}
\label{app:miscgpcoh}

In this section we collect a handful of pertinent statements in
group cohomology that will be used elsewhere.

First, we often use the fact that
for a group 2-cochain $\omega: G \times G \rightarrow U(1)$
(with trivial action on the coefficients),
\begin{equation}
(d \omega)(a,b,c) \: = \:
\frac{\omega(b,c) \omega(a,bc) }{
\omega(ab, c) \omega(a,b) }.
\end{equation}

We will typically use normalized cocycles,
by which we mean that
\begin{equation}
\omega(1,g) \: = \: 1 \: = \: \omega(g,1) 
\end{equation}
for all $g \in G$.
In addition, we can impose two more constraints:
\begin{enumerate}
\item $\omega(g,g^{-1}) = 1 = \omega(g^{-1},g)$,
\item for $\omega$-regular elements $g \in G$, for $h$ commuting with $g$,
\begin{equation}
\omega(h,g) \omega(hg,h^{-1}) \: = \: 1.
\end{equation}
\end{enumerate}

The first of these two conditions can be demonstated as follows.
From $(d \omega)g,g^{-1},g) = 1$ and the normalization condition, we have
\begin{equation}
\omega(g,g^{-1}) \: = \: \omega(g^{-1},g),
\end{equation}
and by picking a coboundary $\mu$ such that
$\mu(1) = 1$, $\mu(g) \mu(g^{-1}) = \omega(g,g^{-1})^{-1}$,
we can replace $\omega$ by $\omega'$ such that $\omega'(g,g^{-1}) = 1$.

The second condition
can be demonstrated as follows.
First, recall that for an element $g \in G$ to be 
$\omega$-regular means that for all $h$ that commute with $g$,
$\omega(g,h) = \omega(h,g)$.
Then, from the cocycle condition $(d\omega)(g,h,h^{-1}) = 1$,
we have that
\begin{equation}
\omega(h,g) \omega(hg,h^{-1}) \: = \: \omega(h,h^{-1})
\frac{ \omega(h,g) }{ \omega(g,h) }.
\end{equation}
From $\omega$-regularity, the right-hand side reduces to $\omega(h,h^{-1})$,
which as already demonstrated can be chosen to equal $1$.


\begin{thebibliography}{199}

\addcontentsline{toc}{section}{References}

\bibitem{Komargodski:2020ved}
Z.~Komargodski, S.~S.~Razamat, O.~Sela and A.~Sharon,
``A nilpotency index of conformal manifolds,''
JHEP \textbf{10} (2020) 183,
{\tt arXiv:2003.04579}.

\bibitem{Rudelius:2020orz}
T.~Rudelius and S.~H.~Shao,
``Topological operators and completeness of spectrum in discrete gauge theories,''
JHEP \textbf{12} (2020) 172,
{\tt arXiv:2006.10052}.

\bibitem{Heidenreich:2021tna}
B.~Heidenreich, J.~McNamara, M.~Montero, M.~Reece, T.~Rudelius and I.~Valenzuela,
``Non-invertible global symmetries and completeness of the spectrum,''
{\tt arXiv:2104.07036}.

\bibitem{Komargodski:2020mxz}
Z.~Komargodski, K.~Ohmori, K.~Roumpedakis and S.~Seifnashri,
``Symmetries and strings of adjoint QCD$_{2}$,''
JHEP \textbf{03} (2021) 103,
{\tt arXiv:2008.07567}.

\bibitem{Thorngren:2019iar}
R.~Thorngren and Y.~Wang,
``Fusion category symmetry I: anomaly in-flow and gapped phases,''
{\tt arXiv:1912.02817}.


\bibitem{Ji:2019jhk}
W.~Ji and X.~G.~Wen,
``Categorical symmetry and noninvertible anomaly in symmetry-breaking and topological phase transitions,''
Phys. Rev. Res. \textbf{2} (2020)  033417,
{\tt arXiv:1912.13492}.


\bibitem{Nguyen:2021yld}
M.~Nguyen, Y.~Tanizaki and M.~\"Unsal,
``Semi-Abelian gauge theories, non-invertible symmetries, and string tensions beyond $N$-ality,''
JHEP \textbf{03} (2021) 238,
{\tt arXiv:2101.02227}.

\bibitem{Nguyen:2021naa}
M.~Nguyen, Y.~Tanizaki and M.~\"Unsal,
``Non-invertible 1-form symmetry and Casimir scaling in 2d Yang-Mills theory,''
{\tt arXiv:2104.01824}.


\bibitem{Bhardwaj:2017xup}
L.~Bhardwaj and Y.~Tachikawa,
``On finite symmetries and their gauging in two dimensions,''
JHEP \textbf{03} (2018) 189,
{\tt arXiv:1704.02330}.

\bibitem{Brunner:2013ota}
I.~Brunner, N.~Carqueville and D.~Plencner,
``Orbifolds and topological defects,''
Commun. Math. Phys. \textbf{332} (2014) 669-712,
{\tt arXiv:1307.3141}. 

\bibitem{Brunner:2013xna}
I.~Brunner, N.~Carqueville and D.~Plencner,
``A quick guide to defect orbifolds,''
Proc. Symp. Pure Math. \textbf{88} (2014) 231-242,
{\tt arXiv:1310.0062}.

\bibitem{Brunner:2014lua}
I.~Brunner, N.~Carqueville and D.~Plencner,
``Discrete torsion defects,''
Commun. Math. Phys. \textbf{337} (2015) 429-453,
{\tt arXiv:1404.7497}.






\bibitem{Gukov:2006jk}
S.~Gukov and E.~Witten,
``Gauge theory, ramification, and the geometric Langlands program,''
{\tt arXiv:hep-th/0612073}.

\bibitem{Gukov:2008sn}
S.~Gukov and E.~Witten,
``Rigid surface operators,''
Adv. Theor. Math. Phys. \textbf{14} (2010)  87-178,
{\tt arXiv:0804.1561}.

\bibitem{Polchinski:2003bq}
J.~Polchinski,
``Monopoles, duality, and string theory,''
Int. J. Mod. Phys. A \textbf{19S1} (2004) 145-156,
{\tt arXiv:hep-th/0304042}.


\bibitem{Craig:2018yvw}
N.~Craig, I.~Garcia Garcia and S.~Koren,
``Discrete gauge symmetries and the weak gravity conjecture,''
JHEP \textbf{05} (2019) 140,
{\tt arXiv:1812.08181}.

\bibitem{Pantev:2005rh}
T.~Pantev and E.~Sharpe, 
``Notes on gauging noneffective group actions,''
{\tt arXiv:hep-th/0502027}.


\bibitem{Pantev:2005wj}
T.~Pantev and E.~Sharpe,
``String compactifications on Calabi-Yau stacks,''
Nucl. Phys. B \textbf{733} (2006) 233-296,
{\tt arXiv:hep-th/0502044}.

\bibitem{Pantev:2005zs}
T.~Pantev and E.~Sharpe,
``GLSM's for gerbes (and other toric stacks),''
Adv. Theor. Math. Phys. \textbf{10} (2006) 77-121,
{\tt arXiv:hep-th/0502053}.

\bibitem{Hellerman:2006zs}
S.~Hellerman, A.~Henriques, T.~Pantev, E.~Sharpe and M.~Ando,
``Cluster decomposition, T-duality, and gerby CFT's,''
Adv. Theor. Math. Phys. \textbf{11} (2007) 751-818,
{\tt arXiv:hep-th/0606034}.

\bibitem{ajt1} E. Andreini, Y. Jiang, H.-H. Tseng,
``On Gromov-Witten theory of root gerbes,''
{\tt arXiv:0812.4477}.

\bibitem{ajt2} E. Andreini, Y. Jiang, H.-H. Tseng,
``Gromov-Witten theory of product stacks,''
{\tt arXiv:0905.2258}.

\bibitem{ajt3} E. Andreini, Y. Jiang, H.-H. Tseng,
``Gromov-Witten theory of etale gerbes, I: root gerbes,''
{\tt arXiv:0907.2087}.

\bibitem{tseng1} H.-H. Tseng,
``On degree zero elliptic orbifold Gromov-Witten invariants,''
Int. Math. Res. Not. IMRN 2011 2444-2468,
{\tt arXiv:0912.3580}.

\bibitem{gt1} A. Gholampour, H.-H. Tseng,
``On Donaldson-Thomas invariants of threefold stacks and gerbes,''
Proc. Amer. Math. Soc. {\bf 141} (2013) 191-203,
{\tt arXiv:1001.0435}.

\bibitem{xt1} X. Tang, H.-H. Tseng,
``Duality theorems of \'etale gerbes on orbifolds,''
Adv. Math. {\bf 250} (2014) 496-569,
{\tt arXiv:1004.1376}.


\bibitem{Caldararu:2007tc}
A.~C\u{a}ld\u{a}raru, J.~Distler, S.~Hellerman, T.~Pantev and E.~Sharpe,
``Non-birational twisted derived equivalences in abelian GLSMs,''
Commun. Math. Phys. \textbf{294} (2010) 605-645,
{\tt arXiv:0709.3855}.

\bibitem{Hori:2011pd}
K.~Hori,
``Duality in two-dimensional (2,2) supersymmetric non-Abelian gauge theories,''
JHEP \textbf{10} (2013) 121,
{\tt arXiv:1104.2853}.


\bibitem{Halverson:2013eua}
J.~Halverson, V.~Kumar and D.~R.~Morrison,
``New methods for characterizing phases of 2D supersymmetric gauge theories,''
JHEP \textbf{09} (2013) 143,
{\tt arXiv:1305.3278}.

\bibitem{Hori:2013gga}
K.~Hori and J.~Knapp,
``Linear sigma models with strongly coupled phases - one parameter models,''
JHEP \textbf{11} (2013) 070,
{\tt arXiv:1308.6265}.

\bibitem{Wong:2017cqs}
K.~Wong,
``Two-dimensional gauge dynamics and the topology of singular determinantal varieties,''
JHEP \textbf{03} (2017) 132,
{\tt arXiv:1702.00730}.

\bibitem{Eager:2020rra}
R.~Eager and E.~Sharpe,
``Elliptic genera of pure gauge theories in two dimensions with semisimple non-simply-connected gauge groups,''
{\tt arXiv:2009.03907}.

\bibitem{Robbins:2021lry}
D.~Robbins, E.~Sharpe and T.~Vandermeulen,
``Anomalies, extensions and orbifolds,''
{\tt arXiv:2106.00693}.

\bibitem{Robbins:2021ibx}
D.~Robbins, E.~Sharpe and T.~Vandermeulen,
``Quantum symmetries in orbifolds and decomposition,''
{\tt arXiv:2107.12386}.

\bibitem{Robbins:2021xce}
D.~Robbins, E.~Sharpe and T.~Vandermeulen,
``Anomaly resolution via decomposition,''
{\tt arXiv:2107.13552}.

\bibitem{Tanizaki:2019rbk}
Y.~Tanizaki and M.~\"Unsal,
``Modified instanton sum in QCD and higher-groups,''
JHEP \textbf{03} (2020) 123,
{\tt arXiv:1912.01033}. 

\bibitem{Sharpe:2019ddn}
E.~Sharpe,
``Undoing decomposition,''
Int. J. Mod. Phys. A \textbf{34} (2020) 1950233,
{\tt arXiv:1911.05080}.

\bibitem{Cherman:2020cvw}
A.~Cherman and T.~Jacobson,
``Lifetimes of near eternal false vacua,''
Phys. Rev. D \textbf{103} (2021)  105012,
{\tt arXiv:2012.10555}.

\bibitem{Robbins:2021ylj}
D.~G.~Robbins, E.~Sharpe and T.~Vandermeulen,
``A generalization of decomposition in orbifolds,''
{\tt arXiv:2101.11619}.

\bibitem{Delmastro:2021otj}
D.~Delmastro, J.~Gomis and M.~Yu,
``Infrared phases of 2d QCD,''
{\tt arXiv:2108.02202}.

\bibitem{Chen:2000cy}
W.~M.~Chen and Y.~B.~Ruan,
``A new cohomology theory for orbifold,''
Commun. Math. Phys. \textbf{248} (2004) 1-31,
{\tt arXiv:math/0004129}.


\bibitem{Frohlich:2006ch}
J.~Frohlich, J.~Fuchs, I.~Runkel and C.~Schweigert,
``Duality and defects in rational conformal field theory,''
Nucl. Phys. B \textbf{763} (2007) 354-430,
{\tt arXiv:hep-th/0607247}.

\bibitem{Chang:2018iay}
C.~M.~Chang, Y.~H.~Lin, S.~H.~Shao, Y.~Wang and X.~Yin,
``Topological defect lines and renormalization group flows in two dimensions,''
JHEP \textbf{01} (2019) 026,
{\tt arXiv:1802.04445}. 

\bibitem{yujitasi2019} Y. Tachikawa, TASI 2019 lectures,
available at
{\tt https://member.ipmu.jp/yuji.tachikawa/lectures/2019-top-anom/tasi2019.pdf}

\bibitem{conlon} S. B. Conlon, ``Twisted group algebras and their
representations,'' 
J. Aust. Math. Soc. {\bf 4} (1964) 152-173.


\bibitem{serrerep} J.-P. Serre, {\it Linear representations of finite groups},
Springer-Verlag, New York, 1977.

\bibitem{cheng} C. Cheng, ``A character theory for projective representations
of finite groups,''
Linear Alg. Appl. {\bf 469} (2015) 230-242.


\bibitem{costache} T. L. Costache, ``On irreducible projective representations
of finite groups,'' Surv. Math. Appl. {\bf 4} (2009) 191-214.

\bibitem{karpilovsky} G. Karpilovsky, {\it Projective representations of
finite groups}, Marcel Dekker, New York, 1985.

\bibitem{schur1} I. Schur, ``\"Uber die Darstellung der endlicher Gruppen
durch gebrochene lineare Substitutionen,''
J. f\"ur Math. {\bf 127} (1904) 20-40.

\bibitem{schur2} I. Schur, ``Untersuchungen \"uber die Darstellung der
endlicher Gruppen durch
gebrochene lineare Substitutionen,''
J. f\"ur Math. {\bf 132} (1907) 85-137.

\bibitem{schur3} I. Schur, ``\"Uber die Darstellung der symmetrischen und der
alternierenden Gruppe durch gebrochene lineare Substitutionen,''
J. f\"ur Math. {\bf 139} (1911) 155-250.

\bibitem{Vafa:1994rv}
C.~Vafa and E.~Witten,
``On orbifolds with discrete torsion,''
J. Geom. Phys. \textbf{15} (1995) 189-214,
{\tt arXiv:hep-th/9409188}.

\bibitem{Thorngren:2021yso}
R.~Thorngren and Y.~Wang,
``Fusion category symmetry II: categoriosities at $c$ = 1 and beyond,''
{\tt arXiv:2106.12577}.

\bibitem{lang} S. Lang, {\it Algebra}, third edition,
Addison-Wesley, Reading, Massachusetts, 1993.

\bibitem{cr81} C. Curtis, I. Reiner,
{\it Methods of representation theory}, volume 1,
John Wiley \& Sons, New York, 1981.


\bibitem{atmac}
M. F. Atiyah, I. G. MacDonald,
{\it Introduction to commutative algebra},
Addison-Wesley, Reading, Massachusetts, 1969.


\bibitem{hirano} Y. Hirano, ``Multiplicative sets of idempotents in
a semilocal ring,''
available at
{\tt http://noetherian.net/conference/Proceedings/msi2.pdf}

\bibitem{lam} T.-Y. Lam, ``Local rings, semilocal rings, and idempotents,''
pp. 293-343 in {\it A first course in noncommutative rings},
Graduate Texts in Math. 131, Springer, New York, 1991.

\bibitem{gks} H. Grover, D. Khurana, S. Singh,
``Rings with multiplicative sets of primitive idempotents,''
Comm. in Alg. 37(8) (2009) 2583-2590.






\bibitem{tHooft:1977nqb}
G.~'t Hooft,
``On the phase transition towards permanent quark confinement,''
Nucl. Phys. B \textbf{138} (1978) 1-25.





\bibitem{ty} D. Tambara, S. Yamagami, ``Tensor categories with fusion rules
of self-duality for finite abelian groups,''
J. Algebra {\bf 209} (1998) 692-707.

\bibitem{fh} W. Fulton, J. Harris,
{\it Representation theory:  a first course},
Springer-Verlag, New York, 1991. 




\bibitem{Hori:2000kt}
K.~Hori and C.~Vafa,
``Mirror symmetry,''
{\tt arXiv:hep-th/0002222}.

\bibitem{Sharpe:2014tca}
E.~Sharpe,
``Decomposition in diverse dimensions,''
Phys. Rev. D \textbf{90} (2014)  025030,
{\tt arXiv:1404.3986}.





\bibitem{Gu:2018fpm}
W.~Gu and E.~Sharpe,
``A proposal for nonabelian mirrors,''
{\tt arXiv:1806.04678}.

\bibitem{Chen:2018wep}
Z.~Chen, W.~Gu, H.~Parsian and E.~Sharpe,
``Two-dimensional supersymmetric gauge theories with exceptional gauge groups,''
Adv. Theor. Math. Phys. \textbf{24} (2020)  67-123,
{\tt arXiv:1808.04070}.

\bibitem{Gu:2019zkw}
W.~Gu, H.~Parsian and E.~Sharpe,
``More non-Abelian mirrors and some two-dimensional dualities,''
Int. J. Mod. Phys. A \textbf{34} (2019)  1950181,
{\tt arXiv:1907.06647}.

\bibitem{Gu:2019byn}
W.~Gu, J.~Guo and E.~Sharpe,
``A proposal for nonabelian (0,2) mirrors,''
{\tt arXiv:1908.06036}.


\bibitem{Gu:2020ivl}
W.~Gu, E.~Sharpe and H.~Zou,
``Notes on two-dimensional pure supersymmetric gauge theories,''
JHEP \textbf{04} (2021) 261,
{\tt arXiv:2005.10845}.

\bibitem{Gu:2020ana}
W.~Gu, J.~Guo and Y.~Wen,
``Nonabelian mirrors and Gromov-Witten invariants,''
{\tt arXiv:2012.04664}.


\bibitem{cr} C. Curtis, I. Reiner,
{\it Representation theory of finite groups and associative algebras},
John Wiley \& Sons, New York, 1962.

\bibitem{collins} M. J. Collins,
{\it Representations and characters of finite groups},
Cambridge University Press, Cambridge, 1990.


\bibitem{georgi} H. Georgi, {\it Lie algebras in particle physics},
second edition, Perseus Books, Reading, Massachusetts, 1999.

\bibitem{Gross:1992tu}
D.~J.~Gross,
``Two-dimensional QCD as a string theory,''
Nucl. Phys. B \textbf{400} (1993) 161-180,
{\tt arXiv:hep-th/9212149}.


\bibitem{Hellerman:2010fv}
S.~Hellerman and E.~Sharpe,
``Sums over topological sectors and quantization of Fayet-Iliopoulos parameters,''
Adv. Theor. Math. Phys. \textbf{15} (2011) 1141-1199,
{\tt arXiv:1012.5999 [hep-th]}.





\end{thebibliography}
\end{document}